\documentclass[floatfix,aps,prd,superscriptaddress,preprint]{revtex4-1}

\usepackage{float}
\usepackage{amsmath,amssymb,bm,bbm}
\usepackage{amsfonts}
\usepackage{graphicx,graphics,color,xcolor,epsfig}
\usepackage[colorlinks=true,linkcolor=blue,urlcolor=blue,citecolor=blue]{hyperref}
\usepackage{soul}
\usepackage{dsfont}
\usepackage{slashed}
\usepackage{mathtools}
\usepackage{mathrsfs}
\usepackage{feynmp-auto}
\usepackage{tikz}
\usepackage[utf8]{inputenc}
\usepackage[normalem]{ulem}
\usepackage{hhline, multirow,tabularx,makecell}
\usepackage{dcolumn}
\usepackage{url}
\newcommand{\lna}{\mbox{\tiny LNA}}

\begin{document}
\preprint{JLAB-THY-20-3300}

\title{Octet and decuplet baryon self-energies \\ in relativistic SU(3) chiral effective theory}

\author{P. M. Copeland}
\affiliation{\mbox{Department of Physics and Astronomy, Clemson University, Clemson, SC 29634, USA}}
\author{Chueng-Ryong Ji}
\affiliation{\mbox{Department of Physics, North Carolina State University, Raleigh, NC 27607, USA}}
\author{W. Melnitchouk}
\affiliation{Jefferson Lab, Newport News, VA 23606, USA}

\begin{abstract}
The self-energies of the full set of flavor SU(3) octet and decuplet baryons are computed within a relativistic chiral effective theory framework.
The leading nonanalytic chiral behavior is derived for the octet and decuplet masses, and a finite-range regularization consistent with Lorentz and gauge invariance is applied to account for the finite size of the baryons.
Using a four-dimensional dipole form factor, the relative importance of various meson-baryon loop contributions to the self-energies is studied numerically as a function of the dipole range parameter and meson mass, and comparison is made between the relativistic results and earlier approximations within the heavy baryon limit.
\end{abstract}

\date{\today} 
\maketitle

\section{Introduction}
\label{eq:intro}

Understanding the structure and interactions of atomic nuclei and their constituents from the fundamental theory of Quantum Chromodynamics (QCD) poses one of the greatest challenges of modern subatomic physics.
In the nearly 50 years since the formulation of QCD, significant progress has been made in describing the high-energy behavior of hadronic cross sections in terms of quark and gluon degrees of freedom, using the tools of perturbation theory to expand around the small value of the QCD coupling at short distances.
In the low-energy realm, however, where the coupling becomes large and these tools are no longer applicable, other methods must be sought to provide approximate solutions.

The most common approach for describing low-energy hadron structure has been the use of effective field theories, in which ``effective'' hadronic degrees of freedom are typically used, constrained by the known symmetries of QCD which any such approximate theory must respect. 
Along with Lorentz invariance and electromagnetic gauge invariance, one of the most crucial symmetries for understanding the dynamics of hadrons and nuclei at low energies is chiral symmetry.
In particular, the spontaneous breaking of chiral symmetry leads to the appearance of nearly massless pseudoscalar Goldstone bosons, which are identified in nature with pions and kaons.
Here effective chiral theories have been constructed, in which the pseudoscalar mesons play a fundamental role, and calculations can be performed based on expansions of observables in powers of (low) momenta or the pion mass relative to the nucleon mass.

The applications of chiral symmetry and its breaking for hadron and nuclear structure are too numerous to list (see, {\it e.g.}, Refs.~\cite{Ericson:1988gk, Thomas:1982kv, Young:2003} for overviews).
One of the important areas where this has received attention in recent years is in the first principles calculation of hadron properties, such as masses, in lattice QCD~\cite{Aoki:2008sm, Bietenholz:2011qq, Durr:2008zz, Alexandrou:2017xwd}.
Considerable progress has been made in pushing lattice calculations closer to the physical region, in terms of the lattice spacing, lattice volume, and quark mass.
Some simulations are now routinely performed at the physical quark mass, $m_q^{\rm phys} \sim m_\pi^2 \approx (140~{\rm MeV})^2$, although extrapolations to the continuum and infinite volume limits still need to be applied.

The role of meson loops in the analysis of lattice data on hadron masses has been stressed by many authors.
In particular, the behavior of baryon masses near the chiral limit, $m_q \to 0$, is known to deviate strongly from the linear $M_{\rm baryon} \sim m_q$ dependence expected at large $m_q$ values.
Expanding the masses in terms of powers of $m_\pi$, the low-$m_\pi$ behavior is characterized by model-independent nonanalytic terms that involve odd powers of $m_\pi$ or logarithms of $m_\pi$~\cite{Li:1971vr}.
Such behavior can only arise from pseudoscalar meson loops, and must be present in any effective treatment of QCD near the chiral limit~\cite{Thomas:1999}.

Although the physics implications of the chiral loops are relatively clear~\cite{Thomas:2002sj}, in the literature various methods have been used to implement them in practical calculations.
In particular, while the long-range structure and dynamics of baryons is characterized by model-independent features of pion loops, the short-range structure depends on nonperturbative dynamics and details of how the ultraviolet behavior of loops is regularized.
Historically, a popular approach has utilized the framework of chiral perturbation theory~\cite{Gasser:1987rb, Becher:1999he, Fuchs:2003qc, Gegelia:1999qt, Pascalutsa:2005nd}, with dimensional regularization to regularize divergences and parametrize the short-distance physics via counter terms. 
Other approaches have emphasized the importance of taking the finite size of baryons into account, regularizing the ultraviolet behavior via form factors or finite-range regulators~\cite{Donoghue:1998bs}.
The latter have been argued to lead to better convergence of the chiral expansion, through a resummation of nominally higher-order terms as relativistic corrections to the leading nonanalytic (LNA) terms~\cite{Young:2003}.

Regardless of the specific approach to the regularization, most of the early efforts have relied on the heavy baryon or nonrelativistic approximations~\cite{Jenkins:1991ts, Leinweber:1999ig, Young:2003}, focusing primarily on the properties of the nucleon, often emphasizing the important role played by the $\Delta$ resonance~\cite{Theberge:1980ye}.
More recently, manifestly covariant formulations of chiral effective theory have been developed, partly in an effort to further improve the convergence properties of the heavy-baryon approaches~\cite{Pascalutsa:2005nd}.

Extensions to baryons other than the nucleon have been made in a number of studies, both in the context of the heavy baryon expansion with finite-range regularization~\cite{Young:2010, Shanahan:2011, Shanahan:2012wh}, as well as in covariant calculations with dimensional regularization~\cite{Banerjee:1994bk, Camalich:2010, Ren:2014}.
Most recently, relativistic loop corrections to masses arising from octet and decuplet intermediate states were computed using a Gaussian form factor~\cite{Ghaderi:2018}, however, only nucleon external states were considered.

In other recent applications, careful treatment of chiral loops has been stressed in connection with the meson--baryon splitting functions needed for the calculation of parton distribution functions~\cite{Burkardt:2012hk, Salamu:2014pka, Wang:2016eoq, Wang:2016ndh, Salamu:2018cny, Salamu:2019dok, Wang:2020hkn}, and in particular the key role played by light-front zero-mode contributions~\cite{Ji:2009jc, Alberg:2012wr, Ji:2012pv, Hecht:2002ej, Oettel:2002cw}.
Elsewhere, the effects of relativistic chiral corrections on electromagnetic form factors~\cite{He:2017viu, He:2018eyz} and transverse momentum distributions~\cite{He:2019fzn} have recently been studied.
Considerable interest has also been devoted to the application of chiral loops to scalar matrix elements and $\sigma$-terms~\cite{Pascalutsa:2005nd, Young:2010, Shanahan:2012wh, Bernard:1993nj} for the nucleon and other baryons, as well as for more exotic hadrons such as the H-dibaryon~\cite{Shanahan:2011}.

In this paper we build upon the previous work on chiral loops to compute the self-energies for the complete set of SU(3) octet and decuplet baryon intermediate and external states, using a fully relativistic chiral effective theory framework with a four-dimensional finite-range regulator.
We begin the discussion in Sec.~\ref{sec:foundations} with a brief review of the basics of the chiral effective theory, and the definitions of the baryon octet and decuplet self-energies.
In Sec.~\ref{sec:FRR-self-energies} we present the results for the self-energies using a four-dimensional dipole regulator, for each of the octet-octet, octet-decuplet, decuplet-octet and decuplet-decuplet transitions, and compare these with some nonrelativistic approximations.
The LNA behavior of the self-energies is derived in Sec.~\ref{sec: Prop. and app. of SEs}, along with the decay widths for channels in which the initial baryon mass is larger than the intermediate baryon mass plus the meson mass.  
A numerical comparison of the self-energies as a function of the regulator mass and the pion mass is presented in Sec.~\ref{sec: results} for all octet and decuplet baryons, along with a direct assessment of the role of relativistic effects.
Finally, in Sec.~\ref{sec: conclusions} we summarize our findings and outline future applications of the results obtained.

\section{Foundations}
\label{sec:foundations}

In this section we briefly summarize the basic elements of the chiral SU(3) effective theory, and introduce the formal definitions of the octet and decuplet baryon self-energies associated with the fluctuations into meson-baryon intermediate states.

\subsection{Chiral SU(3) effective theory}
\label{ssec: Chiral SU(3) effective theory}

The effective chiral SU(3)$_L$ $\times$ SU(3)$_R$ Lagrangian describing the interactions of octet ($B$) and decuplet ($T_\mu$) baryons with psuedoscalar mesons ($\phi$) can be written at leading order as~\cite{Jenkins:1991ts, Ledwig:2014rfa, Bernard:1995dp, Scherer:2012xha}
\begin{eqnarray}
\label{eq:L}
{\cal L}
&=& \frac{f_\phi^2}{4} {\rm Tr} \big[ D_\mu U (D^\mu U)^\dag \big]
 +  {\rm Tr} \big[ \bar B (i\slashed{D} - M_B) B \big]
 +  \big(\overline{T}_\mu\big)^{ijk}
    (i\gamma^{\mu \nu \alpha} D_\alpha - M_T \gamma^{\mu\nu}) 
    \big(T_\nu\big)^{ijk}
\notag\\
&-& \frac{D}{2}\, {\rm Tr} \big[ \bar B \gamma^\mu \gamma_5 \{u_\mu, B\} \big]
 -  \frac{F}{2}\, {\rm Tr} \big[ \bar B \gamma^\mu \gamma_5 [ u_\mu ,B ] \big]
\\
&-& \frac { {\cal C}}{2}
    \left[ \varepsilon^{ijk}\, \big(\overline{T}_\mu\big)^{ilm}\,
           \Theta^{\mu\nu}\, (u_\nu)^{lj} \big(B\big)^{mk} + {\rm h.c.}
    \right]
 -  \frac {\cal H}{2}\,
    \big(\overline{T}_\mu\big)^{ijk} \gamma^{\alpha} \gamma_5\, (u_\alpha)^{kl}\,
	      \big(T^\mu\big)^{ijl},
\notag
\end{eqnarray}
where $M_B$ and $M_T$ are the octet and decuplet baryon masses, and ``h.c'' indicates the Hermatian conjugate.
The coefficients of the various terms in ${\cal L}$ are the pseudoscalar decay constant, $f_\phi = 93$~MeV, the meson-octet baryon coupling constants, $D$ and $F$, and the meson-octet-decuplet and meson-decuplet coupling constants, $\cal C$ and $\cal H$, respectively.
The tensors in Eq.~(\ref{eq:L}) are defined as
    $\gamma^{\mu\nu} = \frac12 [\gamma^{\mu},\gamma^{\nu}]$,
and
    $\gamma^{\mu\nu\alpha} = \frac12 \{\gamma^{\mu\nu},\gamma^{\alpha}\}$
in terms of the Dirac $\gamma$ matrices, and $\varepsilon^{i j k}$ is the antisymmetric tensor in flavor space.

The flavor SU(3) baryon octet fields $B^{ij}$ are comprised of the nucleon $N$ ($= p,n$), $\Lambda$, $\Sigma^{\pm, 0}$ and $\Xi^{-,0}$ hyperon fields, and can be represented in the matrix form,
\begin{eqnarray}
\label{eq:B}
B =
\left(
\begin{array}{ccc}
  \frac{1}{\sqrt 2} \Sigma^0 + \frac{1}{\sqrt 6} \Lambda
& \Sigma^+
& p					\\
  \Sigma^-
&-\frac{1}{\sqrt 2} \Sigma^0 + \frac{1}{\sqrt 6} \Lambda
& n					\\
  \Xi^-
& \Xi^0
&-\frac{2}{\sqrt6} \Lambda
\end{array}
\right).
\end{eqnarray}
The baryon decuplet fields are parametrized in terms of the spin-3/2 Rarita-Schwinger field, and represented by the tensor $\big(T_\mu\big)^{ijk}$, which includes the $\Delta$-isobar, $\Sigma^*$, $\Xi^*$ and the triply-strange $\Omega^-$ fields,
\begin{eqnarray}
\label{eq:Ttensor}
T_\mu &=&
\left\{\!\!
\left(
\begin{array}{ccc}
\Delta^{++} & \frac{1}{\sqrt3}\Delta^+ & \frac{1}{\sqrt3}\Sigma^{*+} \\
\frac{1}{\sqrt3}\Delta^+ & \frac{1}{\sqrt3}\Delta^0 & \frac{1}{\sqrt6}\Sigma^{*0} \\
\frac{1}{\sqrt3}\Sigma^{*+} & \frac{1}{\sqrt6}\Sigma^{*0} & \frac{1}{\sqrt3}\Xi^{*0}\\
\end{array}
\right)\!,\!
\left(
\begin{array}{ccc}
\frac{1}{\sqrt3}\Delta^+ & \frac{1}{\sqrt3}\Delta^0 & \frac{1}{\sqrt6}\Sigma^{*0} \\
\frac{1}{\sqrt3}\Delta^0 & \Delta^- & \frac{1}{\sqrt3}\Sigma^{*-} \\
\frac{1}{\sqrt6}\Sigma^{*0} & \frac{1}{\sqrt3}\Sigma^{*-} & \frac{1}{\sqrt3}\Xi^{*-} \\
\end{array}
\right)\!,\!
\left(
\begin{array}{ccc}
\frac{1}{\sqrt3}\Sigma^{*+} & \frac{1}{\sqrt6}\Sigma^{*0} & \frac{1}{\sqrt3}\Xi^{*0} \\
\frac{1}{\sqrt6}\Sigma^{*0} & \frac{1}{\sqrt3}\Sigma^{*-} & \frac{1}{\sqrt3}\Xi^{*-} \\
\frac{1}{\sqrt3}\Xi^{*0} & \frac{1}{\sqrt3}\Xi^{*-} & \Omega^- \\
\end{array}
\right)\!\!
\right\}. \nonumber \\ 
&&
\end{eqnarray}
Note, however, that the tensor $\big(T_\mu\big)^{ijk}$ contains spurious spin-1/2 components, which must be removed by projecting onto spin 3/2.
This amounts to the replacement on the meson-decuplet-decuplet interaction term,
\begin{equation}
\label{eq:Lttphi}
-\frac {\cal H}{2}\,
 \big(\overline{T}_\mu\big)^{ijk} \gamma^\alpha \gamma_5\, (u_\alpha)^{kl}\, \big(T^\mu\big)^{ijl}\
\rightarrow\
-i\frac {\cal H}{2}\,
 \big(\overline{T}_\mu\big)^{ijk} \epsilon^{\mu\nu\alpha\beta} \gamma_\beta\, (u_\alpha)^{kl}\, \big(T_\nu\big)^{ijl}.
\end{equation}
The mesonic operator $U$ is defined in terms of the matrix of pseudoscalar fields $\phi$,
\begin{equation}
\label{eq:Udef}
    U = u^2, \ \ \ \text{with} \ \ u = \exp\bigg(i \frac{\phi}{\sqrt2 f_\phi} \bigg),
\end{equation}
where the meson field $\phi$ includes the isotriplet $\pi$, isospin-1/2 $K$ and isosinglet $\eta$ mesons,
\begin{eqnarray}
\label{eq:phi}
\phi =
\left(
{\begin{array}{*{20}{c}}
  \frac{1}{\sqrt 2} \pi^0 + \frac{1}{\sqrt 6}\, \eta
& \pi^+
& K^+						\\
  \pi^-
& -\frac{1}{\sqrt 2} \pi^0 + \frac{1}{\sqrt 6}\, \eta
& K^0						\\
  K^-
& \overline{K}^0
& -\frac{\sqrt2}{\sqrt3}\, \eta
\end{array}}
\right).
\end{eqnarray}

The first three terms in the Lagrangian in Eq.~(\ref{eq:L}) represent the meson, octet baryon and decuplet baryon free-field Lagrangians, while the terms proportional to the couplings $D$, $F$, $\cal C$ and $\cal H$ involve interactions between fields.
In particular, for the octet-decuplet transition, the tensor $\Theta^{\mu\nu}$ is defined as
\begin{equation}
\Theta^{\mu\nu}
= g^{\mu\nu} - \big( Z+\tfrac12 \big) \gamma^\mu \gamma^\nu,
\label{eq:Theta}
\end{equation}
where $Z$ is the decuplet off-shell parameter that gives the relative strength of the two terms in Eq.~(\ref{eq:Theta}).
Note that observables, such as masses and cross sections, do not depend on the choice of $Z$, but specific choices of $Z$ may simplify the calculations.
In this analysis we follow the conventional choice and set $Z=\frac12$~\cite{Scherer:2012xha}.

The psuedoscalar mesons couple to the baryon fields via the vector and axial vector combinations defined by
\begin{eqnarray}
\Gamma_\mu 
&=& \frac12(u^\dagger \partial_\mu u + u \partial_\mu u^\dagger)
 - \frac{i}{2} (u^\dagger \lambda^\alpha u + u \lambda^\alpha u^\dagger)\, v^\alpha_\mu,      \\
u_\mu
&=& i(u^\dagger \partial_\mu u - u \partial_\mu u^\dagger) + (u^\dagger \lambda^\alpha u - u \lambda^\alpha u^\dagger)\, v^\alpha_\mu,
\label{eq:u_mu}
\end{eqnarray}
where $v_\mu^\alpha$ is an external vector field, $\lambda^\alpha$ ($\alpha = 1, \ldots, 8$) are the SU(3) Gell-Mann matrices, and $u$ is defined in Eq.~(\ref{eq:Udef}).
The covariant derivatives $D_\mu$ of the octet and decuplet fields in Eq.~(\ref{eq:L}) are defined as
\begin{eqnarray}
D_\mu B^{ij}
&=& \partial_\mu B^{ij}
 + [\Gamma_\mu, B]^{ij}
 - i\langle\lambda^0\rangle\, v_\mu^0\, B^{ij},     \\
D_\mu \big(T_\nu\big)^{ijk} 
&=& \partial_\mu \big(T_\nu\big)^{ijk} 
 + \big(\Gamma_\mu, T_\nu\big)^{ijk} 
 - i\langle\lambda^0\rangle\, v_\mu^0\, \big(T_\nu\big)^{ijk}, 
\label{eq:DmuT}
\end{eqnarray}
where $v_\mu^0$ denotes an external singlet vector field, $\lambda^0$ is the unit matrix, and $\langle ... \rangle$ indicates a trace in flavor space. 
The second term on the right hand side in Eq.~(\ref{eq:DmuT}) denotes the combination
\begin{equation}
(\Gamma_\mu, T_\nu)^{ijk} 
= \big(\Gamma_\mu\big)_l^i\, \big(T_\nu\big)^{ljk}
+ \big(\Gamma_\mu\big)_l^j\, \big(T_\nu\big)^{ilk}
+ \big(\Gamma_\mu\big)_l^k\, \big(T_\nu\big)^{ijl}.
\end{equation}
Finally, the covariant derivative on the psuedoscalar meson fields is given by
\begin{equation}
D_\mu U = \partial_\mu U + (i U\lambda^\alpha - i\lambda^\alpha U)\, v_\mu^\alpha.
\label{eq:DmuU}
\end{equation}
Using Eqs.~(\ref{eq:B})--(\ref{eq:DmuU}), one can expand the chiral Lagrangian (\ref{eq:L}) to leading order in the baryon and meson fields and derive the appropriate set of Feynman rules needed for the calculation of the self-energies of the SU(3) octet and decuplet baryons, as we discuss in the following.

\subsection{Baryon self-energies}
\label{ssec: self-energy operators}

\begin{figure}[t]
\includegraphics[width=0.75\linewidth]{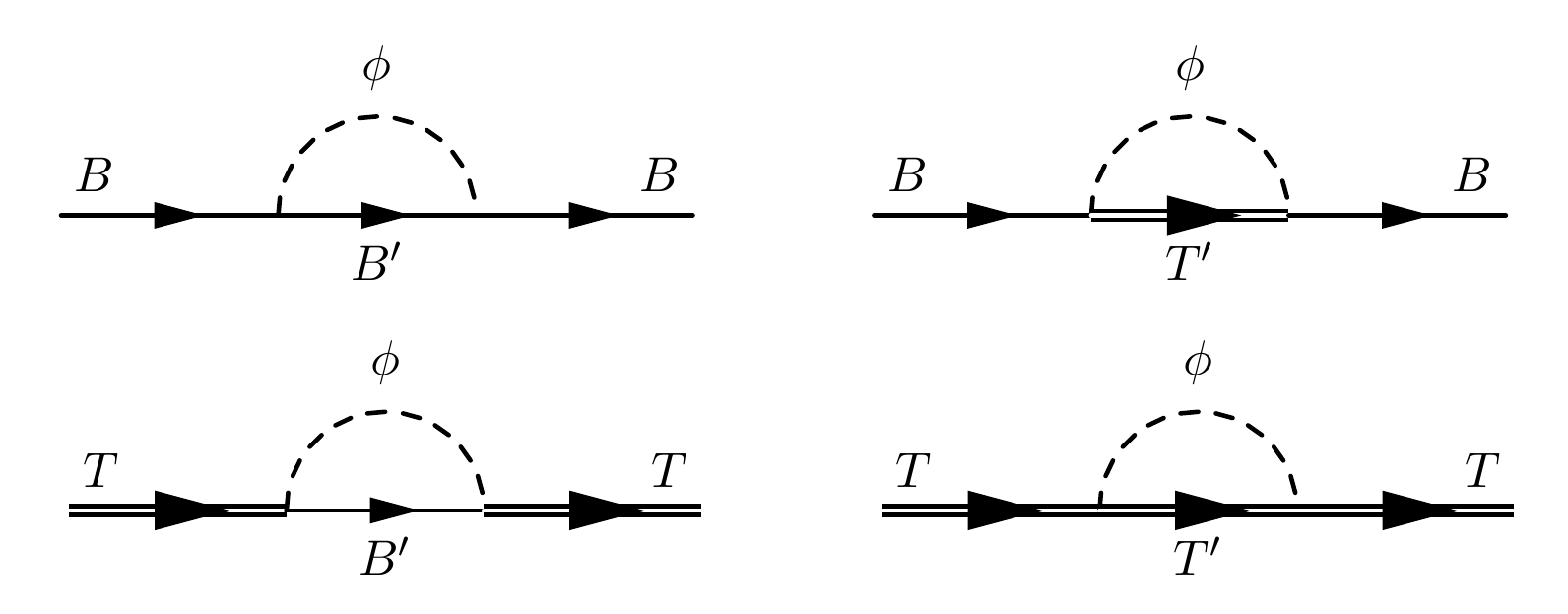}\hspace*{-0.3cm}
\caption{Pseudoscalar meson $\phi$ loop contributions to the self-energies of octet baryons $B$ and decuplet baryons $T$ from loops involving intermediate octet baryons $B'$ and decuplet baryons $T'$.}
\label{fig:loops}
\end{figure}

In this section we introduce the self-energies of the SU(3) octet and decuplet baryons arising from pseudoscalar meson loops, as illustrated in Fig.~\ref{fig:loops}, focusing firstly on octet external states and then on decuplet external states. 

\subsubsection{Octet external states}

The contribution to the self-energy of an octet baryon $B$ (with four-momentum $p$) from the process involving the emission of virtual meson $\phi$ (four-momentum $k$) with an octet baryon $B'$ (four-momentum $p-k$) in the intermediate state is defined by taking the on-shell matrix elements of the $B \to B'\phi$ transition operator $\widehat{\Sigma}_{B \to B'\phi}$,
\begin{equation}
\begin{aligned}
\label{eq: Def. of octet octet SE}
\Sigma_{B \to B'\phi}
&= \frac12 \sum_s \bar{u}_{B}(p,s)\,
   \widehat{\Sigma}_{B \to B'\phi}\, u_{B}(p,s)
 = \frac{1}{4M_B}
   {\rm Tr}\left[ (\slashed{p}+M_B)\, \widehat{\Sigma}_{B \to B'\phi} \right],
\end{aligned}
\end{equation}
where the sum is taken over the spins $s$ of the external octet baryon state, and the Dirac spinor $u_{B}(p,s)$ is normalized such that
$\bar{u}_B(p,s)\, u_B(p,s') = \delta_{ss'}$.
From the terms in the Lagrangian in Eq.~(\ref{eq:L}) involving the couplings $D$ and $F$, the self-energy operator for the octet-octet transition is given by
\begin{equation}
\label{eq:SigBBhat}
\begin{aligned}
\widehat{\Sigma}_{B\to B'\phi}
&= i \bigg(\frac{C_{BB'\phi}}{f_\phi}\bigg)^2
  \int\!\frac{d^4k}{(2\pi)^4}
  \slashed{k} \gamma_5 
  \frac{i(\slashed{p} - \slashed{k} + M_{B'})}{D_{B'}} \gamma_5\slashed{k}
  \frac{i}{D_\phi},
\end{aligned}
\end{equation}
where $C_{BB'\phi}$ is the $BB'\phi$ coupling constant, and we define the meson and octet baryon propagators as
\begin{eqnarray}
\label{eq:propagatorphi}
D_\phi &=& k^2 - m_\phi^2 +i\epsilon,        \\
\label{eq:propagatorB}
D_{B'} &=& (p-k)^2 - M_{B'}^2 + i\epsilon,
\end{eqnarray}
with $m_\phi$ and $M_{B'}$ the masses of the meson and intermediate state octet baryon.
The coupling constants $C_{BB'\phi}$ depend on the couplings $D$ and $F$, and for specific transitions $B \to B' \phi$ are given in Appendix~\ref{app:couplings}.

Similarly, for an octet baryon dressed by a meson loop with a decuplet baryon $T'$ in the intermediate state, the contribution to the self-energy is given by the matrix element of the $B \to T' \phi$ transition operator $\widehat{\Sigma}_{B \to T' \phi}$,
\begin{equation}
\begin{aligned}
\label{eq:DefOctDecSE}
\Sigma_{B \to T' \phi}
= \frac12 \sum_s \bar{u}_B(p,s)\, \widehat{\Sigma}_{B \to T'\phi}\, u_B(p,s)
= \frac{1}{4M_B}
   {\rm Tr}\left[ (\slashed{p} + M_B)\, \widehat{\Sigma}_{B \to T'\phi} \right].
\end{aligned}
\end{equation}
Again, from the terms in the Lagrangian (\ref{eq:L}) involving the coupling $\cal C$, one derives the transition operator
\begin{equation}
\label{eq:SigBThat}
\begin{aligned}
\widehat{\Sigma}_{B \to T' \phi}
&= i \bigg(\frac{C_{BT'\phi}}{f_\phi}\bigg)^2
  \int\!\frac{d^4k}{(2\pi)^4}\,
  \overline{\Theta}^{\mu\nu} k_\nu 
  \frac{-i (\slashed{p} - \slashed{k} + M_{T'})
  \Lambda_{\mu\lambda}(p-k)}{D_{T'}}\, \Theta^{\lambda\sigma} k_\sigma\, \frac{i}{D_\phi},
\end{aligned}
\end{equation}
where the $BT'\phi$ coupling constant $C_{BT'\phi}$ depends on the coefficient $\cal C$, and is given in Appendix~\ref{app:couplings}.
Similarly to Eq.~(\ref{eq:propagatorB}), we define the propagator of a decuplet baryon as
\begin{equation}
\label{eq:propagatorT}
D_{T'} = (p-k)^2 - M_{T'}^2 + i \epsilon .
\end{equation}
Setting the off-shell parameter $Z = \frac12$ in the octet-decuplet transition operator $\Theta^{\mu\nu}$, the spin-3/2 energy projector $\Lambda^{\alpha\beta}$ can be written as
\begin{equation}
\label{eq:3/2projector}
\Lambda_{\alpha\beta}(p)
= g_{\alpha\beta}
- \frac13 \gamma_\alpha \gamma_\beta
- \frac{\gamma_\alpha p_\beta - \gamma_\beta p_\alpha}{3 M_{T'}}
- \frac{2 p_\alpha p_\beta}{3 M_{T'}^2}.
\end{equation}
Note that choices of $Z$ other than $Z=\frac12$ would introduce additional $Z$ dependence into the projector $\Lambda_{\alpha\beta}$.

\subsubsection{Decuplet external states}

Extending the discussion to decuplet external states, $T$, the contribution to the self-energy from intermediate states with octet baryons, $B'$, is defined in terms of the Rarita-Schwinger spin-3/2 spinor-tensor $u^T_\mu(p,s)$,
\begin{equation}
\label{eq:decupletoctettrace}
\Sigma_{T \to B' \phi} 
= \frac14 \sum_s 
  \bar{u}^T_\mu(p,s)\, \widehat{\Sigma}_{T \to B'\phi}^{\mu\nu}\, u^T_\nu(p,s)
= -\frac{1}{8M_T}
  {\rm Tr} \left[ (\slashed{p}+M_T)\, \Lambda_{\nu\mu}(p)\,
                  \widehat{\Sigma}_{T \to B'\phi}^{\mu\nu} 
           \right].
\end{equation}
The spinor-tensor $u^T_\mu(p,s)$ is normalized such that
\begin{equation}
\sum_s \bar{u}^{T}_{\mu}(p,s)\, u^{T}_{\nu}(p,s) 
= - \frac{4}{3} \Big( g_{\mu\nu} - \frac{p_\mu p_\nu}{M_T^2} \Big),
\end{equation}
and the energy projector $\Lambda_{\nu\mu}$ is given in Eq.~(\ref{eq:3/2projector}).
Similarly to Eq.~(\ref{eq:SigBThat}), the $T \to B' \phi$ self-energy operator is given by
\begin{equation}
\label{eq:SigTBhat}
\begin{aligned}
\widehat{\Sigma}_{T\to B'\phi}^{\mu\nu}
&= i \bigg( \frac{C_{TB'\phi}}{f_\phi} \bigg)^2
   \int\!\frac{d^4k}{(2\pi)^4}
   \overline{\Theta}^{\mu\alpha} k_\alpha
   \frac{i (\slashed{p} - \slashed{k} + M_{B'})}{D_{B'}}
   \Theta^{\nu\beta} k_\beta
   \frac{i}{D_\phi},
\end{aligned}
\end{equation}
where $C_{TB'\phi}$ is the $T B' \phi$ coupling, given for specific transitions in Appendix~\ref{app:couplings}.

Finally, for the decuplet--decuplet transition $T \to T'\phi$, the self-energy contribution can be written as
\begin{eqnarray}
\label{eq: decupletdecuplet trace}
\Sigma_{T \to T' \phi}
= \frac14 \sum_s 
  \bar{u}^T_\mu(p,s)\, \widehat{\Sigma}_{T \to T'\phi}^{\mu\nu}\, u^T_\nu(p,s)
&=& -\frac{1}{8M_T}
    {\rm Tr} \left[ (\slashed{p}+M_T)\, \Lambda_{\nu\mu}(p)\,
                    \widehat{\Sigma}_{T \to T' \phi}^{\mu\nu}
             \right],
\end{eqnarray}
where the relevant self-energy operator is given by
\begin{equation}
\label{eq:SigTThat}
\widehat{\Sigma}_{T\to T'\phi}^{\mu\nu}
= i \bigg( \frac{C_{TT'\phi}}{f_\phi} \bigg)^2
  \int\!\frac{d^4k}{(2\pi)^4}
  \epsilon^{\mu\sigma\alpha\beta} \gamma_\beta k_\alpha
  \frac{-i (\slashed{p} - \slashed{k} + M_{T'})\Lambda_{\sigma\lambda}(p-k)}{D_{T'}} 
   \epsilon^{\lambda\nu\rho\delta} \gamma_\delta k_\rho 
  \frac{i}{D_\phi},
\end{equation}
with $C_{TT'\phi}$ the corresponding $T T' \phi$ coupling, and $\epsilon^{\mu \sigma \alpha \beta}$ the Levi-Civita tensor.

The calculation of the baryon self-energies is in principle straightforward, but simple power counting shows that the integrals over the loop momentum $k$ in the self-energy operators in Eqs.~(\ref{eq:SigBBhat}), (\ref{eq:SigBThat}), (\ref{eq:SigTBhat}) and (\ref{eq:SigTThat}) are divergent, and therefore need to be regularized.
In the next section we discuss the computation of the self-energies using finite-range regularization.

\newpage
\section{Self-energies with finite-range regularization}
\label{sec:FRR-self-energies}

As outlined in Sec.~\ref{eq:intro}, various regularization prescriptions have been discussed in the literature in calculations of baryon self-energies.
An important consistency requirement is that the regularization procedure preserves the Lorentz and gauge symmetry of the fundamental QCD theory.
This is satisfied by the commonly used dimensional regularization; however, for applications to particles with finite size, finite-range regularization has been argued to have some advantages regarding the convergence properties of the integrals~\cite{Donoghue:1998bs, Thomas:2002sj}.

Finite-range schemes such as Pauli-Villars regularization satisfy all of the symmetry requirements, and are a special case of four-dimensional form factors applied to the integrands of point-like results.
Lorentz invariance restricts form factors to be functions of the meson virtuality $k^2$ and baryon virtuality $(p-k)^2$.
Following earlier work~\cite{Forkel:1994yx, Musolf:1993fu, Melnitchouk:1991ui, Salamu:2018cny}, in the present analysis we apply a four-dimensional form factor that is a function of $k^2$ only.
In particular, we employ a four-dimensional dipole shape function $F(k, \Lambda)$ with a regulator mass $\Lambda$,
\begin{equation}
\label{eq:dipoleFF}
    F(k, \Lambda) = \bigg(\frac{\widetilde{\Lambda}^2}{D_\Lambda}\bigg)^2,
\end{equation}
where $\widetilde{\Lambda}^2 = \Lambda^2 - m_\phi^2$, and we define, in analogy with Eq.~(\ref{eq:propagatorphi}), 
\begin{equation}
D_\Lambda = k^2 - \Lambda^2 + i\epsilon.
\end{equation}
The form (\ref{eq:dipoleFF}) respects the necessary symmetries of the calculation, and suppresses the divergences in the self-energy integrals.

In the calculations, it will also be convenient to use light-front coordinates, in which a four-vector $v^\mu = (v^+, v^-, {\bm v}_\perp)$ is written in terms of the ``longitudinal'' $v^\pm = v_0 \pm v_z$ components and the transverse component $\bm{v}_\perp^2 = v_x^2 + v_y^2$.
For convenience we define the light-front momentum fraction of the initial state baryon carried by the meson $\phi$ by $y = k^+/p^+$, with a corresponding momentum fraction $\bar y \equiv 1-y = (p^+ - k^+)/p^+$ carried by the intermediate state baryon.
Also, without loss of generality, we choose a frame in which $\bm{p}_\perp = 0$.

In the following, we discuss the evaluation of the self-energies in detail.
We pay particular attention to ensuring that the four-dimensional integrations correctly take into account the end-point contributions~\cite{Salamu:2014pka}, which are associated with $\delta$-function terms in the variable $y$ and can affect the model-independent leading nonanalytic behavior (see Sec.~\ref{ssec:LNA}).
We describe in detail how this is achieved by reducing the integrands to forms where the momentum dependence is contained mostly in the propagator factors with minimal momentum dependence in the numerators.

\subsection{Octet $\to$ octet transitions}
\label{Octet-octet}

We begin with the simplest case of the contribution to the self-energy of an octet baryon $B$ from intermediate states with an octet baryon $B'$ and meson $\phi$.
Substituting the dipole form factor $F(k, \Lambda)$ in Eq.~(\ref{eq:dipoleFF}) into Eq.~(\ref{eq:SigBBhat}) and taking the spin trace, the self-energy can be written as
\begin{equation}
\label{eq:SigBB'int}
\Sigma_{B\to B'\phi}
= -i\bigg( \frac{C_{BB'\phi}}{f_\phi} \bigg)^2
\frac{1}{2M_B}
\int\!\frac{d^4k}{(2\pi)^4} 
\bigg( \frac{\widetilde\Lambda^4}{D_\Lambda^2}\bigg)^{\!2} 
\frac{\big[ 2 M_B \overline{M}_{\!BB'}\, k^2 + 2 p\cdot k\, \big(k^2 - 2 p\cdot k\big)\big]}
     {D_{B'} D_\phi},
\end{equation}
where we introduce the shorthand notation
\begin{eqnarray}
\label{eq:MBB'}
\overline{M}_{\cal BB'} &\equiv M_{\cal B} + M_{\cal B'}, \\
\label{eq:DelBB'}
\Delta_{\cal B'B} &\equiv M_{\cal B'} - M_{\cal B},
\end{eqnarray}
for a generic baryon ${\cal B} = B$ or $T$ (for decuplet states, see below).
Rearranging the propagators in Eqs.~(\ref{eq:propagatorphi}) and (\ref{eq:propagatorB}), we can make the substitutions in the numerator of Eq.~(\ref{eq:SigBB'int}), 
\begin{eqnarray}
\begin{aligned}
\label{eq:k2sub}
k^2\ & \to\ D_\phi + m_\phi^2\ \ \ \ \text{or}\ \ \
            D_\Lambda + \Lambda^2,                  \\
p \cdot k\ 
& \to\  \frac12 \big( D_\phi - D_{B'} + M_B^2 - M_{B'}^2 + m_\phi^2 \big).
\end{aligned}
\end{eqnarray}
With these replacements, the $k$-dependent terms in the numerator can then be reduced to 
\begin{equation}
\begin{aligned}
\label{eq:OctetOctetreuced}
\Sigma_{B\to B'\phi}
=& -\!i \bigg(\frac{C_{BB'\phi}}{f_\phi}\bigg)^{\!2}
   \frac{\widetilde\Lambda^8}{2M_B}\!
   \int\!\!\frac{d^4k}{(2\pi)^4}
   \bigg[ \frac{\big( m_\phi^2 - \Delta_{B'B}^2\big) \overline{M}_{\!BB'}^2}
                {D_\phi D_{B'} D_\Lambda^4} 
 + \frac{\overline{M}_{\!BB'}^2}{D_{B'} D_\Lambda^4}
 + \frac{2 p\cdot k - \overline{M}_{\!BB'} \Delta_{B'B}}
        {D_\phi D_\Lambda^4}
   \bigg].
\end{aligned}
\end{equation}
Note that the first two terms in the brackets of Eq.~(\ref{eq:OctetOctetreuced}) have poles in different half-planes, so that using Cauchy's integral formula one can choose a contour in either the upper or lower half-plane to perform the $k^-$ integration analytically.
As shown in Appendix~\ref{app:integrations}, taking the pole in the baryon propagator 
allows one to evaluate the the first two terms in Eq.~(\ref{eq:OctetOctetreuced}).

For the third term in the brackets of Eq.~(\ref{eq:OctetOctetreuced}), the first part involving $(p\cdot k)$ in the numerator is odd in the pion momentum $k$, and since the four-dimensional Lorentz invariant regulator (\ref{eq:dipoleFF}) does not introduce any additional dependence on $p\cdot k$, this will integrate to zero.
For the second part of the term involving constants and propagators the integral vanishes when $k^+ \neq 0$, since here both the $D_\phi$ and $D_\Lambda$ poles lie on the same half-plane .
When $k^+ = 0$, however, the integral is divergent and the integration must be handled more carefully. This is also outlined in Appendix~\ref{app:integrations}.
Putting all the terms in (\ref{eq:term1}), (\ref{eq:term2}) and (\ref{eq:DphiDLambdaSimp}) together, the $k^-$ integrated expression for the octet-octet self-energy can be written as
\begin{eqnarray}
\label{eq:octet-octetk-}
\Sigma_{B\to B'\phi}
&=& \frac{ C_{BB'\phi}^2}{(4\pi f_\phi)^2}
\frac{\widetilde\Lambda^8\, \overline{M}_{BB'}^2}{2 M_B}
\int_0^1 dy \int_0^\infty\!dk_\perp^2       \nonumber\\
& & \times
\bigg[
    \frac{\bar{y}^4 \big(m_\phi^2 - \Delta_{B'B}^2\big)}{D_{B \phi B'} D_{B\Lambda B'}^4}
  - \frac{\bar{y}^3}{D_{B\Lambda B'}^4}
  + \frac{\Delta_{B'B}}{\overline{M}_{BB'}}
    \int_0^1 dz\, \frac{z^3}{(k_\perp^2+\Omega)^4}\,
    \delta(y)\bigg],
\end{eqnarray}
where the factors $D_{B \phi B'}$ and $D_{B \Lambda B'}$ are the $\phi$ and $\Lambda$ propagators, respectively, taken at the $D_{B'} = 0$ pole and are defined in Eq.~(\ref{eq:props at baryon pole}), and $\Omega$ is defined in Eq.~(\ref{eq:Omega}).

The $dy$ and $dk_\perp^2$ integrations in (\ref{eq:octet-octetk-}) can be performed analytically, although the resulting expressions are rather long and not particularly illuminating, so will not be listed here.
However, the special case of $M_B = M_{B'}$ is interesting since it corresponds to the well known $N \pi$ self-energy of the nucleon. Taking $B=B'=N$, $\phi = \pi$, and $M_B =  M$ to be the mass of the nucleon in Eq.~(\ref{eq:octet-octetk-}), the $N \pi$ loop contribution to the nucleon self-energy is given by the simplified expression,
\begin{eqnarray}
\label{eq:NpiN}
\Sigma_{N\to N\pi}
&=& -\frac{ C_{NN\pi}^2}{8\pi^2f_\pi^2}\,
\big( \widetilde\Lambda^8 M \big)\!
\int_0^1 dy \int_0^\infty\!dk_\perp^2\,
    \frac{\bar{y}^3 \big( k_\perp^2 + M^2 y^2 \big)}
         {D_{N \pi N} D_{N\Lambda N}^4}.
\end{eqnarray}
For a proton external state, summing over the various intermediate nucleon charge states ($p, n$) and using Table~\ref{tab:octet} of Appendix~\ref{app:couplings}, we have for the $NN\pi$ coupling constant
$C_{NN\pi}^2 = C_{pp\pi^0}^2 + C_{pn\pi^+}^2
= \frac34\, (D+F)^2 = \frac34\, g_A^2.$
Evaluating the integrals in Eq.~(\ref{eq:NpiN}) explicitly, one then obtains the final analytic result for the $N \pi$ contribution to the proton self-energy,
\begin{equation}
\begin{aligned}
\label{eq:final-NpiN}
\Sigma_{N\to N\pi}
    =& -\frac{g_A^2}{(8\pi f_\pi)^2}
    \frac{1}{M \Lambda^3 (4 M^2-\Lambda^2)^{5/2}}
    \\
    & \hspace*{-1.5cm} \times 
    \Bigg\{
    \Lambda\sqrt{4 M^2 - \Lambda^2}
    \Bigg[
    M^2 \widetilde\Lambda^4 
    \Big( m_\pi^2 \big(\Lambda^2-10 M^2\big) + 2 \Lambda^2 \big(\Lambda^2-M^2\big) 
    \Big)
    \\
    & \hspace*{-1cm}
    + 3 \Lambda^2 m_\pi^2 \big(4 M^2-\Lambda^2\big)^2 
    \bigg( m_\pi^2 \log\frac{m_\pi^2}{\Lambda^2}
    \\
    &+ 2 m_\pi \sqrt{4 M^2 - m_\pi^2}
    \bigg[
      \tan^{-1}\frac{m_\pi}{\sqrt{4 M^2 - m_\pi^2}}
    + \tan^{-1}\frac{2 M^2-m_\pi^2}{m_\pi \sqrt{4 M^2-m_\pi^2}}
    \bigg]
    \bigg)
    \Bigg]
    \\
    & \hspace*{-1cm} 
    + 6
    \bigg( 2 \Lambda^6 M^4 \big(\Lambda^2-2 M^2\big) 
        - \Lambda^2 m_\pi^4 \big(\Lambda^2-6 M^2\big) 
        \big(\Lambda^4-4 \Lambda^2 M^2 + 6 M^4\big)
    \\ 
    & \hspace*{0cm}
    + 2 \Lambda^4 M^2 m_\pi^2 \big(\Lambda^4 - 10 \Lambda^2 M^2+18 M^4\big) - 4 M^6 m_\pi^6 
    \bigg)
    \\
    &\hspace*{3cm} \times
    \bigg[ \tan^{-1} \frac{\Lambda^2 - 2 M^2}{\Lambda\sqrt{4 M^2-\Lambda^2}} - \tan^{-1} \frac{\Lambda}{\sqrt{4 M^2-\Lambda^2}}
    \bigg]
    \Bigg\}.
\end{aligned}
\end{equation}
We have confirmed that this result coincides with Ref.~\cite{Alberg:2012wr} after accounting for the differences in pseudoscalar and pseudovector couplings.
The expression in Eq.~(\ref{eq:octet-octetk-}) generalizes that result to the case where the initial and intermediate baryons have different mass, $\Delta_{B'B} \not= 0$.

The final result for the $N \to N \pi$ self-energy (\ref{eq:final-NpiN}) can also serve as a reference point for comparing with heavy-baryon chiral expansions.
By expanding in powers of $1/M$ and considering only the first order term, which corresponds to taking the heavy-baryon limit in Eq.~(\ref{eq:final-NpiN}), $M \to \infty$, we obtain the simple result,
\begin{equation}
\label{eq:NpiN-HB}
    \Sigma_{N\to N\pi}^{\rm HB} 
    = -\frac{3 g_A^2 \big(\Lambda - m_\pi\big)^4 
                     \big(\Lambda^2+ 4\Lambda m_\pi + m_\pi^2\big)}
            {512 \pi\, f_\pi^2 \Lambda^3}.
\end{equation}
If the differences in form factors are taken into account, Eq.~(\ref{eq:NpiN-HB}) agrees with the $N \to N \pi$ heavy-baryon result found by Young {\it et al.}~\cite{Young:2003}.
In particular, one can easily verify from Eq.~(\ref{eq:NpiN-HB}) that the coefficient of the ${\cal O}(m_\pi^3)$ LNA term is the familiar result from chiral perturbation theory~\cite{Li:1971vr, Jenkins:1991ts}, 
    $\Sigma_{N\to N\pi}^{\lna} = -(3 g_A^2)/(32 \pi f_\pi^2)\, m_\pi^3$
(see Sec.~\ref{ssec:LNA}).

A further important observation about the relativistic calculation is that, although the $\delta(y)$ term in Eq.~(\ref{eq:octet-octetk-}) does not contribute for the special case of $M_B = M_{B'}$, it is in practice vital to keep terms proportional to $\delta(y)$ to ensure self-consistency of the calculation.
Such terms would arise, for example, if one were to use a pseudoscalar interaction instead of a pseudovector coupling in the calculation of the $N \to N \pi$ self-energy~\cite{Alberg:2012wr} (or indeed for any $B \to B' \phi$ transition).
As discussed in Ref.~\cite{Ji:2012pv}, in that case the light-front zero-modes ($k^+ = 0$) play a crucial role in determining the correct LNA behavior in the chiral limit.
%
%

\subsection{Octet $\to$ decuplet transitions}
\label{Octet-decuplet}

For the contribution to the octet baryon self-energy from intermediate states with decuplet baryons $T'$, using the same dipole form factor (\ref{eq:dipoleFF}) in the self-energy operator in Eq.~(\ref{eq:SigBThat}) and taking the trace as in Eq.~(\ref{eq:DefOctDecSE}), we obtain
\begin{equation}
\begin{aligned}
\label{eq:SigBT'}
\Sigma_{B \to T'\phi}
=& -i \bigg( \frac{C_{BT'\phi}}{ f_\phi} \bigg)^{\!2}
    \frac{1}{12 M_B M_{T'}^2}
    \int\!\frac{d^4k}{(2\pi)^4}
    \biggl( \frac{\widetilde\Lambda^4}{D_\Lambda^2} \biggr)^{\!\!2}\,
    \frac{8 \big( M_B \overline{M}_{\!BT'}-p\cdot k \big)
            \big( M_B^2k^2 - (p\cdot k)^2 \big)}
         {D_\phi D_{T'}}.
\end{aligned}
\end{equation}
From the definitions of the propagators in Eqs.~(\ref{eq:propagatorphi}), (\ref{eq:propagatorB}) and (\ref{eq:propagatorT}), one can reduce the numerator in (\ref{eq:SigBT'}) via the replacements given in Eq. (\ref{eq:k2sub}) and 
\begin{equation}
\label{eq:decuplet k&p subs}
    p\cdot k \to \frac{1}{2}
         \left( D_\phi - D_{T'} + M_B^2-M_{T'}^2 + m_\phi^2 \right).
\end{equation}
The self-energy (\ref{eq:SigBT'}) can then be written in reduced form as 
\begin{eqnarray}
\label{eq:octet-decuplet-reduced}
\Sigma_{B \to T'\phi} 
&=& i \bigg( \frac{C_{BT'\phi}}{f_\phi} \bigg)^2
   \frac{\widetilde\Lambda^8}{12 M_B M_{T'}^2}
   \int\!\frac{d^4k}{(2\pi)^4}   
\Bigg[          
\frac{\big( \Delta_{T'B}^2 - m_\phi^2 \big)
        \big( \overline{M}_{\!BT'}^2 - m_\phi^2 \big)^2}
       {D_\phi D_{T'} D_\Lambda^4}
\nonumber\\
&+& \frac{4(p\cdot k)^2
      - \overline{M}_{\!BT'}^2
        \big( 2 p\cdot k - \overline{M}_{\!BT'} \Delta_{T'B} \big)-2\big(\overline{M}_{\!BT'}^2 - M_B \Delta_{T'B} - p\cdot k\big) m_\phi^2
        + m_\phi^4}
     {D_\phi D_\Lambda^4}
\nonumber\\
&-& \frac{4(p\cdot k')^2 + (\overline{M}_{BT'}^2-M_B M_{T'})-m_\phi^2)^2-5M_B^2M_{T'}^2+ 2 p \cdot k'\, (\overline{M}_{\!BT'}^2-m_\phi^2)}
       {D_{T'} D_\Lambda^4} 
\Bigg],
\end{eqnarray}
where $k' = k - p$, and we define $\overline{M}_{\!BT'}$ and $\Delta_{T'B}$ in analogy with Eqs.~(\ref{eq:MBB'})--(\ref{eq:DelBB'}).
Compared with the octet--octet self-energy case, Eq.~(\ref{eq:octet-decuplet-reduced}) contains two new types of terms, namely, ones proportional to $(p\cdot k')^2$ and to $(p\cdot k)^2$.
The former can be reduced and written in the form
\begin{eqnarray}
\label{eq:pdkprime2}
\frac{4(p\cdot k')^2}{D_{T'} D_{\Lambda}^4}
&=& \frac{\big(M_B^2 + M_T^2 -\Lambda^2 \big)^2}{D_{T'} D_\Lambda^4}
 +  \frac{3 M_B^2 + M_T^2 - 2 p\cdot k - \Lambda^2}{D_\Lambda^4} \nonumber\\
&+& \frac{2 \big( \Lambda^2-M_B^2-M_T^2 \big)}{D_{T'} D_\Lambda^3}
 +  \frac{1}{D_{T'} D_\Lambda^2}
 -  \frac{1}{D_\Lambda^3}.
\end{eqnarray}
using Eqs.~(\ref{eq:k2sub}) and (\ref{eq:decuplet k&p subs}). Each of these terms are straightforward to evaluate in the $d^4k$ integration and are discussed in Appendix \ref{app:integrations}.

The term proportional to $(p\cdot k)^2$ in Eq.~(\ref{eq:octet-decuplet-reduced}) cannot be reduced further and must be evaluated directly.
Following similar steps as those in the Appendix \ref{app:integrations} for the derivation of Eq.~(\ref{eq:DphiDLambdaSimp}), we rewrite this term as
\begin{equation}
\begin{aligned}
\label{eq:pdotk2}
\int d^4k\, \frac{4(p\cdot k)^2}{D_\phi D_\Lambda^4} 
& = \frac12 \int\!dk^+\, d^2{\bm k}_\perp\!
    \int_0^1\!dz\, z^3       
    \int\!dk^-
    \frac{\big(p^+ k^-\big)^2 + 2 M_B^2 k^+ k^- + \big(p^- k^+\big)^2}
         {\big(k^+ k^- - k_\perp^2 - \Omega + i \epsilon\big)^5},
\end{aligned}
\end{equation}
where $\Omega$ is as in Eq.~(\ref{eq:Omega}), and consider each term in the numerator separately.
The term proportional to $(p^+ k^-)^2$ can be written as
\begin{equation}
\begin{aligned}
&  \int dk^+\, d^2\bm{k}_\perp \int_0^1 dz \int dk^- 
    \frac{z^3\, \big(p^+ k^-\big)^2}
         {\big(k^+k^- - k_\perp^2 - \Omega + i\epsilon\big)^5} \\
&= \bigg( \frac{p^+}{2} \bigg)^2 \int\!dk^+\, d^2\bm{k}_\perp
    \int_0^1 dz\, \frac{z^3}{3!} 
    \biggl(\frac{\partial}{\partial k^+}\biggr)^{\!2}
    \biggl(\frac{\partial}{\partial \Omega}\biggr)^{\!2} 
    \biggl[ 2\pi i\log\biggl( \frac{k_\perp^2 + \Omega}{\mu^2} \biggr)\, \delta(k^+) \biggr].
\end{aligned}
\end{equation}
The next step must be handled with care. 
Note that the partial $k^+$ derivative is applied only to the $\delta$ function, giving
\begin{equation}
\label{eq:deltafnder}
    \frac{\partial}{\partial k^+}\, \delta(k^+) = -\frac{1}{k^+}\delta(k^+).
\end{equation}
Since the $\delta$ function is even in $k^+$, $\delta(k^+) = \delta(-k^+)$, integration of this term over all $k^+$ will vanish.
Higher derivatives applied in Eq.~(\ref{eq:deltafnder}) will not modify the null result, so the integral proportional to $(p^+ k^-)^2$ in (\ref{eq:pdotk2}) is zero.

For the term proportional to $k^+ k^-$, following similar steps one can derive
\begin{equation}
\label{eq:nonzero-pdk2}
\int\!dk^+ d^2\bm{k}_\perp \int_0^1\!dz \int\!dk^- 
\frac{z^3\, k^+ k^-}{\big(k^+k^- - k_\perp^2- \Omega +i\epsilon\big)^5}
= \frac{2 \pi^2 i}{3} \int\!dk^+\, dk_\perp^2 
  \int_0^1\!dz\, \frac{z^3}{(k_\perp^2 + \Omega)^3}\, 
  \delta(k^+).
\end{equation}
Finally, for the $(p^- k^+)^2$ term, which has no $k^-$ dependence, after the $k^-$ integration the integral will be proportional to $k^+ \delta(k^+)$, which again for reasons of symmetry will vanish.
The term proportional to $k^-$ is therefore the only part of Eq.~(\ref{eq:pdotk2}) that gives a nonzero contribution, given by Eq.~(\ref{eq:nonzero-pdk2}).
The complete $k^-$ integrated expression for the decuplet intermediate state contribution to the octet baryon self-energy is given by
\begin{eqnarray}
\label{eq:octet-decuplet-k-}
\Sigma_{B \to T'\phi}
&=& \frac{C_{BT'\phi}^2}{(4\pi f_\phi)^2}
\frac{\widetilde\Lambda^8}{12 M_B M_{T'}^2}
\int_0^1 dy \int_0^\infty dk_\perp^2
\bigg[
  \frac{\bar{y}^4 \big( m_\phi^2-\Delta_{T'B}^2 \big) 
                  \big( m_\phi^2-\overline{M}_{\!BT'}^2 \big)^2}
       {D_{B\phi T'} D_{B\Lambda T'}^4}
\nonumber\\
&&
-\ \frac{\bar{y}^3}{D_{B\Lambda T'}^4} 
  \Big( \big(M_B^2 + M_{T'}^2 -\Lambda^2 \big)^2
      + \big(\overline{M}_{BT'}^2-M_B M_{T'}-m_\phi^2 \big)^2
      - 5 M_B^2 M_{T'}^2
  \Big)
\nonumber\\
&&
-\ \frac{2\bar{y}^2}{D_{B\Lambda T'}^3}
  \big( M_B^2 + M_{T'}^2 - \Lambda^2 \big) 
- \frac{\bar{y}}{D_{B\Lambda T'}^2}
- \frac{\big( 3 k_\perp^2 + 6 M_B^2 + 2 M_{T'}^2 + \Lambda^2 \big)}
        {6 \big( k_\perp^2+\Lambda^2 \big)^3}\,
  \delta(y) 
\nonumber\\
&&
+\ \int_0^1 dz\, \frac{z^3}{(k_\perp^2+\Omega)^4}
  \Big( \frac23 M_B^2 \big( k_\perp^2 + \Omega \big)
        - \overline{M}_{\!BT'}^3 \Delta_{T'B}
\nonumber\\
&& \hspace*{3.8cm}
        + 2\, \big( \overline{M}_{\!BT'}^2 - M_B\Delta_{T'B} \big) m_\phi^2
        - m_\phi^4 
  \Big)
  \delta(y)
\biggr].
\end{eqnarray}
Note that in Eq.~(\ref{eq:octet-decuplet-k-}) terms proportional to $p \cdot k$ and $p \cdot k'$ have been omitted, since these are odd in $k$ and $k'$, respectively, and hence vanish after integration.
The fully integrated expression for the octet $\to$ decuplet self-energy is quite lengthy, but can be easily obtained by evaluating the $y$, $k_\perp^2$, and $z$ integrals in Eq.~(\ref{eq:octet-decuplet-k-}).

A similar chiral effective theory calculation of the $N \to \Delta \pi$ contribution to the proton self-energy was performed in Ref.~\cite{Alberg:2012wr}, although with some important differences compared to our result in Eq.~(\ref{eq:octet-decuplet-k-}), which affect the resulting LNA behavior.
These differences can be traced back to the treatment of the light-front zero-modes and the handling of the light-front energy $k^-$ integration in Eq.~(\ref{eq:octet-decuplet-reduced}).
In particular, Eq.~(13) of Ref.~\cite{Alberg:2012wr} is the result of taking the pole in which the intermediate $\Delta$ is on its mass shell in the light-front energy $k^-$ integration.
However, the absence of terms proportional to $\delta(y)$ in Eq.~(13) of Ref.~\cite{Alberg:2012wr} suggests that the light-front zero-mode ($k^+=0$) contribution has not been included.

In order to ensure that the integration captures the $k^+=0$ contribution, we first reduce the numerator of Eq.~(\ref{eq:SigBT'}) using Eq.~(\ref{eq:decuplet k&p subs}) to decompose the total amplitude into terms proportional to $1/(D_\pi D_\Delta)$, $1/D_\pi$ and $1/D_\Delta$, along with the overall factor $1/D_\Lambda^4$ from the regulating function $F(k,\Lambda)$.
Each individual denominator term is then computed separately, as shown in Eqs.~(\ref{eq:octet-decuplet-reduced})--(\ref{eq:octet-decuplet-k-}).

While it is certainly legitimate to take the $D_\Delta=0$ pole in the upper $k^-$ half-plane for terms such as $1/(D_\pi D_\Delta D_\Lambda^4)$ that also have poles in the lower $k^-$ half-plane, it would not be correct to take $D_\Delta=0$ pole for the terms such as $1/(D_\pi D_\Lambda^4)$ that have poles only in the lower $k^-$ half-plane.
One could mistakenly do this, for instance, by directly taking the $D_\Delta=0$ pole in Eq.~(\ref{eq:SigBT'}) before reducing the momentum dependence in the numerator.
As shown in Eqs.~(\ref{eq:pdotk2})--(\ref{eq:nonzero-pdk2}), terms with poles only in the lower $k^-$ half-plane are survived by the light-front zero-mode ($k^+=0$) contributions, resulting in the terms proportional to $\delta(y)$ in Eq.~(\ref{eq:octet-decuplet-k-}).
Without correctly capturing the $k^+=0$ contribution, the result in Eq.~(13) of~\cite{Alberg:2012wr} yields an LNA term of the form $m_\pi^2\, \log m_\pi^2$, in addition to the standard $m_\pi^4\, \log m_\pi^2$ behavior.
In contrast, our LNA results, discussed more fully in Sec.~\ref{ssec:LNA} below, include the light-front zero-mode ($k^+=0$) contribution, and Eq.~(\ref{LNA-Delta}) for $\Delta_{\cal B'B} > m_\phi$ reproduces the standard expression
    $\sim m_\pi^4/\Delta_{\Delta N} \log\, m_\pi^2$
for the LNA behavior of the $N \to \Delta \pi$ transition~\cite{Leinweber:1999ig} (see also Eq.~(2.17) of Ref.~\cite{Bernard:1993nj}).

In the same vein as the light-front zero-mode $\delta(y)$ contribution, we also note the appearance of $\delta({\bar y})$ terms from the light-front end-point singularity in the point-like limit, $F(k,\Lambda) \to 1$ with $\Lambda \to \infty$.
Taking the $D_\Delta = 0$ pole for the terms proportional to the denominator $1/D_\Delta$ alone would in this case lead to incorrect results.
Just as terms with poles only in the lower $k^-$ half-plane are survived by the zero-mode ($k^+ = 0$) contribution, so too the terms proportional to $1/D_\Delta$ with the pole only in the upper $k^-$ half-plane are survived by the end-point ($k^+ = p^+$) contribution, leading to the terms proportional to $\delta({\bar y})$.
The appearance of these terms can be traced back to the ${\bar y}^{(n-1)}/{D_{B\Lambda T'}^n}$ terms with the sole denominator $D_{B\Lambda T'}$ for $2 \leq n \leq 4$ in Eq.~(\ref{eq:octet-decuplet-k-}). 
As shown by Salamu {\it et al.}~\cite{Salamu:2018cny}, one can identify these terms in the $\Lambda \to \infty$ limit with terms proportional to $\delta({\bar y})$,
\begin{eqnarray}
\label{pointlike-NDeltapi}
\frac{\widetilde\Lambda^8\, {\bar{y}^3}}{D_{B\Lambda T'}^4}
&&\ \underset{\Lambda\to\infty}
{\longrightarrow}\
\mathop{\lim}\limits_{\Omega_0 \to \infty}
  \left.
  \int_{\Omega_0}^{\Omega_{T'}}\, dt
  \frac{-4 y \bar{y}^3\, \widetilde\Lambda^8}
       {\big(y\, t - y\, \bar{y} M_B^2 + \bar{y}\, \Omega_\Lambda\big)^5}
  \right|_{\Lambda\to\infty}			\notag\\
&&\ \underset{\Lambda\to\infty}
{\longrightarrow}\
   -\log\frac{\Omega_{T'}}{\Omega_0}\, \delta(\bar y)\
=\ \left(1- \log\frac{\Omega_{T'}}{\mu^2}\right) \delta(\bar{y}),
\end{eqnarray}
where
    $\Omega_\Lambda = k_\perp^2 + \Lambda^2$
and $\Omega_{T'} = k_\perp^2 + M_{T'}^2$,
with $\Omega_0$ a $\Lambda$-independent constant, and $\mu$ is defined such that
    $\log(\Omega_{T'}/\mu^2) = \log(\Omega_{T'}/\Omega_0) + 1$.
Taking the $\Lambda \to \infty$ limit of Eq.(\ref{eq:octet-decuplet-k-}) for the $N \to \Delta\pi$ transition, the self-energy for the point-like case is then given by
\begin{eqnarray}
\label{point-limit-Delta-Nucleon}
\Sigma_{N \to \Delta\pi}^{\textrm{point}}
&=& \frac{C_{N\Delta\pi}}{(4\pi f_\pi)^2}
\frac{1}{12 M M_\Delta^2}
\int_0^1 dy \int_0^\infty dk_\perp^2\,
\Bigg\{
  \frac{\big( m_\pi^2-\Delta_{\Delta N}^2 \big)
        \big( m_\pi^2-\overline{M}_{N\Delta}^2 \big)^2}
       {D_{N\pi\Delta}} 
\nonumber\\
&&
-\, \log\frac{k_\perp^2+m_\pi^2}{\mu^2}\,
\Big[ \overline{M}_{\!N\Delta}^3 \Delta_{\Delta N}
   - 2 m_\pi^2 \big( \overline{M}_{\!N\Delta}^2 - M\Delta_{\Delta N} \big)
   + m_\pi^4
\Big]\, \delta(y)
\nonumber\\
&&
-\,
\bigg[
\log\frac{k_\perp^2+M_\Delta^2}{\mu^2}
\Big( (\overline{M}_{\!N\Delta}^2-M M_{\Delta}-m_\pi^2)^2-5M^2 M_\Delta^2 \Big)
\nonumber\\
&&\hspace{2.6cm} 
-\ \Big( 1 - \log\frac{k_\perp^2+M_\Delta^2}{\mu^2} \Big)\,
    2 M^2 \big( k_\perp^2 + M_\Delta^2 \big)
\bigg]\,
\delta(\bar{y})
\Bigg\}.
\end{eqnarray}
The same result can also be obtained using the manifestly covariant dimensional regularization method.
The result (\ref{point-limit-Delta-Nucleon}) can also be contrasted with the point-like limit of Eq.~(13) in Ref.~\cite{Alberg:2012wr} by setting the form factor ``$F_\Delta(-t)$'' there to unity, where
    $t = \big( k_\perp^2 + y (M_\Delta^2-M^2) + y^2 M^2 \big) / \bar{y}$. 
The terms in Eq.~(13) of \cite{Alberg:2012wr} with higher powers of $t$ are obtained by taking the $D_\Delta = 0$ pole in the $k^-$ integration; however, as illustrated above, one should not take the $D_\Delta = 0$ pole for terms proportional to the sole denominator $1/D_\pi$ or $1/D_\Delta$.
These terms are survived by the light-front zero-mode ($k^+ = 0$) and end-point ($k^+ = p^+$) contributions, leading to the $\delta(y)$ and $\delta(\bar{y})$ terms, respectively, in Eq.~(\ref{point-limit-Delta-Nucleon}).
In particular, the $\delta(y)$ contribution is crucial for obtaining the correct LNA behavior $\sim (m_\pi^4/\Delta_{\Delta N}) \log m_\pi^2$ \cite{Bernard:1993nj, Leinweber}, as mentioned above.
We will discuss the LNA coefficients in more detail in Sec.~\ref{ssec:LNA} below.
%
%

\subsection{Decuplet $\to$ octet transitions}
\label{Decuplet-octet}

The derivation of the $T \to B' \phi$ loop contribution to the self-energy of the decuplet baryon $T$ follows closely that of the octet $\to$ decuplet transitions in Sec.~\ref{Octet-decuplet}.
Starting from the expression in Eq.~(\ref{eq:decupletoctettrace}), and applying the dipole regulator of Eq.~(\ref{eq:dipoleFF}), we find
\begin{equation}
\label{eq:TtoB'pho}
\begin{aligned}
\Sigma_{T \to B'\phi}
=& -i \bigg( \frac{C_{TB'\phi}}{f_\phi} \bigg)^2
\frac{1}{24 M_{T}^3}
\int\!\frac{d^4k}{(2\pi)^4}
\biggl( \frac{\widetilde\Lambda^4}{D_\Lambda^2} \biggr)^{\!2}\,
\frac{8 \big( M_T\overline{M}_{B'T}-p\cdot k \big)
        \big( M_T^2 k^2 - (p\cdot k)^2 \big)}
     {D_\phi D_{B'}}.
\end{aligned}
\end{equation}
Using the replacement 
\begin{eqnarray}
\begin{aligned}
\label{decuplet k&p subs}
p \cdot k\ &\to\ \frac12 \big( M_B^2 - M_{T'}^2 + D_\phi - D_{B'} + m_\phi^2 \big)
\end{aligned}
\end{eqnarray}
in the numerator factors, Eq.~(\ref{eq:TtoB'pho}) can be reduced to the form
\begin{equation}
\begin{aligned}
\label{eq:decuplet-octet-reduced}
\Sigma_{T \to B'\phi} 
&=  i \bigg(\frac{C_{TB'\phi}}{ f_\phi}\bigg)^2
   \frac{\widetilde\Lambda^8}{24 M_T^3}
   \int\frac{d^4k}{(2\pi)^4}
\bigg[           
\frac{\big( \Delta_{B'T}^2 - m_\phi^2 \big)
        \big( \overline{M}_{\!BT'}^2 - m_\phi^2 \big)^2}
       {D_\phi D_{B'} D_\Lambda^4}
\\
    &
+
\frac{4(p\cdot k)^2
      - \overline{M}_{\!TB'}^2
        \big( 2 p\cdot k - \overline{M}_{\!TB'} \Delta_{B'T} \big)-2\big(\overline{M}_{\!TB'}^2 - M_B \Delta_{B'T} - p\cdot k\big) m_\phi^2
        + m_\phi^4}
     {D_\phi D_\Lambda^4} \\
    &
- \frac{4(p\cdot k')^2 + \big(\overline{M}_{TB'}^2-M_T M_{B'}-m_\phi^2\big)^2
        - 5 M_T^2 M_{B'}^2 + 2 p \cdot k'\, \big(\overline{M}_{\!TB'}^2-m_\phi^2\big)}
       {D_{B'} D_\Lambda^4} \bigg]. 
\end{aligned}
\end{equation}
Note that the decuplet-octet self-energy is almost identical to the octet-decuplet self-energy om Eq.~(\ref{eq:octet-decuplet-reduced}), with the external decuplet baryon mass and the internal octet baryon mass switched.

We can follow closely the steps and expressions in the previous Sec.~\ref{Octet-decuplet}, with the substitutions $M_B \to M_T$, $M_{T'} \to M_{B'}$, and $D_{T'} \to D_{B'}$. 
Using the integral relations in Eqs.~(\ref{eq:term1}), (\ref{eq:term2}) and (\ref{eq:DphiDL4}) in Appendix~\ref{app:integrations}, together with the identities  (\ref{eq:pdkprime2}), and (\ref{eq:nonzero-pdk2}), we arrive at the $k^-$ integrated expression for the decuplet-octet self-energy,
\begin{eqnarray}
\label{eq:decuplet-octet-k-}
\Sigma_{T \to B'\phi}
&=& \frac{C_{TB'\phi}^2}{(4\pi f_\phi)^2}
\frac{\widetilde\Lambda^8}{24 M_T M_{B'}^2}
\int_0^1 dy \int_0^\infty dk_\perp^2
\bigg[
  \frac{\bar{y}^4 \big( m_\phi^2-\Delta_{BT'}^2 \big) 
                  \big( m_\phi^2-\overline{M}_{\!TB'}^2 \big)^2}
       {D_{B\phi T'} D_{B\Lambda T'}^4}
\nonumber\\
&&
-\ \frac{\bar{y}^3}{D_{B\Lambda T'}^4} 
  \Big( \big(M_T^2 + M_{B'}^2 -\Lambda^2 \big)^2
      + \big(\overline{M}_{TB'}^2-M_T M_{B'}-m_\phi^2 \big)^2
      - 5 M_T^2 M_{B'}^2
  \Big)
\nonumber\\
&&
-\ \frac{2\bar{y}^2}{D_{B\Lambda T'}^3}
  \big( M_T^2 + M_{B'}^2 - \Lambda^2 \big) 
- \frac{\bar{y}}{D_{B\Lambda T'}^2}
- \frac{\big( 3 k_\perp^2 + 6 M_T^2 + 2 M_{B'}^2 + \Lambda^2 \big)}
        {6 \big( k_\perp^2+\Lambda^2 \big)^3}\,
  \delta(y) 
\nonumber\\
&&
-\ \int_0^1 dz\, \frac{z^3}{(k_\perp^2+\Omega)^4}
  \Big( \frac23 M_T^2 \big( k_\perp^2 + \Omega \big)
        - \overline{M}_{\!BT'}^3 \Delta_{BT'}
\nonumber\\
&& \hspace*{3.8cm}
        + 2\, \big( \overline{M}_{\!BT'}^2 - M_T\Delta_{BT'} \big) m_\phi^2
        - m_\phi^4 
  \Big)
  \delta(y)
\biggr],
\end{eqnarray}
where again terms involving $p \cdot k$ and $p \cdot k'$ have been dropped as discussed in Sec.~\ref{Octet-decuplet}.

Note that for the $T \to B' \phi$ transitions, the negative baryon mass difference $-\Delta_{B'T} = M_T-M_{B'}$ can be larger than the meson mass, $m_\phi$.
In particular, the specific transitions $\Delta \to N \pi$, $\Sigma^* \to \Lambda \pi$, $\Sigma^* \to \Sigma \pi$ and $\Xi^* \to \Xi \pi$ are all kinematically allowed in the physical region.
In this case the corresponding self-energies will develop imaginary parts, which are related to the decay rates $\Gamma_{T \to B'\pi}$.
Similar to the optical theorem relating the total cross section with the imaginary part of forward scattering amplitudes, the decay rate $\Gamma$ for the physical transition is related to the imaginary part of the corresponding self-energy,
    $\Gamma = -2 \, \Im {\rm m}\, \Sigma$.
In Sec.~\ref{sec: Decay rates} we discuss the basic features of this relation and the general results for the various transitions, and in Appendix~\ref{decay-rate-derivation} illustrate the derivation with the explicit example of the decay of a spin-1/2 excited state to a nucleon and pion, $N' \to N \pi$.

\subsection{Decuplet $\to$ decuplet transitions}

To complete this section, we present the results for the contribution to the self-energy of decuplet baryons from loops involving decuplet baryons and mesons.
Using the definition given in Eq. (\ref{eq: decupletdecuplet trace}), and applying the dipole form factor we get 
\begin{equation}
\begin{aligned}
\Sigma_{T \to T' \phi} 
=& -i\biggl( \frac{C_{TT'\phi}}{f_\phi} \biggr)^2
\frac{1}{72M_T^3 M_{T'}^2} \int\!\frac{d^4k}{(2\pi)^4}
\biggl( \frac{\widetilde\Lambda^2}{D_\Lambda}\biggr)^{\!4} 
\frac{8}{D_\phi D_{T'}} 
\\
\times&
\biggl[
  2 \overline{M}_{TT'}^2 \big( p\cdot k\big)^3
+ 2 M_T \big( \Delta_{T'T}^3 - 3\overline{M}_{TT'} M_T M_{T'} + M_{T'}^3 \big)
    \big( p\cdot k\big)^2
\\
&
+ M_T^2 \big( 9 M_{T'}^2 - 2\overline{M}_{TT'}^2 \big) p\cdot k\, k^2
+ M_T^3 \overline{M}_{TT'} \big( 2\overline{M}_{T'T}^2 + 3M_{T'}(\Delta_{T'T} - M_T)\big) k^2
\biggr].
\end{aligned}
\end{equation}
The rearranged propagators are then used to reduce the momentum dependence from the numerator factors, as in the other self-energy calculations above.
Namely, making the substitution
\begin{equation}
\label{decuplet k&p subs}
p \cdot k\, \to\, \frac{1}{2} \Big( M_T^2-M_{T'}^2+D_\phi - D_{T'} + m_\phi^2 \Big),
\end{equation}
we arrive at the expression
\begin{eqnarray}
\label{eq:decuplet-decuplet-reduced}
\Sigma_{T \to T' \phi} 
&=& -i \biggl( \frac{C_{TT'\phi}}{f_\phi} \biggr)^2
\frac{\widetilde\Lambda^8}{36 M_T^3 M_{T'}^2}
\int\!\frac{d^4k}{(2\pi)^4}
\nonumber\\
& & \hspace*{-1cm} \times
\Bigg[
\frac{\overline{M}^2_{TT'}}{D_\phi D_{T'} D_\Lambda^4}
\bigg(
  \big( m_\phi^2 - \Delta_{T'T}^2 \big)
  \Big[ \big( \Delta_{T'T}^2 - m_\phi^2 + M_T M_{T'} \big)^2 + 9 M_T^2 M_{T'}^2 \Big]
\bigg)
\nonumber\\
& & \hspace*{-0.5cm} +\,
\frac{\overline{M}^2_{TT'}}{D_{T'} D_\Lambda^4}
\bigg( 2 p\cdot k' \big[ m_\phi^2 - \Delta_{T'T}^2 + 2 M_T M_{T'} \big]
     + \big( m_\phi^2- \Delta_{T'T}^2 \big)^2
     + 10 M_T^2 M_{T'}^2
     - 4 (p\cdot k')^2
\bigg)
\nonumber\\
& & \hspace*{-0.5cm} -\,
\frac{1}{D_\phi D_\Lambda^4}
\bigg (4 \overline{M}^2_{TT'} (p\cdot k)^2
     + 2 p\cdot k 
       \Big[ \overline{M}^2_{TT'} m_\phi^2 - \overline{M}_{TT'}^4
            + 6 M_T M_{T'} \big( \Delta_{T'T}^2 - M_T M_{T'} \big)
       \Big]  
\nonumber\\
&&\hspace{1.5cm}
+\, \overline{M}_{TT'} 
    \Big[ \overline{M}_{TT'} m_\phi^4 + 2m_\phi^2(M_T M_{T'}^2-M_{T'}^3-2M_T^3)
\nonumber\\
&&\hspace{3cm}
+\, \Delta_{T'T} 
    \Big( \big( \Delta_{T'T}^2 + M_T M_{T'} \big)^2 + 9 M_T^2 M_{T'}^2 \Big)
    \Big] 
\bigg)
\Bigg].
\end{eqnarray}
All the different types of terms in Eq.~(\ref{eq:decuplet-decuplet-reduced}) have already been discussed in previous sections.
Using the results in Appendix~\ref{app:integrations} and making the substitution $M_B \to M_T$ in Eq.~(\ref{eq:nonzero-pdk2}) to obtain the relation
\begin{equation}
\int d^4k \frac{4 \big( p\cdot k\big)^2}{D_\phi D_\Lambda^4}
= \pi^2 i \int\!dk^+ \int\!dk_\perp^2  
    \int_0^1 dz \frac{2z^3}{3} \frac{M_T^2}{\big(k_\perp^2 + \Omega\big)^3}\, \delta(k^+),
\end{equation} 
we finally arrive at the $k^-$ integrated result for the decuplet-decuplet self-energy,
\begin{equation}
\begin{aligned}
\label{eq:decuplet-decuplet-k-}
\Sigma_{T \to T' \phi}
&= \frac{C_{TT'\phi}^2}{(24\pi f_\phi)^2}
\frac{\widetilde\Lambda^8\, \overline{M}_{TT'}^2}{M_T^3\, M_{T'}^2}
\int_0^1 dy \int dk_\perp^2
\\
& \hspace*{-1cm} \times
\Bigg\{
\frac{\bar{y}^4}{D_{T\phi T'} D_{T\Lambda T'}^4}
\Big( 
  \big( m_\phi^2 - \Delta_{T'T}^2 \big)
  \big[ \big( \Delta_{T'T}^2 - m_\phi^2 + M_T M_{T'} \big)^2 + 9 M_T^2 M_{T'}^2 
  \big]
\Big)
\\
& \hspace*{-0.5cm}
- \frac{\bar{y}^3}{D_{T\Lambda T'}^4}
  \Big( 
    \big( M_T^2 + M_{T'}^2 - \Lambda^2 \big)^2
  + \big( m_\phi^2 - \Delta_{TT'}^2 \big)^2 + 10 M_T^2 M_{T'}^2
  \Big)
\\
& \hspace*{-0.5cm}
- \frac{2 \bar{y}^2}{D_{T\Lambda T'}^3}
  \big( M_T^2 + M_{T'}^2 - \Lambda^2 \big)
- \frac{\bar{y}}{D_{T\Lambda T'}^2} 
- \frac{\big( 3 k_\perp^2 + 6 M_T^2 + 2M_{T'}^2 + \Lambda^2 \big)}
       {6 \big( k_\perp^2+\Lambda^2 \big)^3}\,
  \delta(y) 
\\
& \hspace*{-0.5cm}
- \int_0^1 dz \frac{z^3}{3(k_\perp^2+\Omega)^4}
  \bigg[ 2 M_T^2\, \big( k_\perp^2+\Omega \big) 
       - \frac{3}{\overline{M}_{T'T}}
         \Big[ 2 m_\phi^2 \big( M_T M_{T'}^2 - 2 M_T^3 - M_{T'}^3 \big)
\\
& \hspace{3cm} 
             + m_\phi^4 \overline{M}_{T'T} 
             + \Delta_{T'T} \Big( \big( \Delta_{T'T}^2+M_T M_{T'} \big)^2
                                + 9 M_T^2 M_{T'}^2 \Big)
         \Big] 
  \bigg] \delta(y)
\Bigg\},
\end{aligned}
\end{equation}
where, as before, we have dropped the $p\cdot k$ and $p\cdot k'$ terms which vanish after integration.

Having derived the full set of results for the self-energies of octet and decuplet baryons, we will proceed to systematically study their effects numerically.
Before doing so, however, we first make a brief aside to discuss the calculation of some of their analytic properties and applications.

\section{Analytic properties of self-energies}
\label{sec: Prop. and app. of SEs}

In this section we discuss several important analytic properties of the self-energies derived in Sec.~\ref{sec:FRR-self-energies}, namely, their LNA behavior in the chiral limit and the decay widths for their various decay channels.

\subsection{Leading nonanalytic behavior}
\label{ssec:LNA}

Expanding the baryon self-energies in a series about the chiral limit, $m_\phi \to 0$, it is well known that coefficients of certain terms in the series are model independent.
In particular, while the coefficients of terms that are analytic in the pseudoscalar meson mass squared ($m_\phi^2 \sim m_q$, according to the Gell-Mann--Oakes--Renner relation~\cite{GellMann:1968rz}) depend on details of short-distance, model-dependent physics, the coefficients of the nonanalytic terms are determined by the long-distance properties of meson loops and are therefore model independent.
The LNA terms in particular can serve as an important check of consistency of any calculation of a hadronic observable with the chiral properties of QCD.

Generally, the LNA results for the self-energies are common for both SU(3) octet and decuplet baryons, depending only on differences between the baryon masses.
For simplicity we therefore adopt the notation ${\cal B}$ and ${\cal B}'$ to indicate external and internal baryon states, respectively, with ${\cal B}, {\cal B}' = B'$ or $T'$.
To derive the LNA behaviors of the self-energies
    $\Sigma_{{\cal B} \to {\cal B}' \phi}$
we write the internal baryon mass as
    $M_{\cal B'} =  \Delta_{\cal B'B} + M_{\cal B}$,
and expand the self-energies in powers of $1/M_{\cal B'}$ and $m_\phi$.
Defining
    $R_{\cal B'B} \equiv \sqrt{\Delta_{\cal B' B}^2 - m_\phi^2}$
and 
    $\overline{R}_{\cal B'B} \equiv \sqrt{m_\phi^2-\Delta_{\cal B' B}^2}$,
the nonanalytic terms in the expansion of the self-energy, denoted by
    $\Sigma^{\lna}_{{\cal BB'}\phi}$,
are found to be
\begin{equation}
\Sigma^{\lna}_{{\cal B'B}\phi}
= \left\{
\begin{array}{l}
\begin{aligned}
\frac{{\cal C}^{\lna}_{\cal BB'}}{\pi^2 f_\phi^2}\,
& \bigg[ 
  \Delta_{\cal B'B} \big( 3 m_\phi^2 - 2 \Delta_{\cal B'B}^2 \big) 
  \log m_\phi^2\,
-\, 2 \overline{R}_{{\cal B'B}}^3 
  \Big( \pi-2\arctan\frac{\Delta_{\cal BB'}}{\overline{R}_{\cal B'B}} \Big)
\bigg],
\hspace*{0.6cm} \Delta_{\cal B'B} < m_\phi ,
\\
\\
\frac{{\cal C}^{\lna}_{\cal BB'}}{\pi^2 f_\phi^2}\,
& \bigg[
  \Delta_{\cal B'B} \big( 3 m_\phi^2 - 2\Delta_{\cal B'B}^2 \big) \log m_\phi^2\,
+\, 2 R_{{\cal B'B}}^3 
    \log{\frac{\Delta_{\cal B'B}-R_{\cal B'B}}{\Delta_{\cal B'B}+R_{\cal BB'}}}
\bigg],
\hspace*{1.3cm} \Delta_{\cal B'B} > m_\phi ,
\end{aligned}
\end{array}
\right. 
\label{eq:sigNA}
\end{equation}
where the coefficients ${\cal C}^{\lna}_{\cal BB'}$ are given in Table~\ref{table: LNA} in terms of the coupling constants $C_{{\cal BB'}\phi}$, which themselves are given in Tables~\ref{tab:octet} and \ref{tab:decuplet} of Appendix~\ref{app:couplings} in terms of the couplings defined in the Lagrangian (\ref{eq:L}).

Considering the two scenarios in (\ref{eq:sigNA}), for the $\Delta_{\cal B'B} < m_\phi$ case the mass difference $\Delta_{\cal B'B}$ approaches zero first in the chiral limit, which then leads to the resulting LNA expression
\begin{equation}
\Sigma^{\lna}_{\cal BB'\phi}
= -\frac{2\, {\cal C}^{\lna}_{\cal BB'}}{\pi f_\phi^2}\, m_\phi^3\, ,
\hspace*{2.5cm} \Delta_{\cal BB'} < m_\phi.
\end{equation}
This displays the characteristic $\sim m_\phi^3$ behavior which has been known, and for the ${\cal BB'\phi} = NN\pi$ case agrees with the well-known result~\cite{Li:1971vr, Jenkins:1991ts},
\begin{equation}
\Sigma^{\lna}_{NN\pi} = - \frac{3 g_A^2}{32 \pi f_\pi^2}\, m_\pi^3,
\end{equation}
where $g_A = D+F$ is the axial vector charge.

\begin{table}[t]
\caption{Coefficients ${\cal C}^{\lna}_{\cal BB'}$ of the LNA terms in the expansion of the self-energies for the various octet ($B, B'$) and decuplet baryon ($T, T'$) initial (${\cal B}=B, T$) and intermediate (${\cal B}'=B', T'$) states around $m_\phi=0$. The ${\cal BB'\phi}$ coupling constants are given in Tables~\ref{tab:octet} and \ref{tab:decuplet} of Appendix~\ref{app:couplings} for specific hadronic states.}
\begin{tabularx}{0.5\textwidth}{ 
   >{\centering\arraybackslash}X 
  |>{\centering\arraybackslash}X 
   >{\centering\arraybackslash}X  }
 \hline
 ${\cal C}^{\lna}_{\cal BB'}$ & $B'$  & $T'$ \\
 \hline
 $B$  & $\frac{1}{16} C^2_{BB'\phi}$  & $\frac{1}{24} C^2_{BT'\phi}$ \\
$T$   & $\frac{1}{48} C^2_{TB'\phi}$      & $\frac{5}{144} C^2_{TT'\phi}$ \\
\hline
\end{tabularx}
\label{table: LNA}
\end{table}

For the $\Delta_{\cal B'B} > m_\phi$ case, expanding the terms in powers of the meson mass gives the result,
\begin{eqnarray}
\label{LNA-Delta}
\Sigma^{\lna}_{\cal BB'\phi}
&=& \frac{{\cal C}^{\lna}_{\cal BB'}}{\pi^2 f_\phi^2}
\bigg[ \big( 3 m_\phi^2 \Delta_{\cal B'B} - 2\Delta_{\cal B'B}^3 \big) \log m_\phi^2
    +2 \Big( \Delta_{\cal B'B}^3 - \frac{3 m_\phi^2 \Delta_{\cal B'B}}{2}
            + \frac{3m_\phi^4}{8\Delta_{\cal B'B}}
       \Big) \log m_\phi^2
\bigg]          \nonumber\\
&=& \frac{3\, {\cal C}^{\lna}_{\cal BB'}}{4 \pi^2 f_\phi^2}
    \frac{m_\phi^4}{\Delta_{\cal B'B}} \log m_\phi^2\, ,
\hspace*{2.5cm} \Delta_{\cal B'B} > m_\phi.
\end{eqnarray}
For the phenomenologically relevant case of ${\cal B} = N$, ${\cal B'} = \Delta$, from Tables~\ref{table: LNA} and \ref{tab:octet} the coupling is given by
    ${\cal C}^{\lna}_{N\Delta} 
    = (1/24)\, C^2_{N\Delta\pi} 
    = (1/24)\, {\cal C}^2$,
and using the SU(6) relation ${\cal C} = (6/5)\, g_A$, we have the familiar behavior
\begin{eqnarray}
\label{NDeltapiLNA}
\Sigma^{\lna}_{N\Delta\pi}
&=& \frac{{\cal C}^2}{32 \pi^2 f_\pi^2}
    \frac{m_\pi^4}{\Delta_{\Delta N}} \log m_\pi^2\,
=\, \frac{3\, g_A^2}{16 \pi^2 f_\pi^2}
    \frac{32}{25} \frac{3 m_\pi^4}{8\Delta_{\Delta N}}\, \log m_\pi\, ,
\end{eqnarray}
which is next-to-leading nonanalytic for the nucleon mass.
We have verified that for the general case our expressions (\ref{eq:sigNA}) agree with the results derived in Ref.~\cite{Pascalutsa:2005nd} using dimensional regularization, where the $N\Delta\pi$ coupling $h_A$ defined there is given by $h_A = \sqrt2\, {\cal C}$.

As noted previously in Sec.~\ref{Octet-decuplet}, it is crucial to capture the light-front zero-mode ($k^+=0$) contribution in order to obtain the correct $m_\pi^4 \log m_\pi$ LNA behavior for the $N \to \Delta \pi$ transition in Eq.~(\ref{NDeltapiLNA}). 
The result given by Eq.~(13) of Ref.~\cite{Alberg:2012wr} omits the $\delta(y)$ terms of Eq.~(\ref{eq:octet-decuplet-k-}) and finds an LNA behavior of the form $m_\pi^2 \log m_\pi^2$.
Such a term is in fact cancelled exactly by the light-front zero-mode ($k^+=0$) contribution.
Furthermore, for the point-like limit ($F_\Delta=1$) in Eq.~(13) of \cite{Alberg:2012wr}, one should also recover the $\delta(\bar{y})$ term from the light-front end-point ($k^+=p^+$) contribution, as discussed in Sec.~\ref{Octet-decuplet}.
The result in Ref.~\cite{Alberg:2012wr} should therefore be corrected by taking into account the light-front zero-mode ($k^+=0$) as well as the light-front end-point ($k^+=p^+$) contributions.

\subsection{Decay rates}
\label{sec: Decay rates}

For a general transition ${\cal B} \to {\cal B}' \phi$ in which the decay channel is open, the baryon decay rate can be computed from the imaginary part of the self-energy,
    $\Gamma_{{\cal B B}' \phi} = -2 \, \Im {\rm m}\, \Sigma_{{\cal B} \to {\cal B}' \phi}$.
As discussed in Appendix~\ref{decay-rate-derivation}, imaginary contributions to the self-energy are generated from $1/(D_\phi D_{\cal B'} D_\Lambda^4)$ type terms, where the propagators of both the internal meson ($D_\phi$) and baryon ($D_{\cal B'}$) co-exist, so that the $k^-$ and $k_\perp$ integration of $1/(D_\phi D_{\cal B'} D_\Lambda^4)$ produces the logarithmic term,
\begin{equation}
\label{eq: negative Log}
    \int_0^1 dy\, \log(-\widetilde{D}),
\end{equation}
where $\widetilde{D} = y \bar{y} M_{\cal B}^2 - y M_{\cal B'}^2 - \bar{y} m_\phi^2$. 
No such negative logarithm term can arise from terms in the self-energy proportional to $1/(D_\phi D_\Lambda^4)$ or $1/(D_{\cal B'} D_\Lambda^4)$ alone.
To find the imaginary part of the self-energy, one can look for only the contribution from the region of $y$ integration where the condition $\widetilde{D}>0$ is satisfied, namely, $y_{\text{min}} < y < y_{\text{max}}$ with
\begin{equation}
y_{\rm min}
= \frac{-\sqrt{ \big( \Delta_{\cal B'B}^2-m_\phi^2 \big) 
                \big( \overline{M}_{\cal BB'}^2 - m_\phi^2 \big)}
              + M_{\cal B}^2 - M_{\cal B'}^2 + m_\phi^2}
       {2 M_{\cal B}^2}
\end{equation}
and
\begin{equation}
y_{\rm max} 
= \frac{\sqrt{ \big( \Delta_{\cal B'B}^2 - m_\phi^2 \big)    
               \big( \overline{M}_{\cal BB'}^2 - m_\phi^2 \big)} 
             + M_{\cal B}^2 - M_{\cal B'}^2 + m_\phi^2}
       {2M_{\cal B}^2}. 
\end{equation}
In this region, one can isolate the $\log(-1) = i\pi$ term in Eq.~(\ref{eq: negative Log}) and evaluate the $y$ integration, 
\begin{equation}
\label{eq: SE im cont}
\int_{y_{\rm min}}^{y_{\rm max}} dy\, \log(-1)
= i\pi \frac{\sqrt{ (\Delta_{\cal B'B}^2 - m_\phi^2)
                    (\overline{M}_{\cal BB'}^2 - m_\phi^2)}}
            {M_{\cal B}^2}.
\end{equation}
The condition for the existence of the non-vanishing $\Im { m}\, \Sigma$ coincides with the kinematic constraint $-\Delta_{\cal B'B} > m_\phi$ allowing the physical decay process ${\cal B} \to {\cal B}'\phi$.
The imaginary terms of the self-energy will therefore be the coefficients of the $1/(D_\phi D_{\cal B'} D_\Lambda^4)$ term multiplied by the result in Eq.~(\ref{eq: SE im cont}), together with a factor $\pi^2 i/\widetilde{\Lambda}^8$ from the $k^-$ and $k_\perp$ integrations.

Applying these relations to the specific ${\cal B} \to {\cal B}' \phi$ channels, for ${\cal B} = $ octet $B$ or decuplet $T$ baryon, we have
\begin{eqnarray}
\Gamma_{BB'\phi}
&=& \frac{C_{BB'\phi}^2}{16\pi f_\phi^2} 
    \frac{\overline{M}_{\!BB'}^2}{M_B^3}
    \big(\Delta_{B'B}^2-m_\phi^2\big)^{3/2}
    \big(\overline{M}_{\!BB'}^2 - m_\phi^2\big)^{1/2}
\\
\Gamma_{BT'\phi}
&=& \frac{C_{BT'\phi}^2}{72\pi f_\phi^2} 
    \frac{1}{M_B^3 M_{T'}^2}
    \big(\Delta_{T'B}^2-m_\phi^2\big)^{3/2}
    \big(\overline{M}_{\!BT'}^2 - m_\phi^2\big)^{5/2}
\\
\Gamma_{TB'\phi} 
&=& \frac{C_{TB'\phi}^2}{192\pi f_\phi^2} 
    \frac{1}{M_T^5}
    \big(\Delta_{B'T}^2-m_\phi^2\big)^{3/2}
    \big(\overline{M}_{\!TB'}^2 - m_\phi^2\big)^{5/2}
\\
\Gamma_{TT'\phi}
&=& \frac{C_{TT'\phi}^2}{288\pi f_\phi^2}
    \frac{\overline{M}_{T'T}^2}{M_T^5 M_{T'}^2}
    \big(\Delta_{T'T}^2-m_\phi^2\big)^{3/2}
    \big(\overline{M}_{\!TT'}^2 - m_\phi^2\big)^{1/2}
\nonumber\\
& & \hspace*{2.5cm} \times
    \Big[ (\Delta_{T'T}^2 - m_\phi^2 + M_T M_{T'})^2 + 9 M_T^2 M_{T'}^2 \Big].
\end{eqnarray}
Comparing these expressions with experimental decay rates, one can then determine the numerical values of the coupling constants, as done in Ref.~\cite{Pascalutsa:2005nd} for the $\Delta \to N \pi$ transition. 
As discussed in Sec.~\ref{Decuplet-octet}, for decuplet to octet baryon transitions the negative baryon mass difference $-\Delta_{B'T} = M_T-M_{B'}$ can be larger than the meson mass, $m_\phi$, even without considering baryon excited states, and physically the transitions $\Delta \to N \pi$, $\Sigma^* \to \Lambda \pi$, $\Sigma^* \to \Sigma \pi$ and $\Xi^* \to \Xi \pi$ are all kinematically allowed.

In the next section we study the baryon self-energies numerically, including their dependence on the pion mass.
For the open decay channels this reveals the rather distinctive curvature as one cross the kinematic thresholds arising from the development of a non-vanishing imaginary part of the self-energy.
%
%

\section{Numerical results}
\label{sec: results}

Having derived the full set of analytical results for the octet and decuplet self-energies, in this section we perform a comprehensive numerical study of the sizes and magnitudes of the various intermediate state contributions, both as a function of the dipole regulator mass, $\Lambda$, and of the pion mass squared, $m_\pi^2$.
At the end of the section we also compare the relativistic calculation of the proton self-energy with results of the heavy baryon approximation, for a common choice of regulator.

\subsection{Octet baryon self-energies}

For the self-energies of the octet baryons, we can analytically evaluate the $k_\perp^2$, $y$ and $z$ integrals in Eqs.~(\ref{eq:octet-octetk-}) and (\ref{eq:octet-decuplet-k-}) for the octet and decuplet intermediate state contributions, respectively. 
For the numerical calculation, we use for the SU(3) couplings $D = 0.85$ and $F = 0.41$, so that the axial vector charge $g_A = D+F = 1.26$, and use the SU(6) value for the meson-octet-decuplet coupling ${\cal C} = (6/5)\, g_A$.
The relations for the various meson-baryon coupling constants $C_{BB'\phi}$ and $C_{BT'\phi}$ in terms of these are given in Appendix~\ref{app:couplings}.

\begin{figure}[t]
\centering
\includegraphics[width=0.45\linewidth]{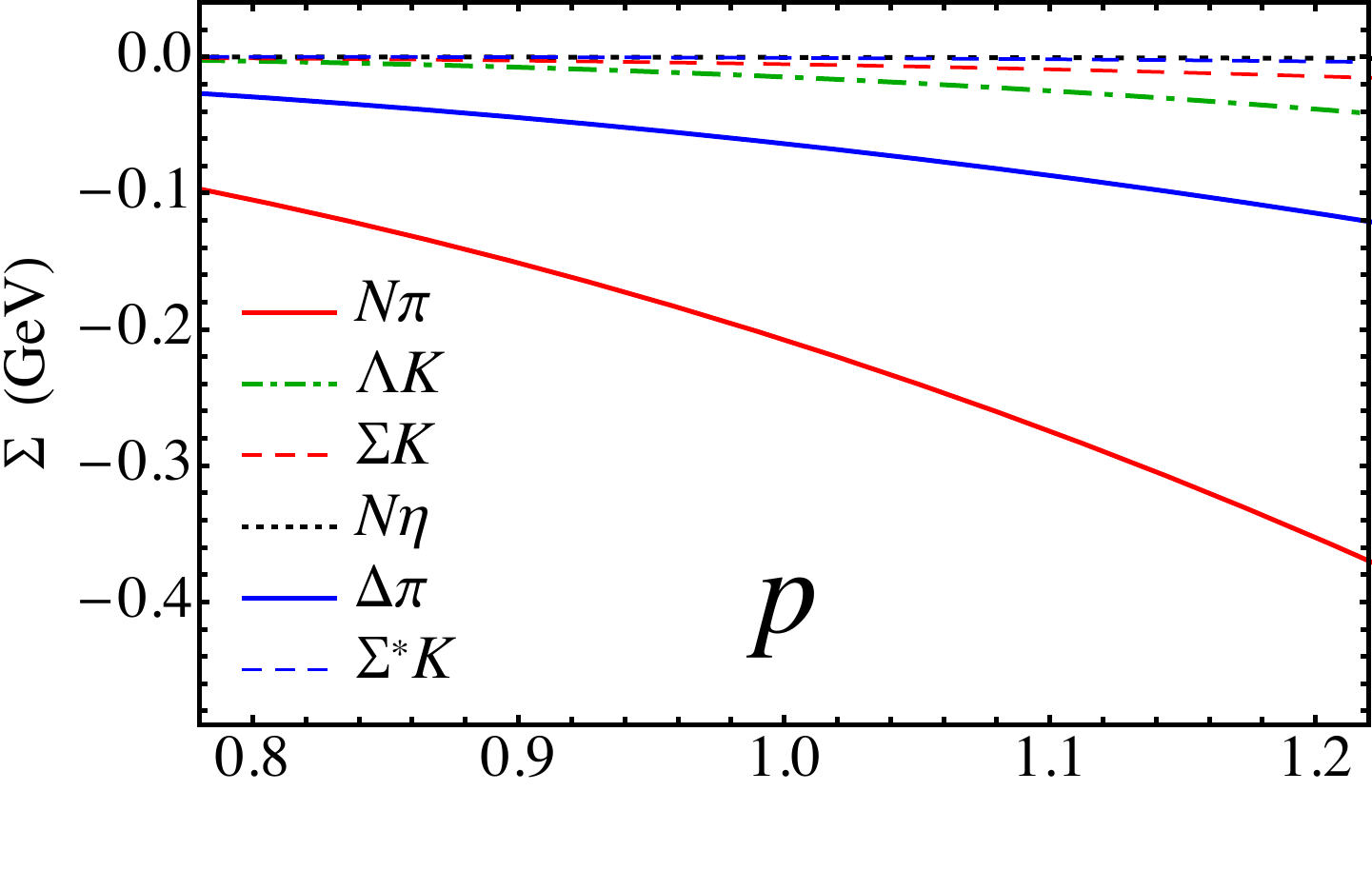}\hspace*{-0.3cm}
\includegraphics[width=0.46\linewidth]{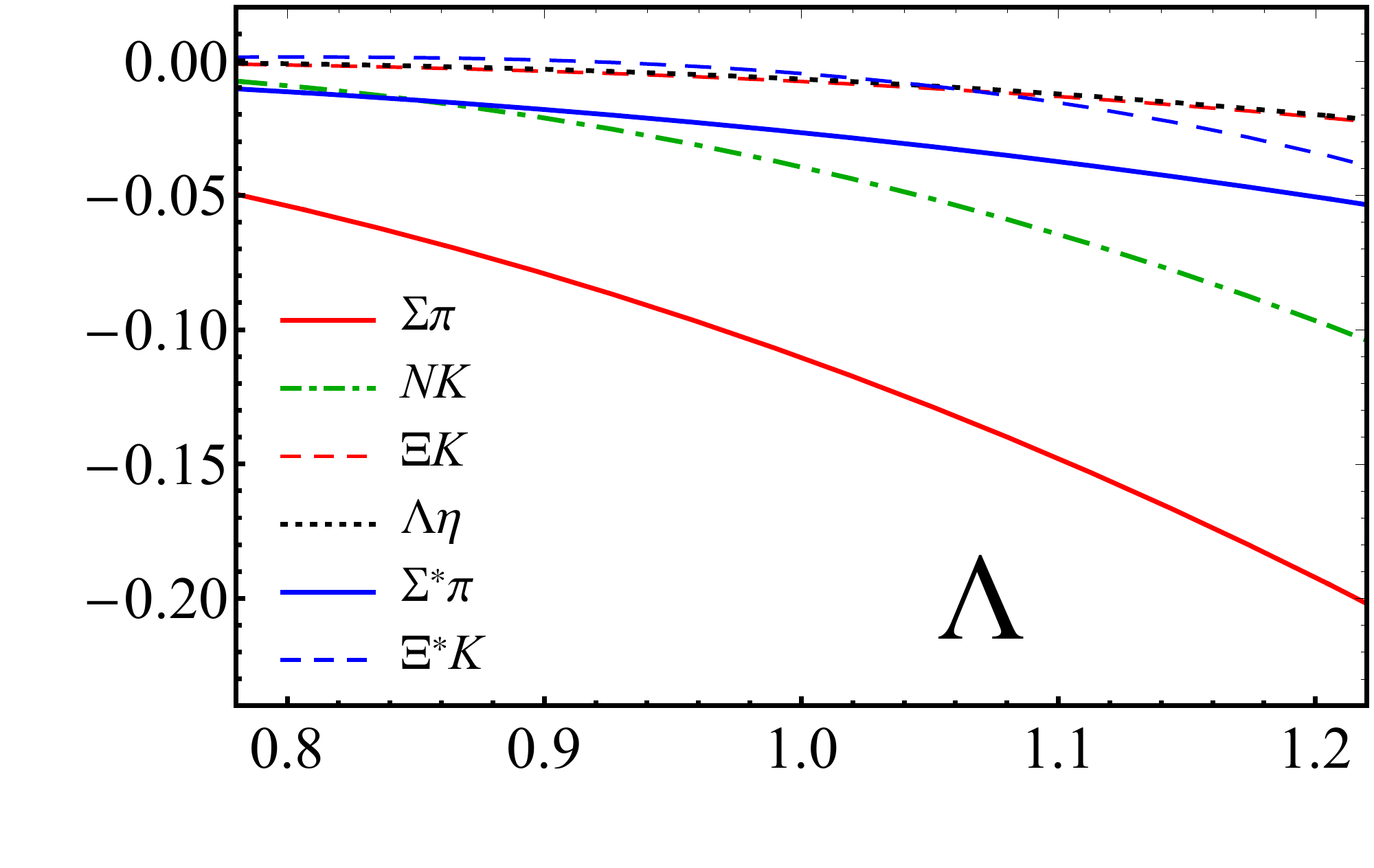}\\
\hspace*{-0.25cm}
\includegraphics[width=0.46\linewidth]{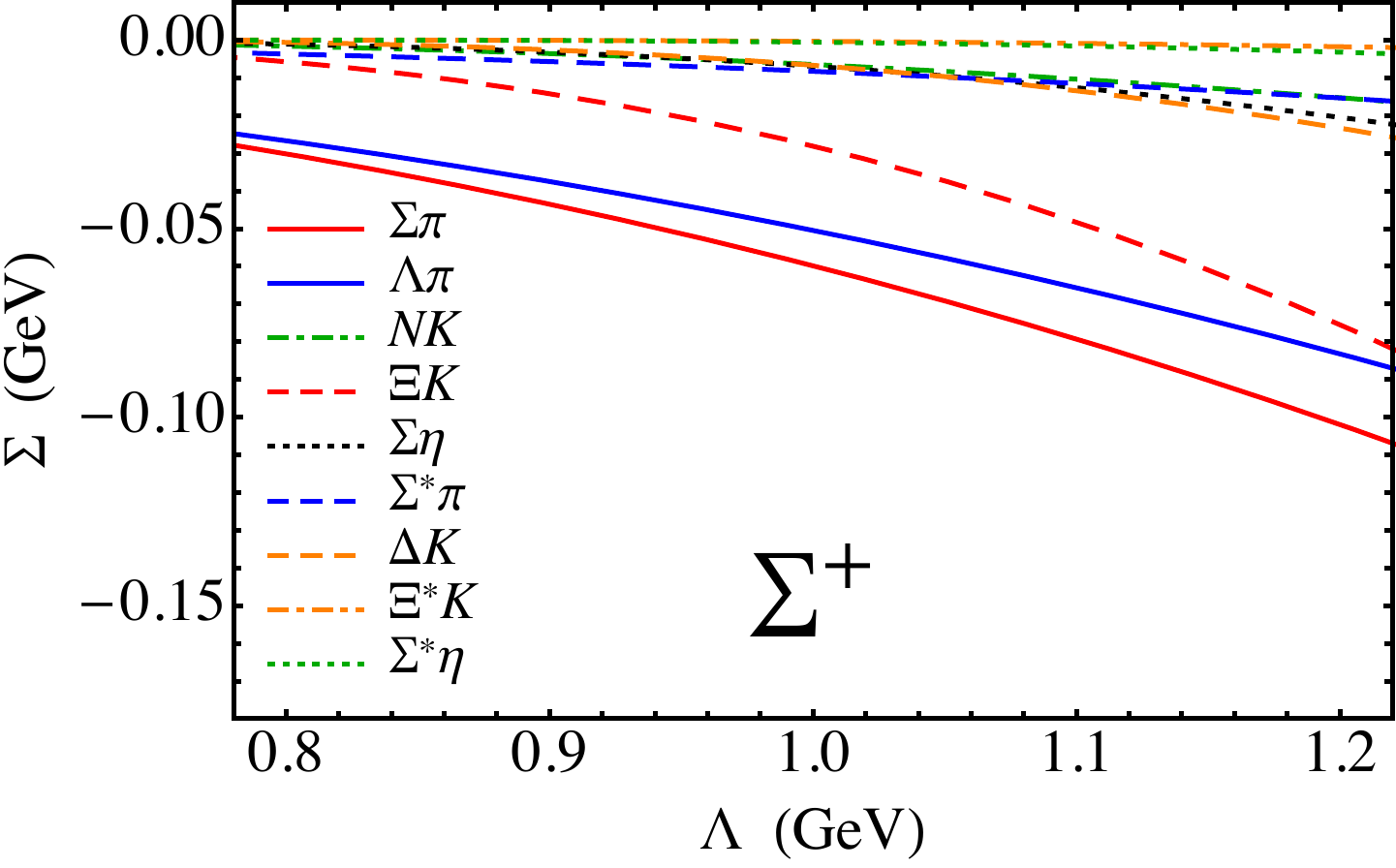}\hspace*{-0.3cm}
\includegraphics[width=0.46\linewidth]{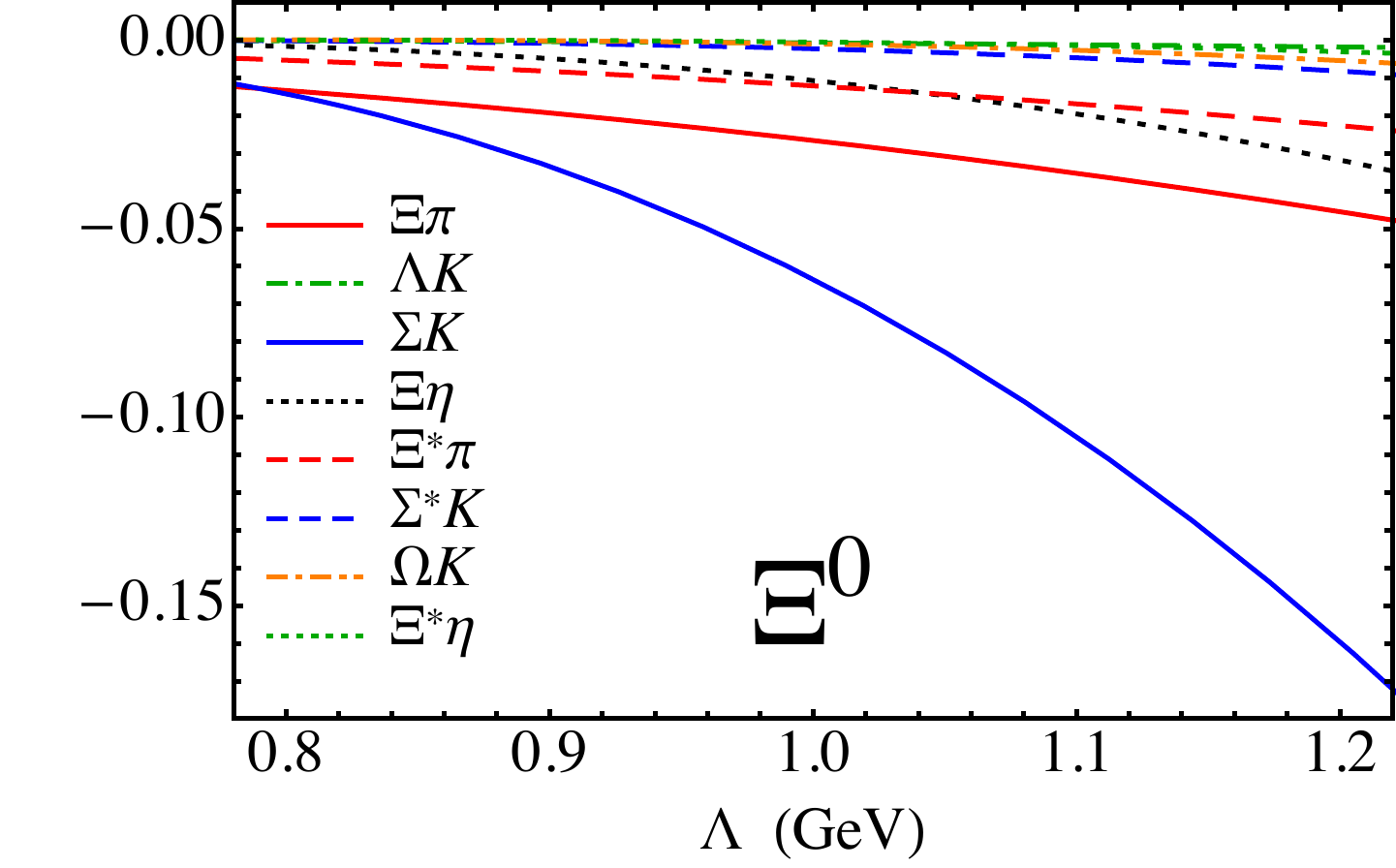}
\caption{Contributions to the self-energies of octet baryons from various meson-baryon intermediate states as a function of the dipole regulator mass parameter $\Lambda$, for the proton $p$, $\Lambda$, $\Sigma^+$, and $\Xi^0$ hyperons.}
\label{fig:B_Lambda}
\end{figure}

The results for the proton, $\Lambda$, $\Sigma^+$ and $\Xi^0$ hyperon self-energies are shown in Fig.~\ref{fig:B_Lambda} as a function of the dipole regulator, $\Lambda$, over a typical range $0.8 \lesssim \Lambda \lesssim 1.2$~GeV, at physical values of the meson and baryon masses.
Naturally, the magnitude of each of the self-energies increases with increasing $\Lambda$, and the general tendency is for the magnitude to decrease with increasing mass of the external baryon.
As an overall trend, the contributions from intermediate states with higher masses have a somewhat stronger variation with $\Lambda$, and the heavier external baryons receive significant contributions from a larger number of intermediate states.
Furthermore, the $K$ loop contributions become more significant for external baryons with larger strangeness, and, along with $\eta$ loops, play a slightly increasing role than the $\pi$ loop contributions for larger regulator masses.

As far as specific external states, for the proton self-energy, $\Sigma_p$, the $N \pi$ intermediate states make the greatest overall contribution, followed by the $\Delta \pi$ state.
Contributions involving kaons and hyperons are generally much smaller than those from nonstrange states, but become relatively more significant with increasing cutoff mass, and those involving the $\eta$ meson are negligible.

For the $\Lambda$ hyperon external state, the most significant contribution to the self-energy is from the $\Sigma \pi$ intermediate state, which is a factor $\approx 2$ smaller in magnitude than the most significant ($N \pi$) contribution to the proton self-energy.
Contributions from $N K$ and $\Sigma^* \pi$ intermediate states are $\approx 2-3$ smaller for the given range of regulator masses, with kaon loops generally playing a greater role than for the proton.

For the $\Sigma^+$, the contributions from the $\Sigma\pi$ and $\Lambda\pi$ intermediate state configurations are similar, but about a factor 2 smaller than the largest ($\Sigma\pi$) contribution to the $\Lambda$ hyperon for a regulator mass $\sim 1$~GeV.
The $\Xi K$ contribution is next largest, and in fact becomes comparable to the pionic contributions for masses $\gtrsim 1.2$~GeV, although at such values the one-loop approximation becomes more questionable.
The remaining contributions from the other intermediates states are mostly negligible.

In contrast to the other hyperons, for the $\Xi^0$ baryon the $\Sigma K$ loop makes the dominant contribution to the self-energy, which is much larger in magnitude than from any of the other states, with the exception perhaps of the $\Xi \pi$ state.
Interestingly, the largest self-energy contribution to the $\Xi^0$ baryon is not the diagonal ($\Xi \pi$) case, but the off-diagonal $\Sigma K$ channel.

\begin{figure}[t]
\centering
\includegraphics[width=0.46\linewidth]{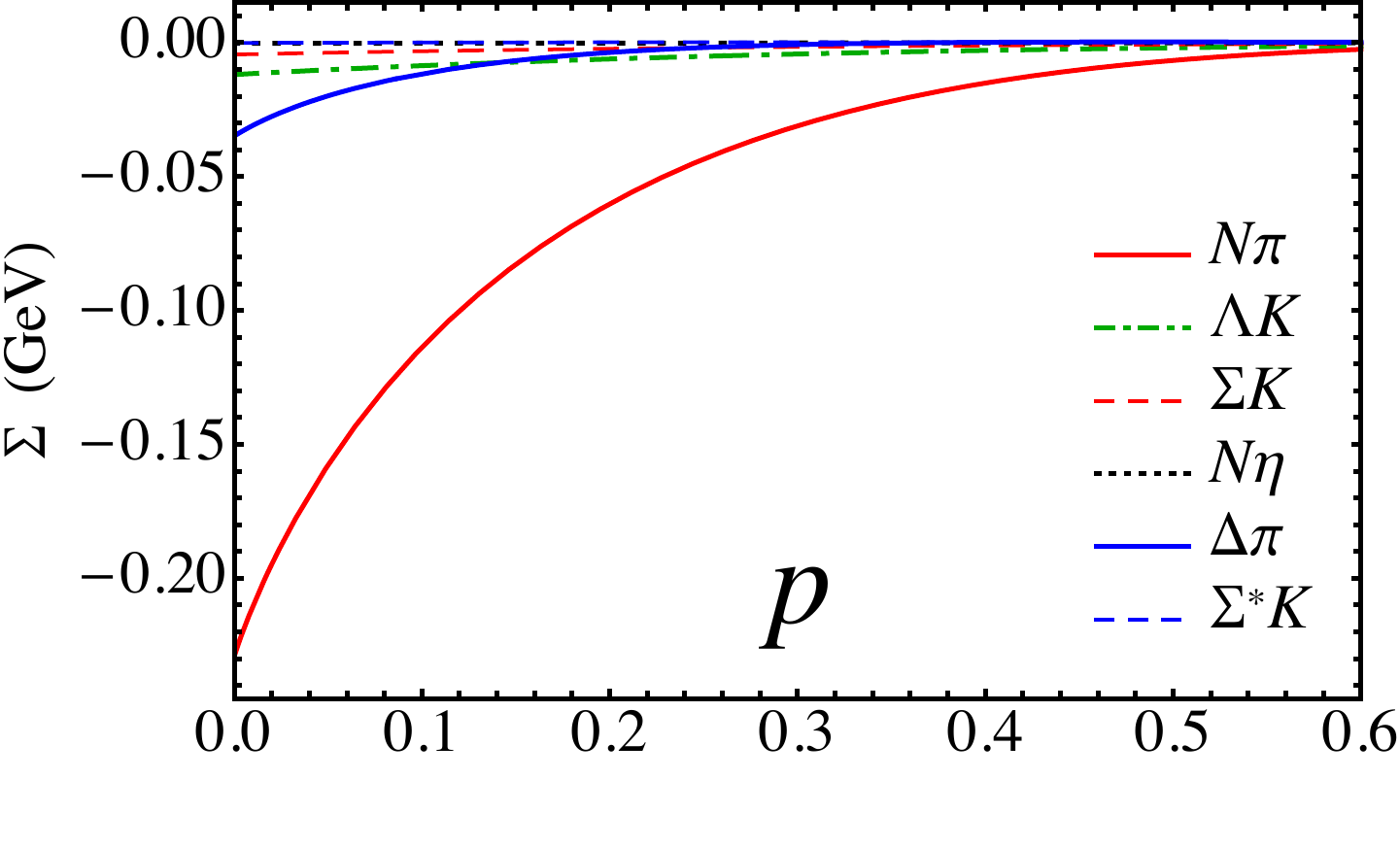}
\includegraphics[width=0.46\linewidth]{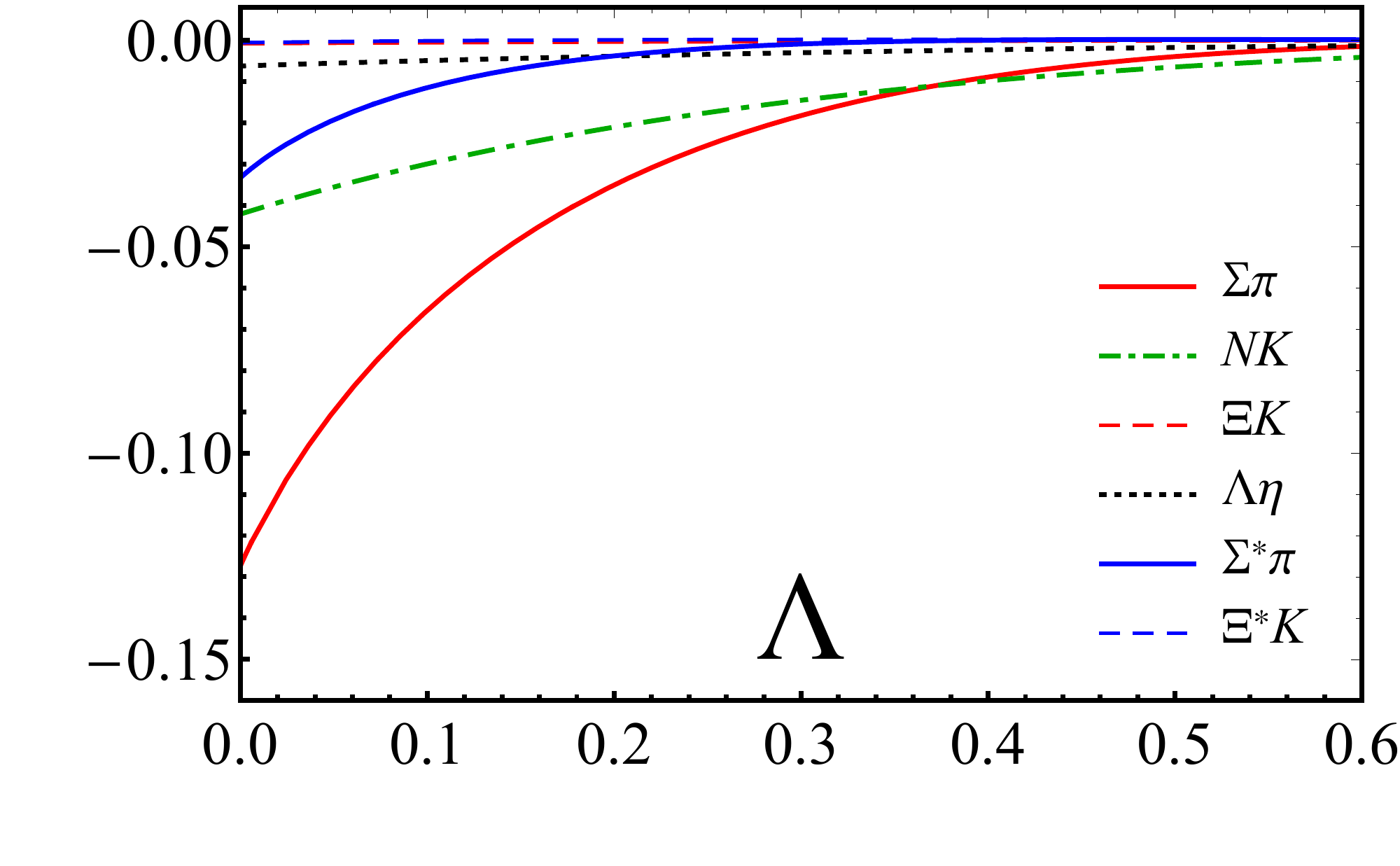}
\includegraphics[width=0.46\linewidth]{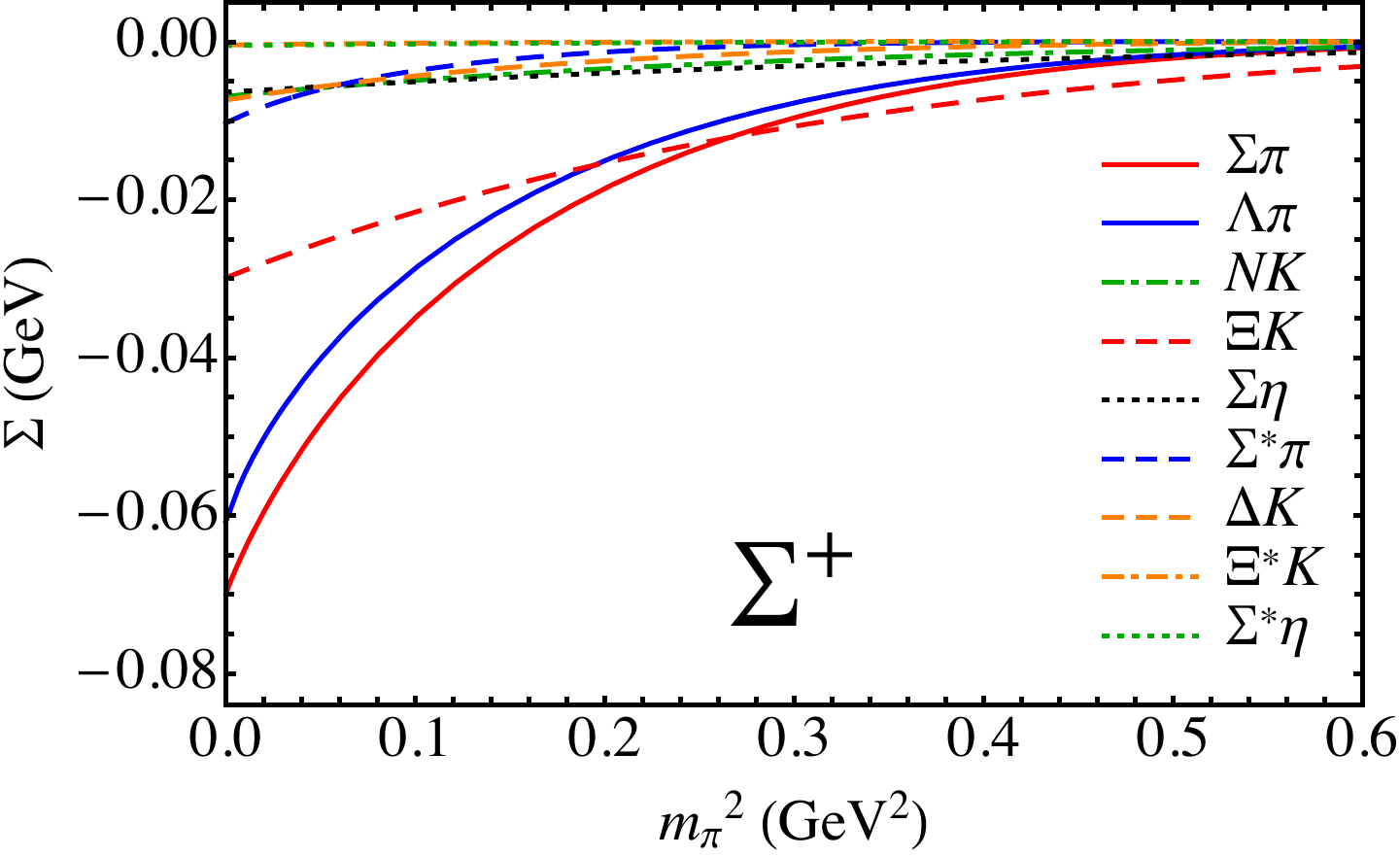}
\includegraphics[width=0.46\linewidth]{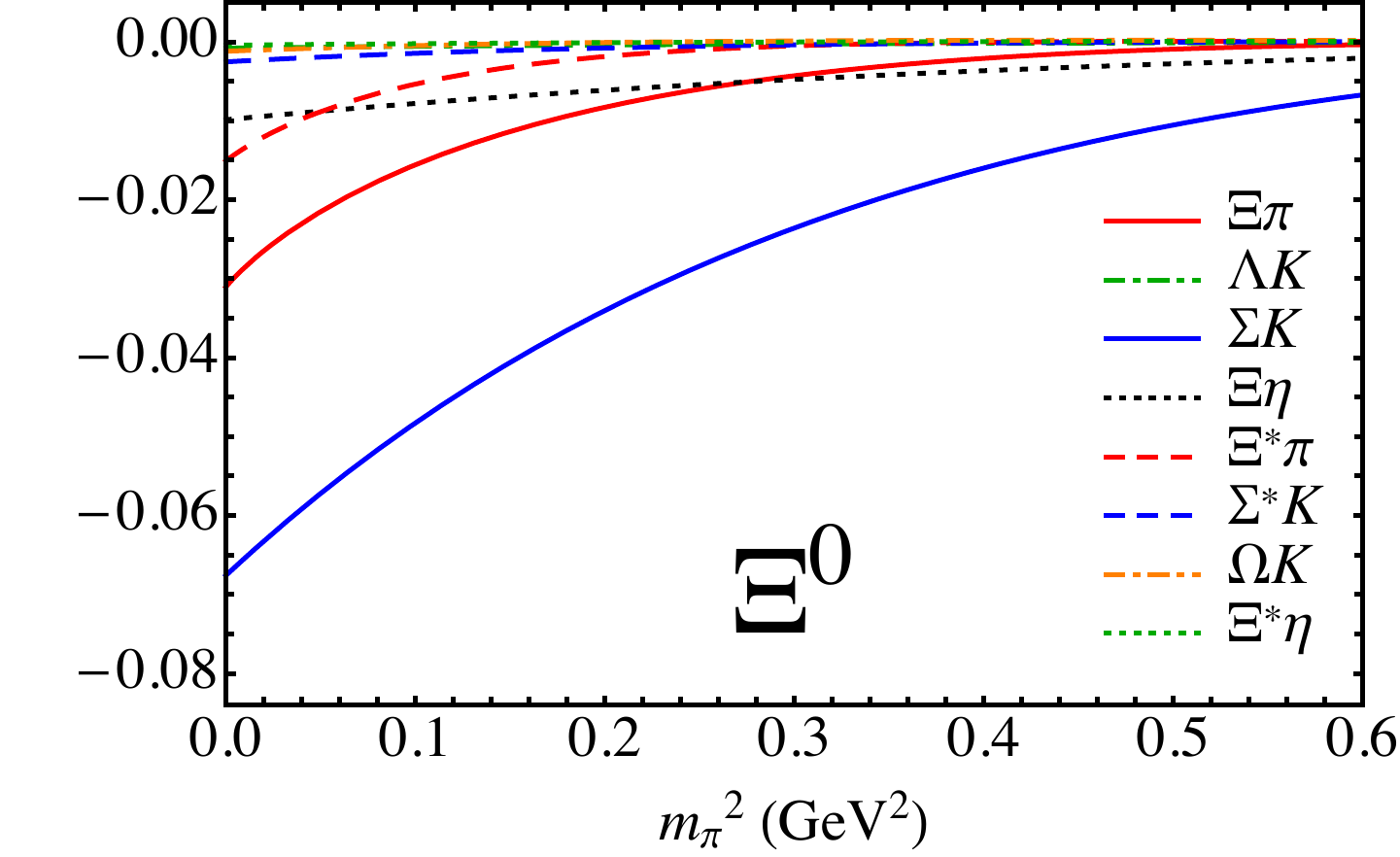}
\caption{Contributions to the self-energies of octet baryons from various meson-baryon intermediate states as a function of the pion mass squared, $m_\pi^2$, for the proton $p$, $\Lambda$, $\Sigma^+$, and $\Xi^0$ hyperons. For transitions involving $K$ and $\eta$ loops, the meson masses are written as functions of $m_\pi^2$ using Eq.~(\ref{eq: meson-pion relations}).}
\label{fig:B_mpi}
\end{figure}

It is also instructive to examine the dependence of the baryon self-energies on the pion mass squared, $m_\pi^2$, which we illustrate in Fig.~\ref{fig:B_mpi} for a fixed value of the regulator mass $\Lambda = 1$~GeV.
While nature provides us with only one physical value for $m_\pi$, the $m_\pi$ dependence can in principle be studied within lattice QCD, where any chosen value of the quark mass can be dialed and the simulation performed also at an unphysical pion mass.
The calculated mass dependence can then be compared with that expected from the QCD chiral analysis, or the latter can be utilized to extrapolate the lattice data from unphysically large masses to the physical ones.
In the numerical analysis of the pion mass dependence, for states involving $K$ and $\eta$ mesons we use the relations~\cite{Walker-Loud:2005}
\begin{equation}
\begin{aligned}
\label{eq: meson-pion relations}
    m_K^2    &=\, \frac{ 4 \lambda}{  f_\pi^2}\, m_s + \frac{m_\pi^2}{2},         \\
    m_\eta^2 &=\, \frac{16 \lambda}{3 f_\pi^2}\, m_s + \frac{m_\pi^2}{3},
\end{aligned}
\end{equation}
where $m_s$ is the strange quark mass and $\lambda$ is a fitting parameter.
Fitting the PACS-CS lattice QCD data for the baryon masses~\cite{Aoki:2008sm}, we find this parameter to be $\lambda = 0.00748$~GeV$^3$ and use $m_s = 0.0674$ GeV to reproduce the experimental kaon mass at the physical value of the pion mass.

At low values of the pion mass, the baryon masses can be expanded in a power series in $m_\pi^2$, with terms $\sim c_0 + c_2 m_\pi^2 + \cdots$ that are analytic in $m_\pi^2$, as well as terms that are nonanalytic.
The latter can only arise from pseudoscalar loops, so it is therefore instructive to examine deviations of the self-energies from linearity at small $m_\pi^2$.
The results in Fig.~\ref{fig:B_mpi} clearly indicate nonlinearity in the self-energies for $m_\pi^2 \lesssim 0.2-0.3$~GeV$^2$ for all the octet baryons, as observed in existing lattice simulations~\cite{Aoki:2008sm}, with greatest nonlinearity apparent for the proton and least for the $\Xi^0$.
For large values of $m_\pi^2$ the various meson--baryon contributions to the self-energies rapidly decrease, and in the limit $m_\phi \to \infty$ these vanish.
Closer inspection of the mass dependence of the individual intermediate states indicates that the contributions from $K$ and $\eta$ loops decrease in magnitude more slowly than the pion loop contributions, which can be understood from the pion mass dependence of $m_K^2$ and $m_\eta^2$ in Eq.~(\ref{eq: meson-pion relations}).

\subsection{Decuplet baryon self-energies}

For the self-energies of the decuplet baryons, as for the octet case, we evaluate the $k_\perp^2$, $y$ and $z$ integrals in Eqs.~(\ref{eq:decuplet-octet-k-}) and (\ref{eq:decuplet-decuplet-k-}) analytically and study the decuplet self-energies numerically as a function of the regulator mass parameter and pion mass squared.
In the numerical calculations we use the same value for the meson-octet-decuplet coupling as above, ${\cal C} = (6/5)\, g_A$, and for the meson-decuplet-decuplet coupling use the SU(6) result ${\cal H} = (9/5)\, g_A$.
The relations for the various meson-baryon coupling constants $C_{TB'\phi}$ and $C_{TT'\phi}$ are given in Table~\ref{tab:decuplet} of Appendix \ref{app:couplings}.

\begin{figure}[t]
\centering
\includegraphics[width=0.46\linewidth]{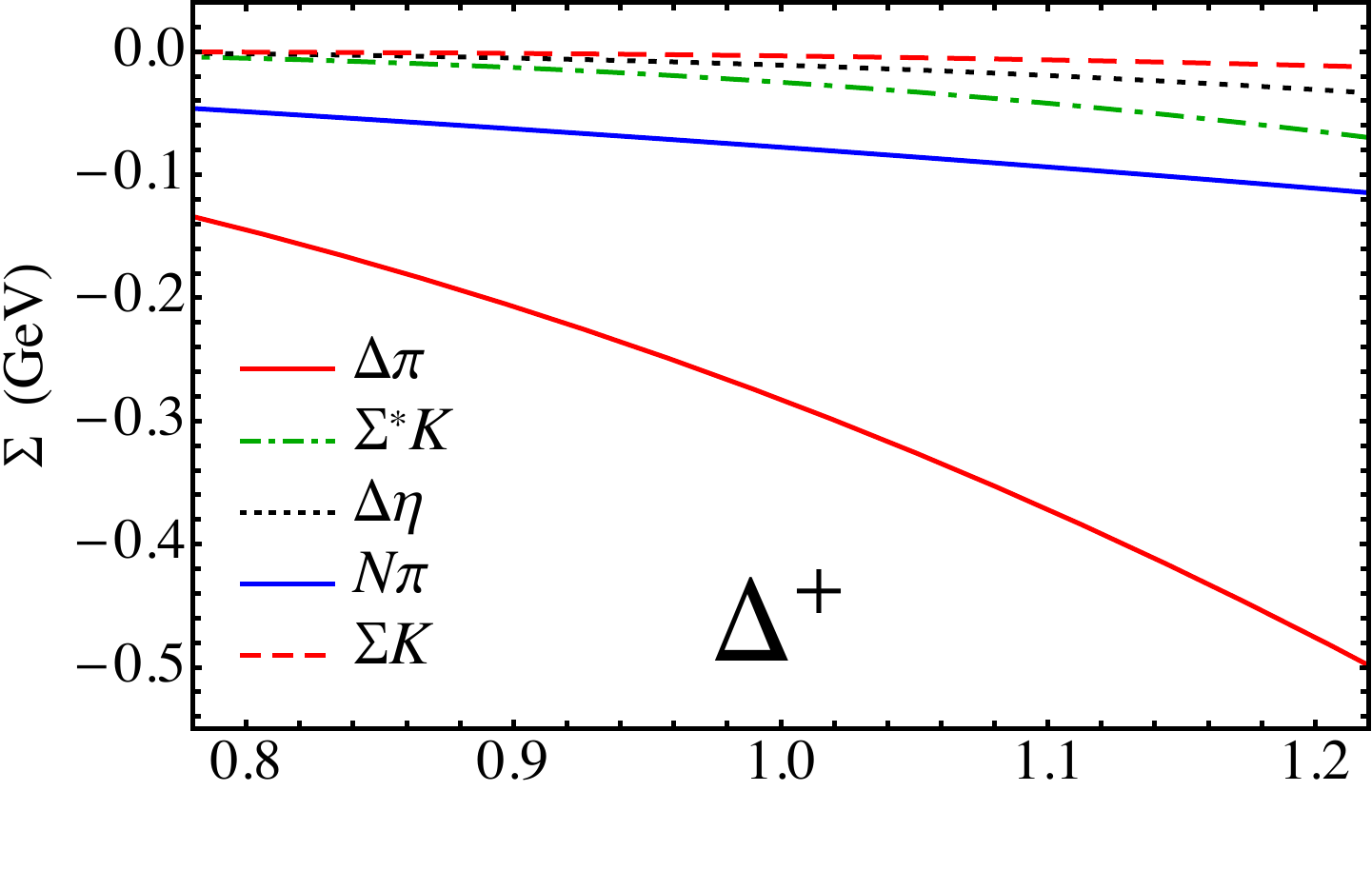}
\includegraphics[width=0.46\linewidth]{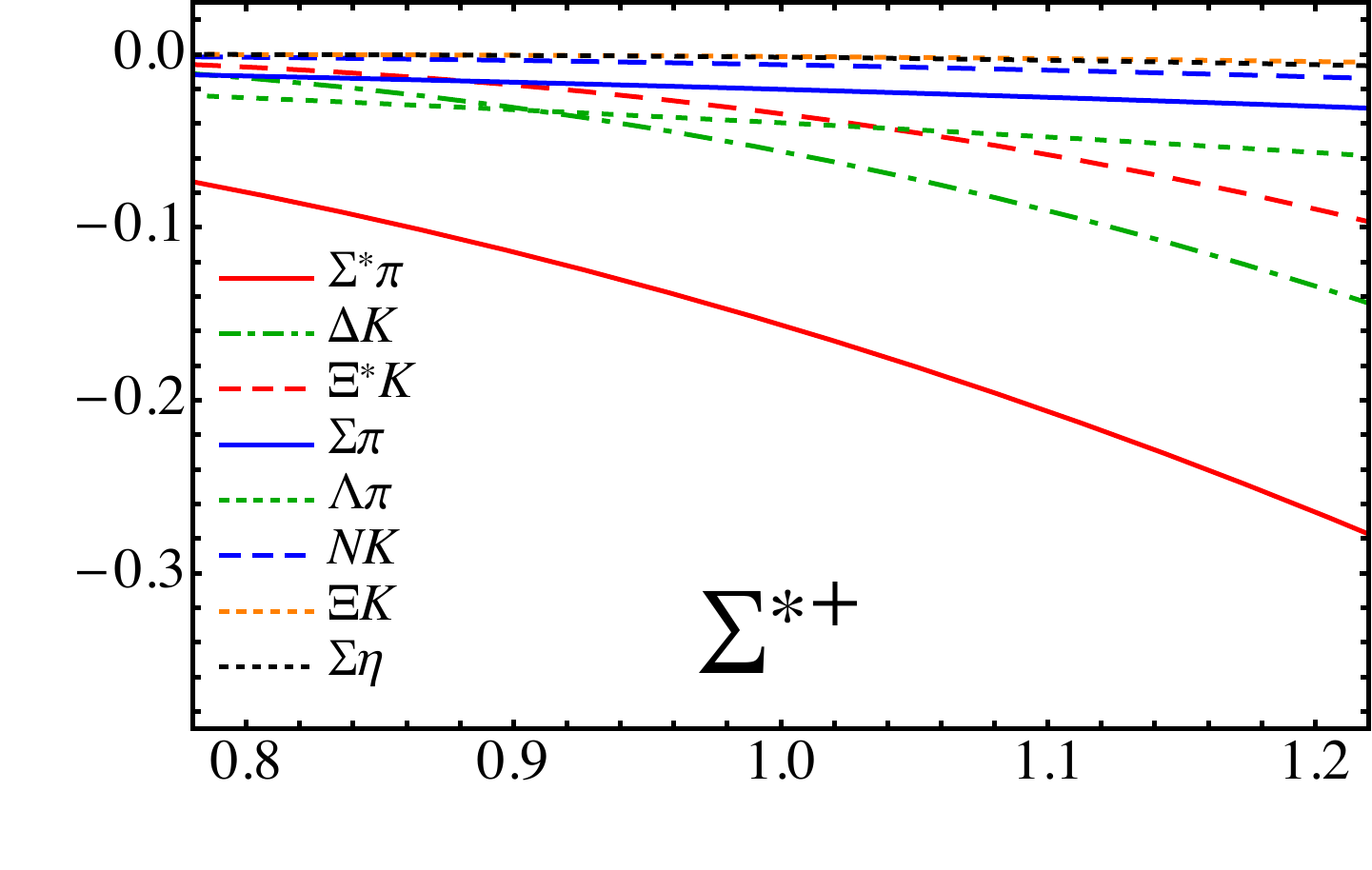}
\includegraphics[width=0.46\linewidth]{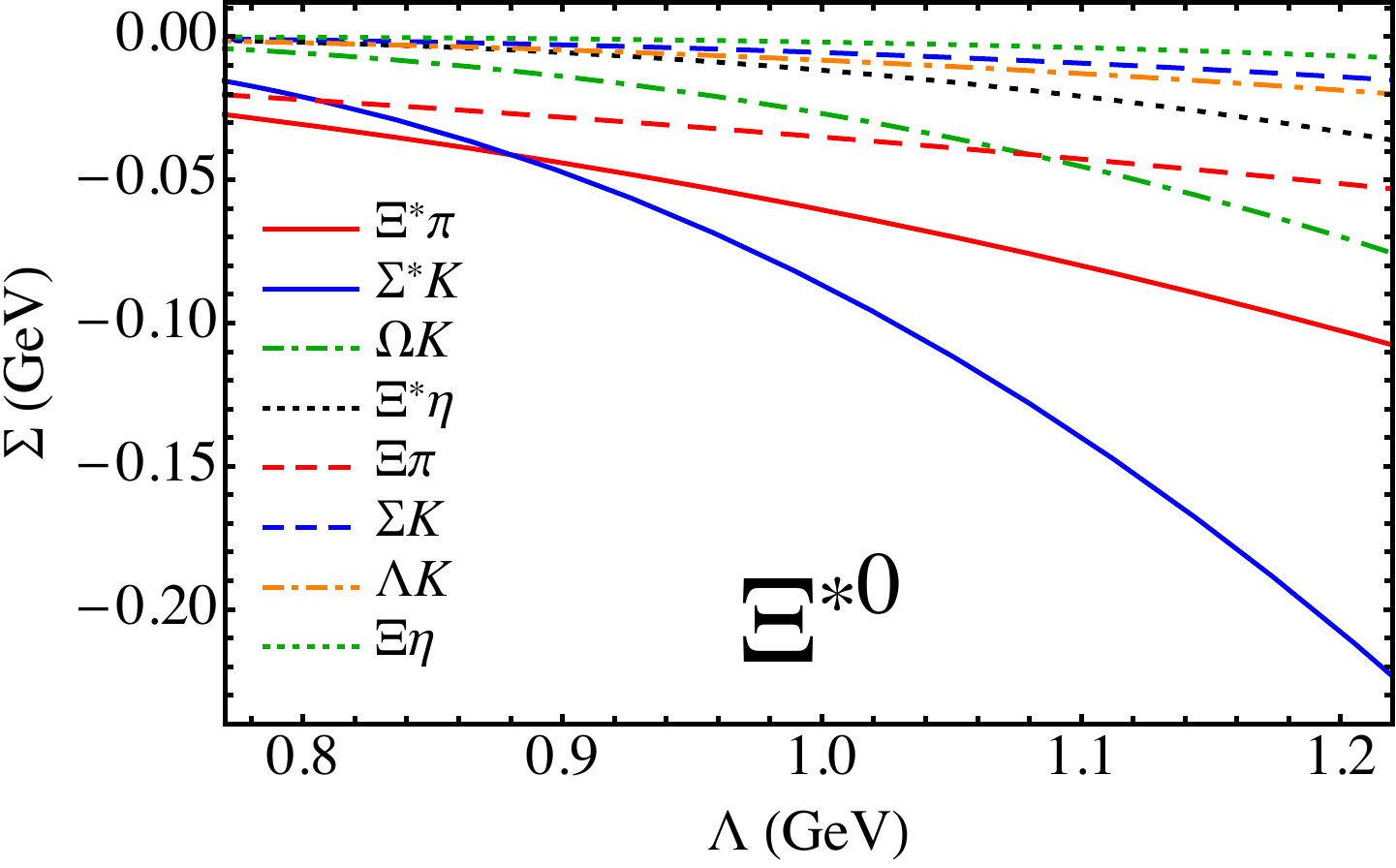}
\includegraphics[width=0.46\linewidth]{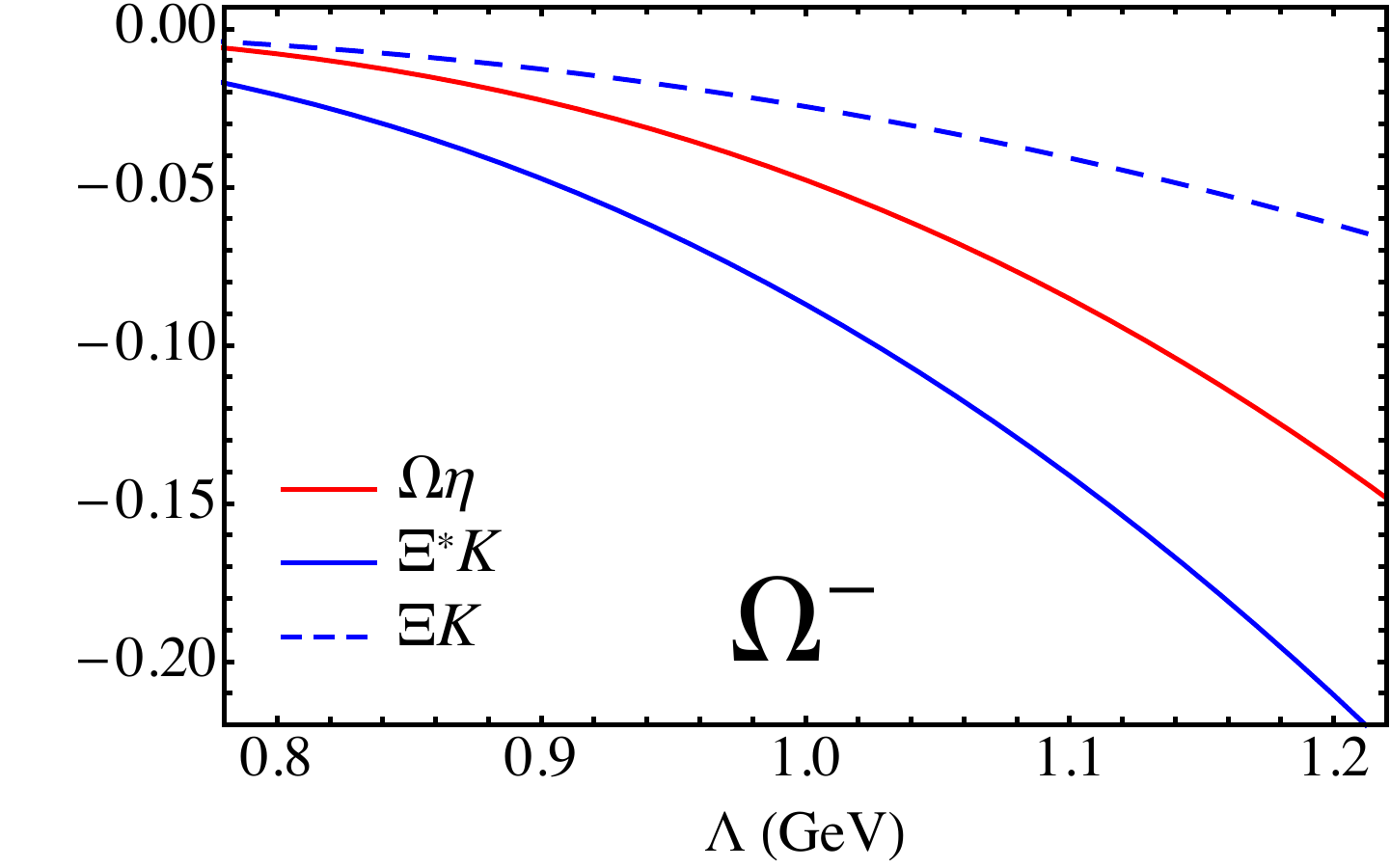}
\caption{Contributions to the self-energies of decuplet baryons from various meson-baryon intermediate states as a function of the dipole regulator mass parameter $\Lambda$, for the $\Delta^+$ isobar, $\Sigma^{*+}$, $\Xi^{*0}$ and $\Omega^-$ hyperons.}
\label{fig:T_L}
\end{figure}

The behavior observed for the decuplet baryon self-energies is qualitatively similar to that for the octet baryons, but with some unique features.
The dependence on the regulator mass is displayed in Fig.~\ref{fig:T_L}, where again the magnitudes of the self-energies are seen to increase with $\Lambda$.
The largest contributions generally arise from intermediate states involving decuplet baryons, and contributions from intermediate states with kaons increase at a somewhat faster rate than their pion counterparts.

For the $\Delta^+$ external state, the diagonal $\Delta \to \Delta \pi$ transition dominates over all other contributions over the entire range of $\Lambda$ considered.
The $N \pi$ intermediate state gives the second largest contribution, while those involving $K$ and $\eta$ loops are relatively insignificant.
For larger $\Lambda$ values the magnitude of the $\Sigma^* K$ contribution increases at a faster rate, and eventually exceeds the $N \pi$.
However, as mentioned previously, the behavior at large regulator masses is questionable because of the increasing importance of higher order terms that are not included in this analysis.

The diagonal $\Sigma^* \pi$ contribution to the $\Sigma^{*+}$ self-energy is also significantly larger than other terms, although a number of other states, such as the $\Delta K$, $\Xi^* K$, $\Lambda \pi$ and $\Sigma \pi$, make non-negligible contributions.
At smaller values of $\Lambda \lesssim 1$~GeV the contributions from the (octet baryon) $\Lambda \pi$ and $\Sigma \pi$ states are actually larger than those from the (decuplet baryon) $\Delta K$ and $\Xi^* K$, however, at larger values of $\Lambda \gtrsim 1$~GeV the latter increase rapidly and become more prominent.

The $\Xi^{*0}$ baryon self-energy displays an interesting feature in that the largest two contributions, namely, from the $\Xi^* \pi$ and $\Sigma^* K$ intermediate states, switch their order at $\Lambda \approx 0.9$~GeV, with the latter becoming much larger at higher values of $\Lambda$.
For lower $\Lambda$ values, the $\Xi \pi$ state also makes an important contribution, but its relative impact decreases with increasing $\Lambda$.
This feature, as evident also for the $\Sigma^{*+}$ self-energies, illustrates the general trend of the contributions involving decuplet baryons playing a more prominent role than those involving octet baryons at larger $\Lambda$.
This is of course expected from the fact that decuplet baryon propagators involve higher powers of the loop momentum, which are less suppressed for larger values of $\Lambda$.

Finally, for the triply strange $\Omega$ baryon there are only three intermediate states that preserve strangeness which can contribute to the self-energy, namely, two involving decuplet baryons, $\Xi^* K$ and $\Omega \eta$, and one involving the octet $\Xi K$.
The $\Xi^* K$ gives the largest contribution, followed by the diagonal $\Omega \eta$, while the octet contribution $\Xi K$ has the smallest magnitude.

\begin{figure}[t]
\centering
\includegraphics[width=0.46\linewidth]{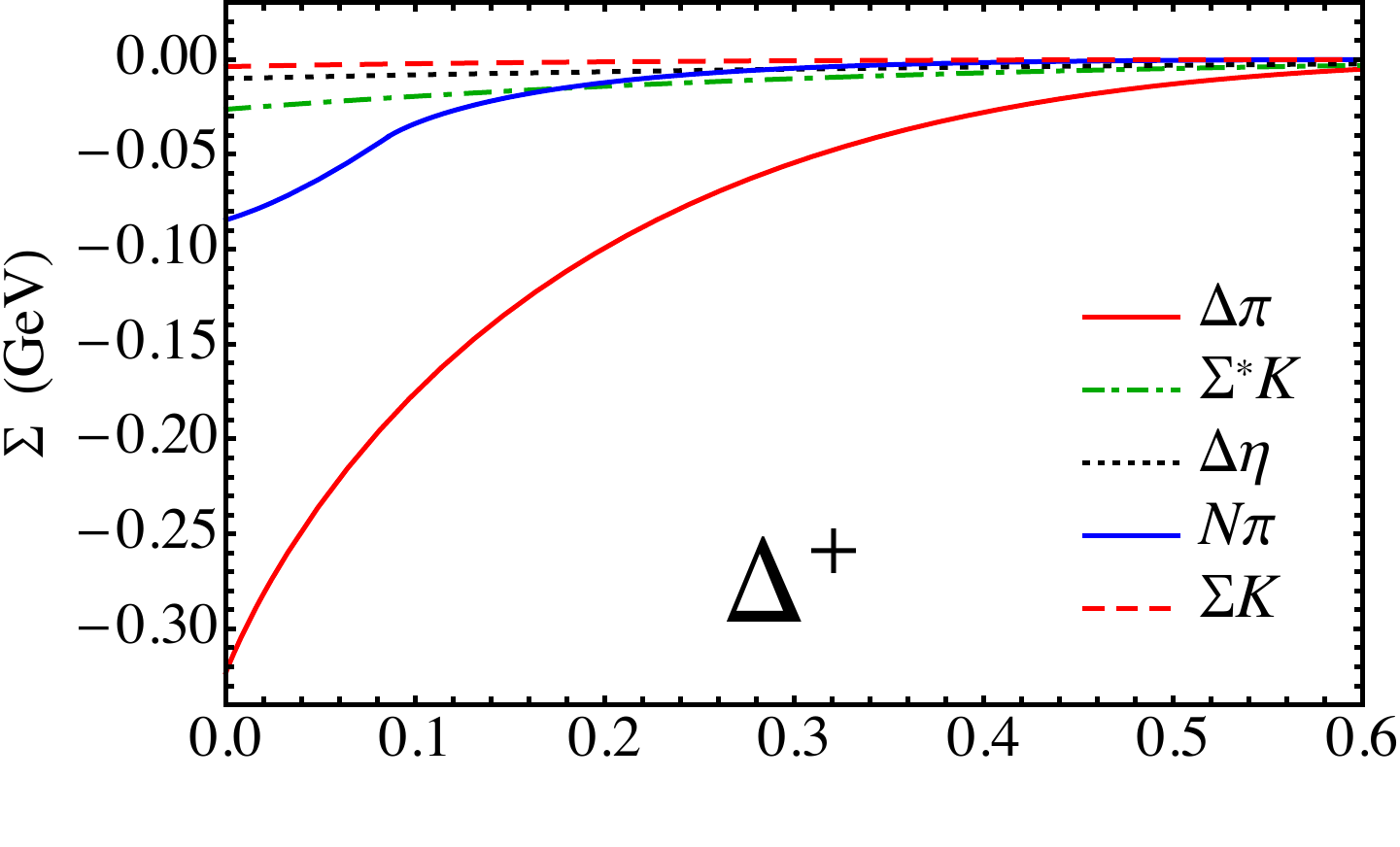}
\includegraphics[width=0.46\linewidth]{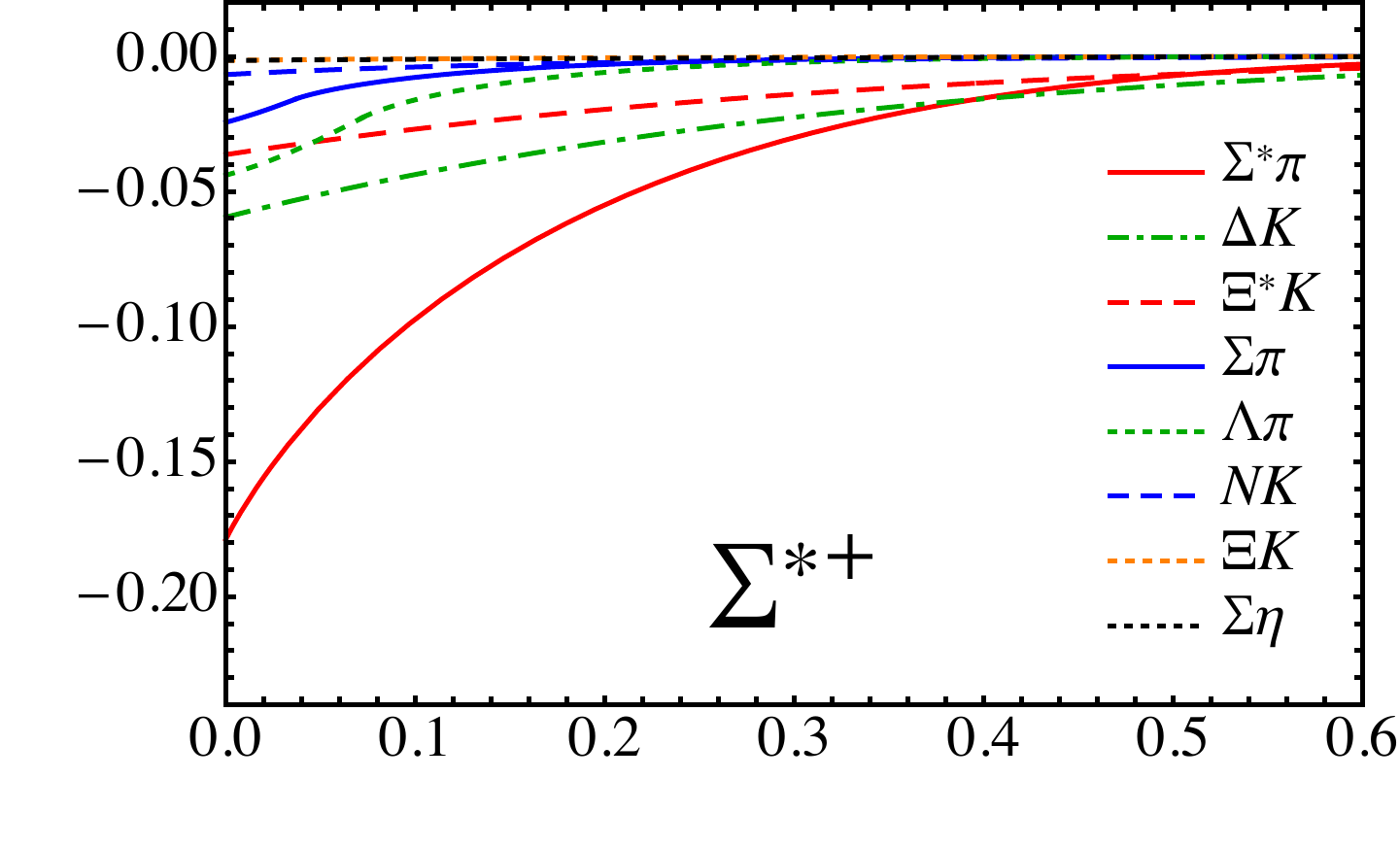}
\includegraphics[width=0.45\linewidth]{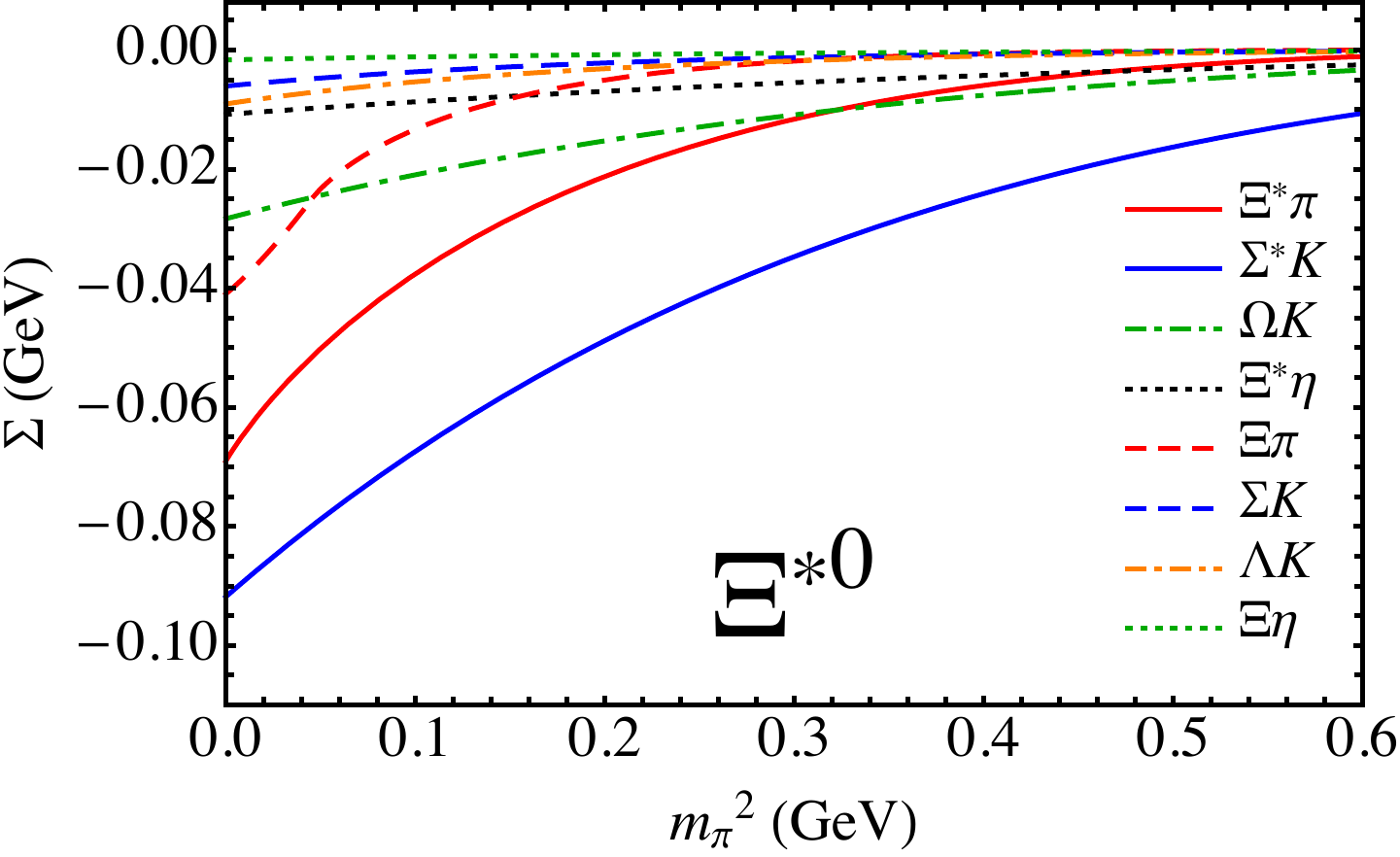}
\includegraphics[width=0.45\linewidth]{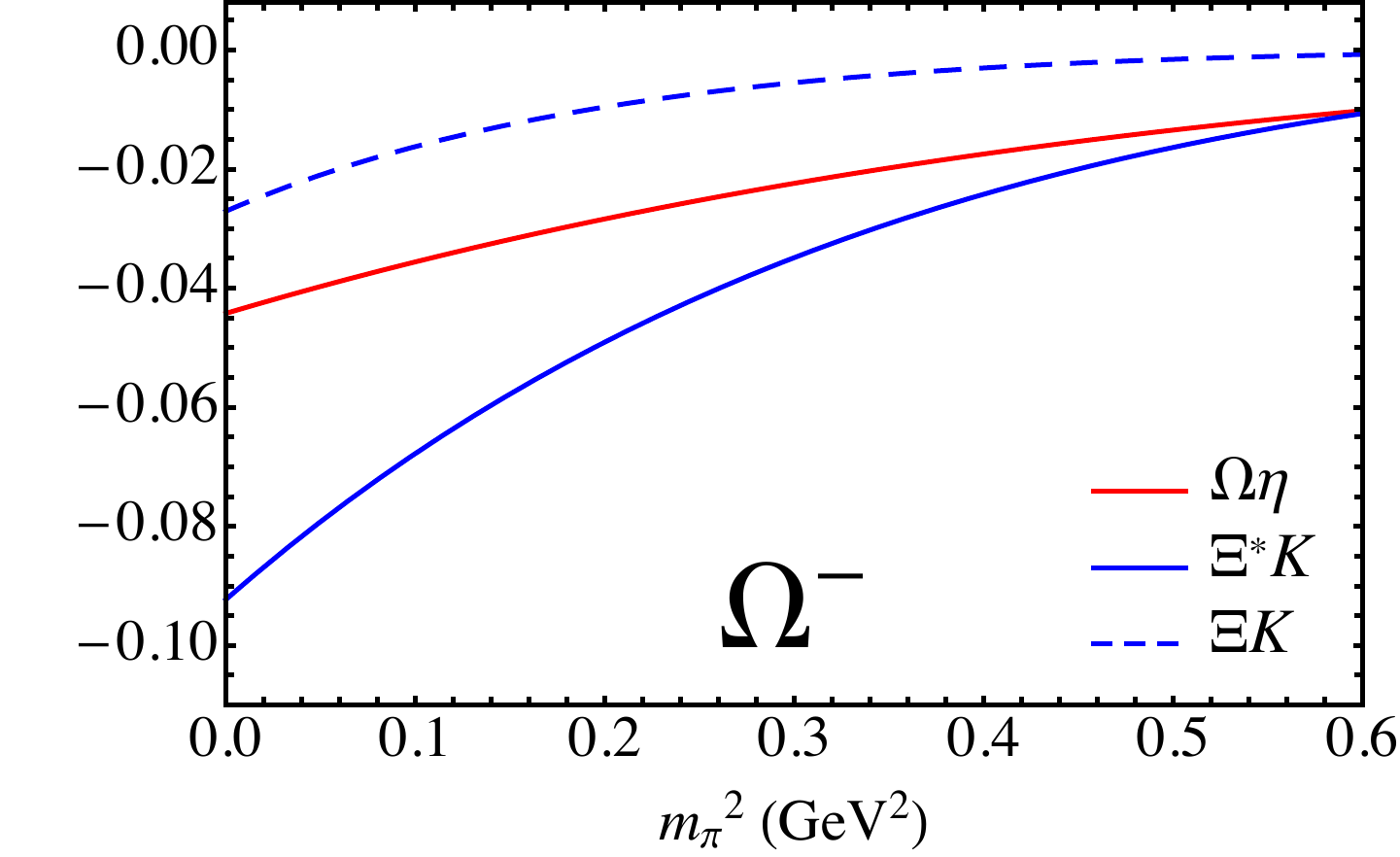}
\caption{Contributions to the self-energies of decuplet baryons from various meson-baryon intermediate states as a function of the pion mass squared, $m_\pi^2$, for the $\Delta^+$, $\Sigma^{*+}$, $\Xi^{*0}$ and $\Omega^-$. For transitions involving $K$ and $\eta$ loops, the meson masses are written as functions of $m_\pi^2$ using Eq.~(\ref{eq: meson-pion relations}).}
\label{fig:T_mpi}
\end{figure}

In a similar manner to the octet self-energies, it is also instructive to examine the variation of the decuplet baryon self-energies with respect to the pion mass squared.
Choosing again the nominal value $\Lambda = 1$~GeV for the finite range regulator mass, we illustrate the octet baryon and decuplet baryon intermediate state contributions to the decuplet self-energies in Fig.~\ref{fig:T_mpi}, using the relations in Eq.~(\ref{eq: meson-pion relations}) to express the $K$ and $\eta$ masses in terms of the pion mass. 
Generally similar behavior is observed for the decuplet baryon self-energies to that for the octet self-energies when varying $m_\pi$, albeit with one rather striking difference.
For mass differences between the external and internal baryons
    ($M_T - M_{\cal B'} = -\Delta_{{\cal B'}T}$) 
greater than the meson mass, the self-energies acquire an imaginary part, as discussed above in Sec.~\ref{Decuplet-octet} and Sec.~\ref{sec: Decay rates}.
This results in noticeable kinks in the self-energies when $m_\phi = -\Delta_{{\cal B'}T}$, below which the decay channels become open.

For the $\Delta \to N \pi$ self-energy, for example, the branch point is clearly seen at $m_\pi \approx 0.293$~GeV, corresponding to the difference between the nucleon and $\Delta$ masses.
For the $\Sigma^{*+}$ baryon, branch points are observed for the $\Lambda \pi$ channel at $m_\pi = 0.267$~GeV and for the $\Sigma \pi$ channel at $m_\pi = 0.19$~GeV.
For the $\Xi^{*0}$ baryon the $\Xi \pi$ intermediate state has a clearly visible branch point at $m_\pi = 0.213$~GeV, while the $\Omega^-$ has no branch points when varying with respect to the pion mass.

Note also that since these results show the variation with respect to $m_\pi^2$ and the $K$ and $\eta$ masses are written using Eq.~(\ref{eq: meson-pion relations}), there are no observable branch points for any intermediate states involving kaons or $\eta$ mesons.
This can be understood from Eq.~(\ref{eq: meson-pion relations}) by noting that at $m_\pi=0$ the $K$ mass is given by
    $m_K = \sqrt{(4\lambda/f_\pi^2)\, m_s} = 0.482$~GeV
and the $\eta$ mass by
    $m_\eta = \sqrt{(16\lambda/3 f_\pi^2)\, m_s} = 0.557$~GeV,
which are larger than the largest baryon mass difference,
$-\Delta_{\Sigma^* N} = M_{\Sigma^*} - M = 0.444$~GeV.

\subsection{Analysis}

\begin{figure}[t]
\centering
\includegraphics[width=0.46\linewidth]{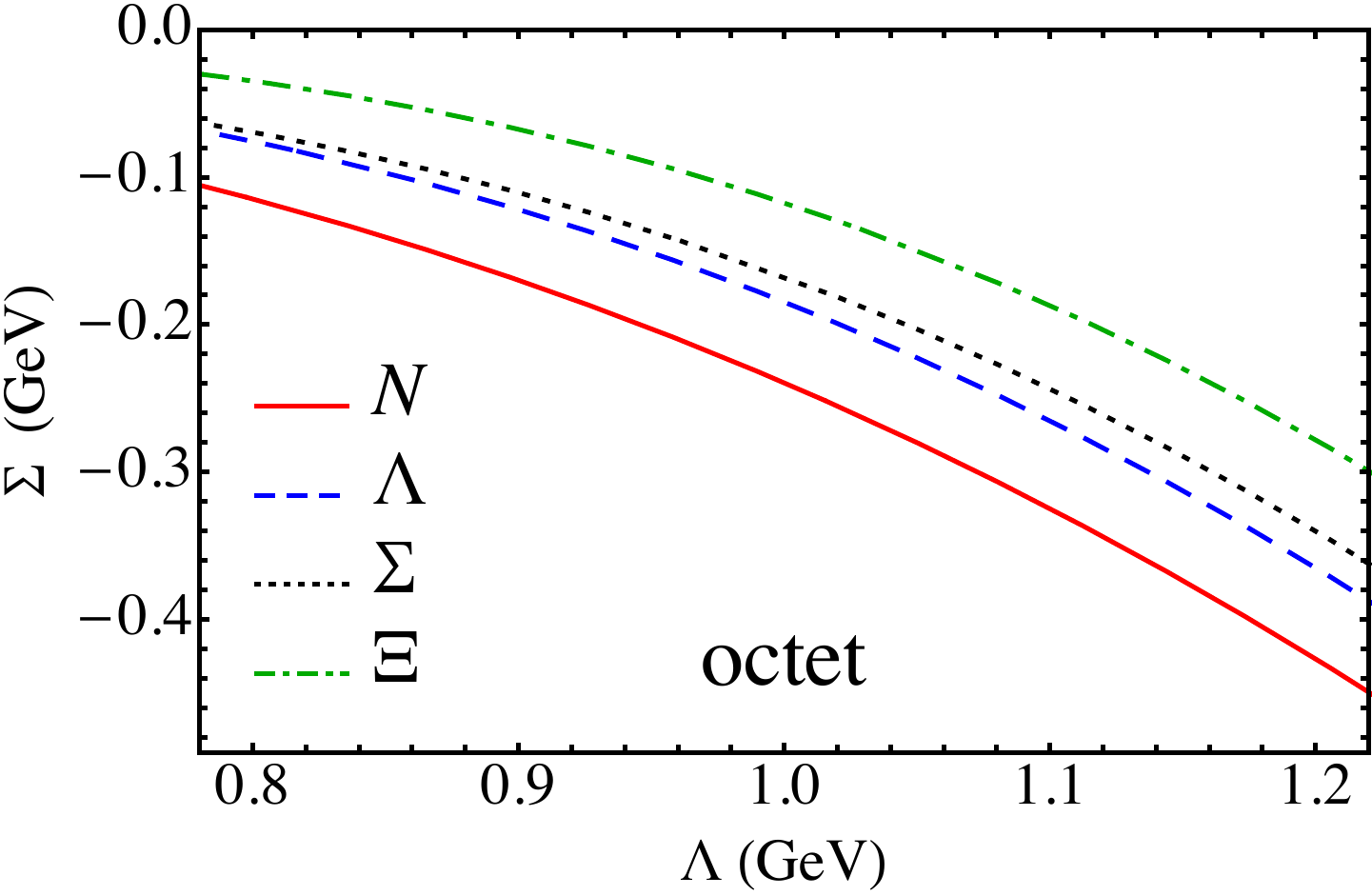}
\includegraphics[width=0.46\linewidth]{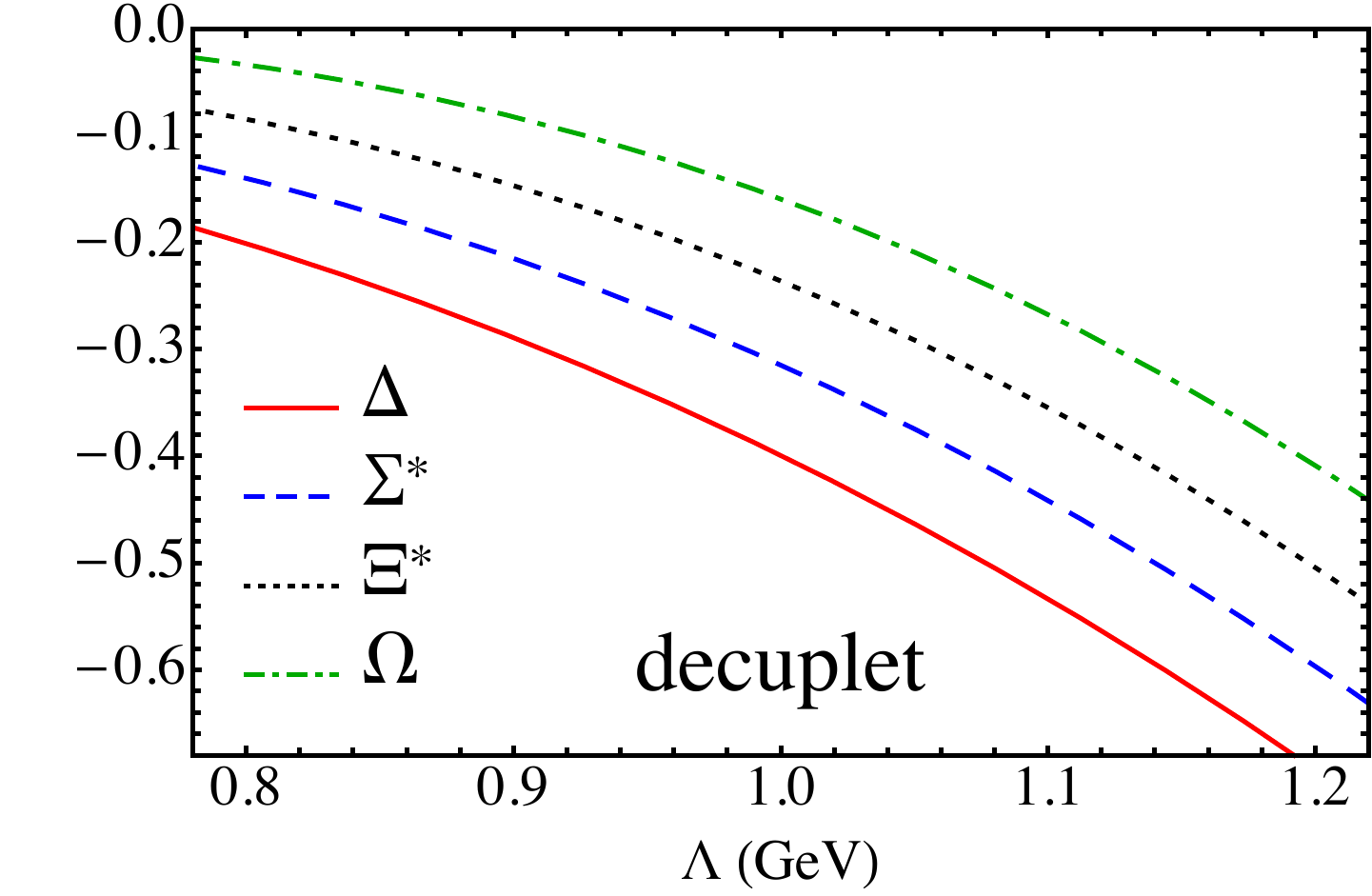}

\vspace*{0.5cm}
\includegraphics[width=0.47\linewidth]{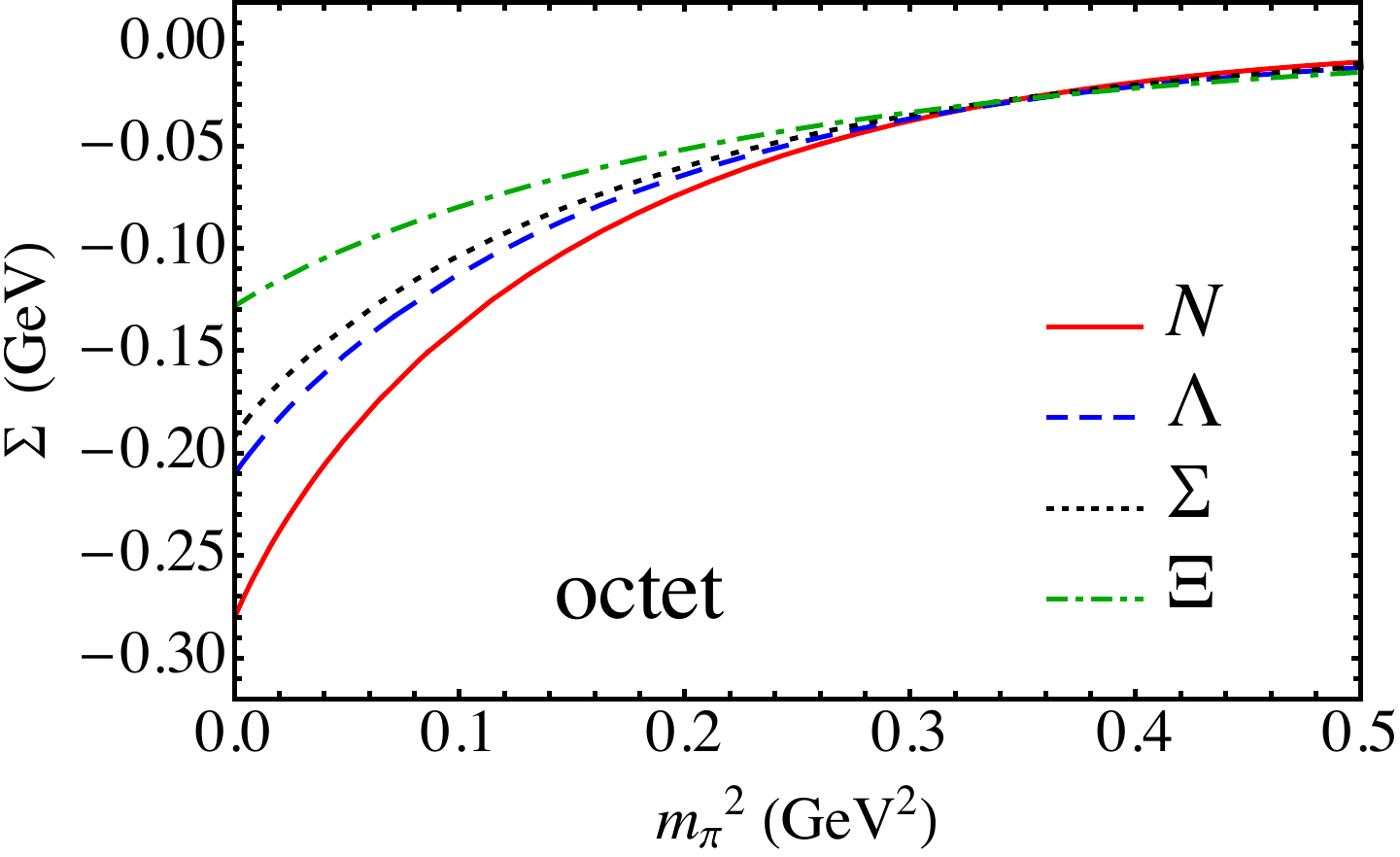}
\includegraphics[width=0.46\linewidth]{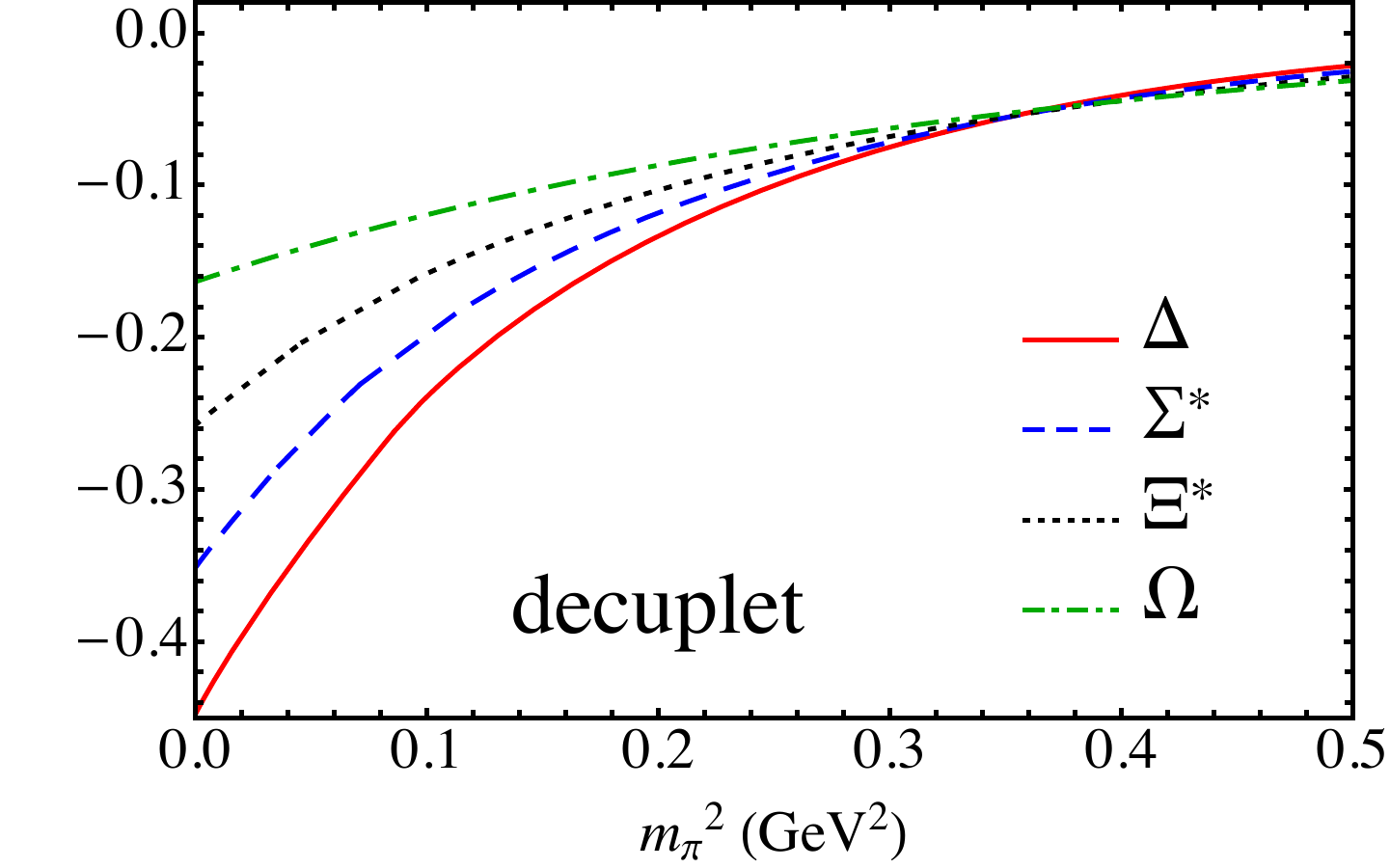}
\caption{Total self-energies for octet (left) and decuplet (right) baryons as functions of $\Lambda$ (top) and $m_\pi^2$ (bottom).}
\label{fig:Total_SEs}
\end{figure}

Putting all these results together, in Fig.~\ref{fig:Total_SEs} we show a comparison of the total self-energies for all the octet and decuplet baryons versus the regulator mass parameter and versus the pion mass squared, as in Figs.~\ref{fig:B_Lambda}--\ref{fig:T_mpi}.
The general trend is for the magnitude of the self-energies to become larger the lighter the baryon is.
For the same value of the regulator mass, the decuplet baryon masses are slightly larger than the corresponding octet baryon masses, for the same value of the strangeness.
For example, the $\Delta$ baryon self-energy is $\approx -0.4$~GeV at $\Lambda=1$~GeV, compared with the nucleon's $\approx -0.2$~GeV, while the $\Xi^*$ baryon self-energy of $\approx -0.2$~GeV is about twice as large in magnitude as the $\Xi$ self-energy at the same value of the regulator mass.

In the comparison of the total decuplet self-energies with varying $m_\pi^2$, the kinks at low $m_\pi$ values are clearly noticeable, and correspond to the points where the decay channels open and the self-energies develop an imaginary part.
For larger $m_\pi$ values, each of the four octet and each of the four decuplet state self-energies cross over at the SU (3) symmetry point, where the masses of the strange and nonstrange quarks coincide.

It should be noted that closer inspection of the intersect region does reveal some slight SU(3) symmetry breaking in our case, which arises from differences between the various baryon masses used in the propagators and spin trace factors in the expressions for the self-energies.
If one were to use the same external mass and mass difference for all possible transitions, SU(3) symmetry would be exact and all lines would intersect at the same point, as observed in Refs.~\cite{Shanahan:2011, Shanahan:2012wh} for example.

We also stress that if all possible transitions were not considered for each octet and decuplet baryon, the intersections of the various curves in Fig.~\ref{fig:Total_SEs} would not occur.
This further emphasizes the importance of a comprehensive approach including all octet and decuplet baryon intermediate state contributions to the self-energies. 

\begin{figure}[t]
\centering
\includegraphics[width=0.45\linewidth]{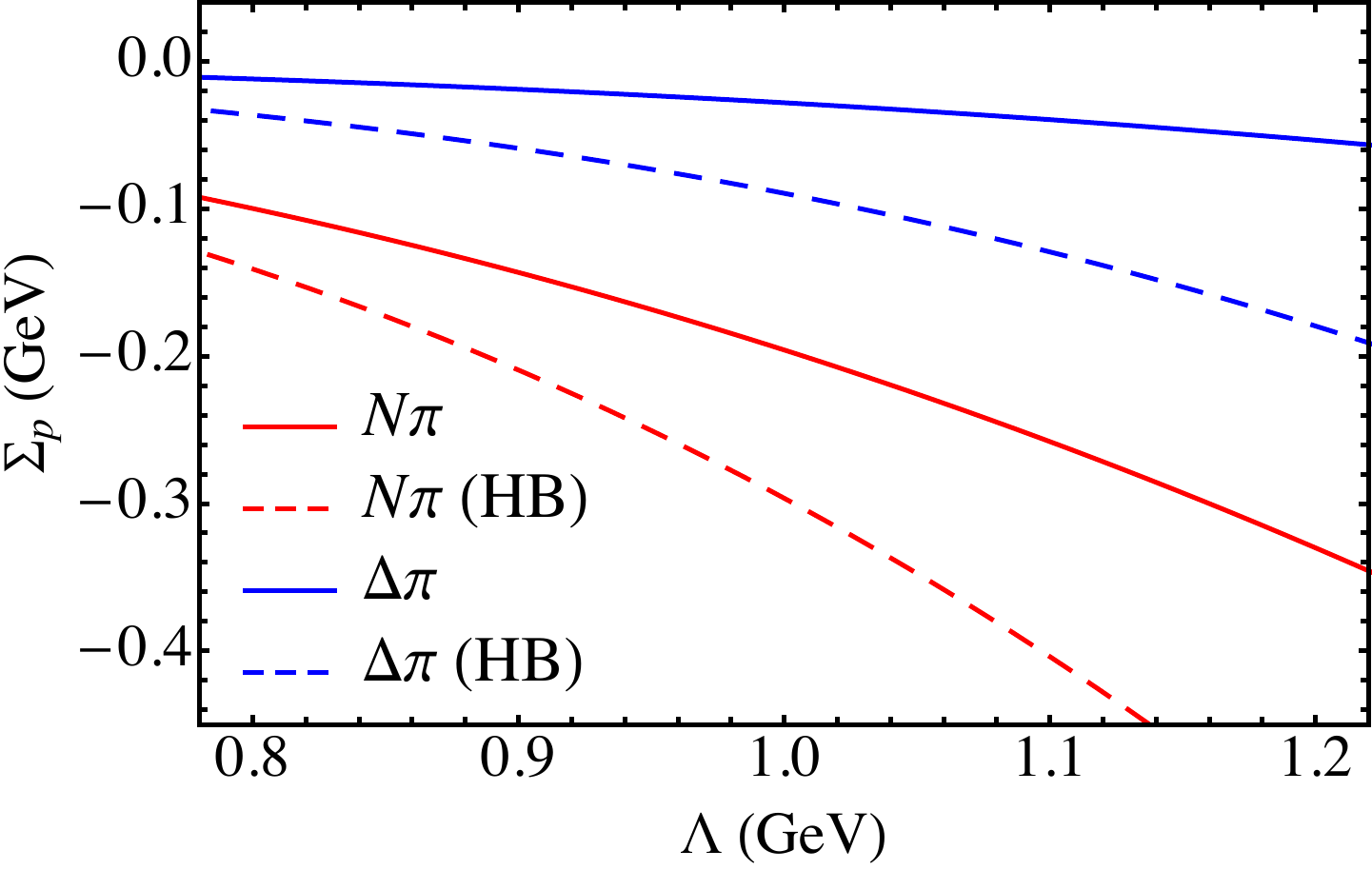}
\includegraphics[width=0.461\linewidth]{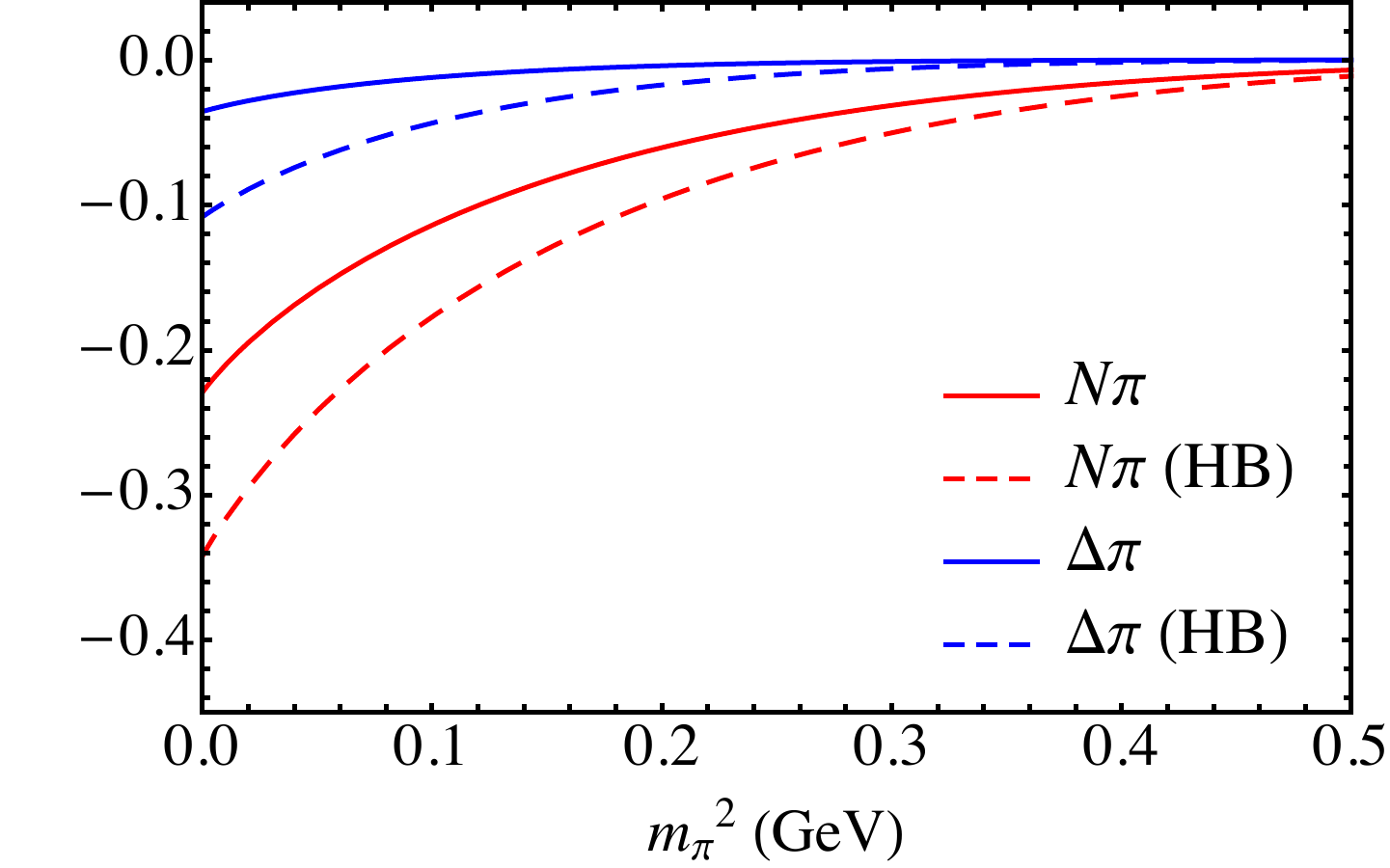}
\caption{Comparison of $N \pi$ (red lines) and $\Delta \pi$ (blue lines) intermediate state contributions to the proton self-energy, $\Sigma_p$, versus the regulator mass parameter $\Lambda$ and versus $m_\pi^2$, for our full relativistic calculation (solid lines) and the heavy baryon approximation (dashed lines).}
\label{fig:ProtonHBandRel}
\end{figure}

Finally, before concluding our discussion it is instructive to compare the results of our relativistic calculations with those obtained in the heavy baryon limit, which has been used in many previous calculations.
For illustration, we focus on the proton external state, and consider the contributions from the $N \pi$ and $\Delta \pi$ intermediate states.
The integrated expressions for the $N \to N \pi$ case for the full relativistic and approximated heavy baryon results are given in Eqs.~(\ref{eq:final-NpiN}) and (\ref{eq:NpiN-HB}), respectively.
In Fig.~\ref{fig:ProtonHBandRel}, we show the $N \pi$ and $\Delta \pi$ intermediate state contributions to the proton self-energy, both as a function of the regulator mass parameter $\Lambda$ and versus $m_\pi^2$, for the relativistic and heavy baryon calculations.

The heavy baryon results for the $N \pi$ contribution are $\approx 50\%$ larger in magnitude for the same value of the regulator $\Lambda \sim 1$~GeV, with the difference remaining relatively stable for varying $\Lambda$ and $m_\pi^2$.
This can be understood by considering the $1/M$ expansion of the relativistic expression in Eq.~(\ref{eq:final-NpiN}).
While Eq.~(\ref{eq:NpiN-HB}) is the leading order term in the heavy baryon approximation, the next to leading order correction in the expansion is positive and $\approx +0.15$~GeV in the chiral limit.
This amounts to $\approx 40\%$ of the magnitude of the leading term, which is $\approx -0.35$~GeV for the same regulator mass, and accounts for most of the difference between the relativistic and heavy baryon results.
The relativistic effects for the $\Delta \pi$ contribution are even more sizeable, with a reduction of the self-energy correction of $\sim 2/3$ from the heavy baryon result in the chiral limit.
These results suggest that relativistic effects in the baryon mass expansion can play a significant role in the self-energies.

\section{Conclusions}
\label{sec: conclusions}

In this paper we have, for the first time, evaluated the self-energies of all baryons in the octet and decuplet representations of flavor SU(3) within a relativistic chiral effective theory, and using a four-dimensional finite-range regulator with a dipole shape.
The use of the four-dimensional regulator ensures that the calculation preserves the necessary Lorentz, gauge and chiral symmetries of the fundamental QCD theory.
Furthermore, we derive the leading nonanalytic behavior of all the self-energies in the chiral limit, which provides an important consistency check for phenomenological model calculations~\cite{Thomas:1998df}.

We studied the dependence of the baryon self-energies numerically as a function of the regulator mass, $\Lambda$, and identified the most important channels for each baryon external state.
To allow for comparisons with lattice QCD data we also considered the dependence of the self-energies on the pion mass, illustrating the characteristic nonlinear behavior with $m_\pi^2$ near the chiral limit.

At larger pion masses, the SU(3) symmetry point, where all self-energies are equal, is identified, with small symmetry-breaking corrections arising from the use of physical baryon masses in the self-energy equations.
This illustrates the importance of accounting for all SU(3) octet and decuplet baryons in the intermediate states, without which the symmetry point would not be observed.
Comparison with the nonrelativistic or heavy baryon approximation for the $N \pi$ and $\Delta \pi$ contributions to the proton mass suggests that relativistic effects are significant, and reduce the magnitude of the (negative) correction by some 40\%--50\% compared with the heavy baryon result.

A natural future extension of our results will be to explore the applications of the self-energies to baryon masses and $\sigma$-terms.
In chiral effective theory, baryon masses can be expanded in powers of the quark mass, including contributions from the self-energies computed here.
Parameters of terms in the expansion that are analytical in the quark mass can be determined from fitting to lattice QCD data on baryon masses, but once these are determined the baryon masses can be used to derive $\sigma$-terms using the Feynman-Hellmann theorem,
    $\sigma_{{\cal B}q} = m_q\, \partial M_{\cal B}/\partial m_q$.
A detailed analysis of the $\sigma$-terms will be presented elsewhere~\cite{CJM-sigmaterms}.

Further applications involve coupling the meson--baryon system to external currents, such as photons, to study the effects of meson loops on electromagnetic elastic and transition form factors.
Moreover, coupling the meson--baryon states to nonlocal currents can provide information about the physical origin of the flavor asymmetries in parton distribution functions, as well as about sea quark contributions to generalized parton distributions.

\acknowledgments

We thank J.~J.~Ethier for helpful discussions and communications.
This work was supported by the US Department of Energy contract DE-AC05-06OR23177, under which Jefferson Science Associates, LLC operates Jefferson Lab and DOE Contract No. DE-FG02-03ER41260.

\appendix
\section{Coupling constants}
\label{app:couplings}

In this appendix, we summarize for convenience the full set of coupling constants for octet and decuplet baryon transitions to meson-baryon intermediate states, in terms of the couplings defined in the Lagrangian in Eq.~(\ref{eq:L}).
For transitions from octet baryon external states, the coupling constants $C_{BB'\phi}$ and $C_{BT'\phi}$ are given in Table~\ref{tab:octet}, while for decuplet baryon external states the coupling constants $C_{TT'\phi}$ and $C_{TB'\phi}$ are given in Table~\ref{tab:decuplet}.

\begin{table}[H]
\caption{Coupling constants $C_{BB'\phi}$ and $C_{BT'\phi}$ for transitions from octet external states $B$ to intermediate states with octet ($B'$) and ($T'$) decuplet baryons, respectively, and a pseudoscalar meson $\phi$, in terms of the couplings $D$, $F$ and ${\cal C}$ defined in Eq.~(\ref{eq:L}). \\}
\centering
\begin{tabular}{c|lr|lr|lr|lr}
\multicolumn{9}{c}{{\bf external\, state}\ \ $\bm{B}$}\\
\hline
\multicolumn{1}{c}{}
& \multicolumn{2}{c}{$\bm{p}$} 
& \multicolumn{2}{c}{$\bm{\Lambda}$}
& \multicolumn{2}{c}{$\bm{\Sigma^+}$}
& \multicolumn{2}{c}{$\bm{\Xi^0}$} 
\\
\hline 
\multirow{8}{*}{($\bm{B'\phi}$)} 
& ~($p\pi^0)$ & $\frac{(D+F)}{2}$~~
& ~($pK^-$) & $-\frac{(D+3F)}{\sqrt{12}}$~~ 
& ~($p\overline{K}^0$) & $\frac{(D-F)}{\sqrt2}$~~ 
& ~($\Sigma^0\overline{K}^0$) & $\frac{(D+F)}{2}$~~
\\ 
& ~($n\pi^+$) & $\frac{(D+F)}{\sqrt{2}}$~~ 
& ~($nK^0$) & $-\frac{(D+3F)}{\sqrt{12}}$~~
& ~($\Xi K^+$) & $\frac{(D+F)}{\sqrt{2}}$~~ 
& ~($\Sigma^+ K^-$) & $\frac{(D+F)}{\sqrt{2}}$~~ 
\\ 
& ~($\Sigma^+ K^0$)  & $\frac{(D-F)}{\sqrt{2}}$~~ 
& ~($\Sigma^+ \pi^-$) & $\frac{1}{\sqrt{3}} D$~~ 
& ~($\Sigma^+\pi^0$) & $F$~~ 
& ~($\Xi^0\pi^0$) & $\frac{(D-F)}{2}$~~
\\ 
& ~($\Sigma^0 K^+$) & $\frac{(D-F)}{2}$~~
& ~($\Sigma^0 \pi^0$) & $\frac{1}{\sqrt{3}} D$~~
& ~($\Sigma^0 \pi^+$) & $-F$~~ 
& ~($\Xi^- \pi^-$) & $\frac{(D-F)}{\sqrt{2}}$~~ 
\\ 
& ~($\Lambda K^+$) & $-\frac{(D+3F)}{\sqrt{12}}$~~
& ~($\Sigma^-\pi^-$) & $\frac{1}{\sqrt{3}} D$~~ 
& ~($\Sigma^+ \eta$) & $\frac{1}{\sqrt{3}} D$~~ 
& ~($\Xi^0 \eta$) & $-\frac{(D+3F)}{\sqrt{12}}$~~
\\
& ~($p\eta$) & $-\frac{(D-3F)}{\sqrt{12}}$~~ 
& ~($\Xi^0 K^0$) & $-\frac{(D-3F)}{\sqrt{12}}$~~ 
& ~($\Lambda \pi^+$) & $\frac{1}{\sqrt{3}} D$~~ 
& ~($\Lambda \overline{K}^0$) & $-\frac{(D-3F)}{\sqrt{12}}$~~ 
\\ 
&  &  
& ~($\Xi^- K^+$ & $-\frac{(D-3F)}{\sqrt{12}}$~~ 
&  &  
&  & 
\\ 
&  & 
& ~($\Lambda\eta$) & $\frac{1}{\sqrt3} D$~~
&  &  
&  & 
\\ 
\hline 
\multirow{6}{*}{($\bm{T'\phi}$)}
& ~($\Delta^{++}\pi^-$) & $\frac{1}{\sqrt2} {\cal C}$~~ 
& ~($\Sigma^{*+}\pi^-$) & $\frac12 {\cal C}$~~ 
& ~($\Delta^{++}K^-$) & $-\frac{1}{\sqrt2} {\cal C}$~~ 
& ~($\Sigma^{*+}K^-$) & $-\frac{1}{\sqrt6} {\cal C}$~~ 
\\ 
& ~($\Delta^+ \pi^{0}$) & $-\frac{1}{\sqrt3} {\cal C}$~~ 
& ~($\Sigma^{*-}\pi^+$) & $\frac12 {\cal C}$~~ 
& ~($\Delta^+ \overline{K}^0$) & $-\frac{1}{\sqrt6} {\cal C}$~~
& ~($\Sigma^{*0} \overline{K}^0$) & $-\frac{~1}{\sqrt{12}} {\cal C}$~~
\\ 
& ~($\Delta^0 \pi^+$) & $-\frac{1}{\sqrt6} {\cal C}$~~
& ~($\Sigma^{*0}\pi^0$) & $\frac12 {\cal C}$~~ 
& ~($\Sigma^{*+}\pi^0$) & $\frac{~1}{\sqrt{12}} {\cal C}$~~ 
& ~($\Xi^{*0} \pi^0$) & $-\frac{~1}{\sqrt{12}} {\cal C}$~~ 
\\
& ~($\Sigma^{*+} K^0$) & $\frac{1}{\sqrt6} {\cal C}$~~
& ~($\Xi^{*0}K^0$) & $-\frac12 {\cal C}$~~ 
& ~($\Sigma^{*0}\pi^+$) & $\frac{~1}{\sqrt{12}} {\cal C}$~~ 
& ~($\Xi^{*-} \pi^+$) & $\frac{1}{\sqrt6} {\cal C}$~~ 
\\ 
& ~($\Sigma^{*0} K^+$) & $-\frac{~1}{\sqrt{12}} {\cal C}$~~ 
& ~($\Xi^{*-}K^+$) & $\frac12 {\cal C}$~~ 
& ~($\Xi^{*0} K^+$) & $\frac{1}{\sqrt6} {\cal C}$~~ 
& ~($\Xi^{*0} \eta$) & $-\frac12 {\cal C}$~~ 
\\ 
&  &  
&  & 
& ~($\Sigma^{*+}\eta$) & $\frac12 {\cal C}$~~
& ~($\Omega^- K^+$) & $\frac{1}{\sqrt2} {\cal C}$~~
\\
\end{tabular}
\label{tab:octet}
\end{table}

\begin{table}[H]
\caption{Coupling constants $C_{TB'\phi}$ and $C_{TT'\phi}$ for transitions from decuplet external states $T$ to intermediate states with octet ($B'$) and ($T'$) decuplet baryons, respectively, and a pseudoscalar meson $\phi$, in terms of the couplings ${\cal C}$ and ${\cal H}$ defined in Eq.~(\ref{eq:L}). \\}
\centering
\begin{tabular}{r|lr|lr|lr|lr}
\multicolumn{9}{c}{{\bf external\, state}\ \ $\bm{T}$}\\
\hline
\multicolumn{1}{c}{} & 
\multicolumn{2}{c}{$\bm{\Delta^+}$}  & 
\multicolumn{2}{c}{$\bm{\Sigma^{*+}}$}& 
\multicolumn{2}{c}{$\bm{\Xi^{*0}}$} & 
\multicolumn{2}{c}{$\bm{\Omega^-}$} \\
\hline
\multirow{6}{*}{($\bm{B'\phi}$)} 
& ~($p\pi^0$)           & $\frac{1}{\sqrt{6}}{\cal C}$~~ 
& ~($p \overline{K}^0$) & $\frac{1}{\sqrt{6}}{\cal C}$~~ 
& ~($\Sigma^+ K^-$)     & $\frac{1}{\sqrt{6}}{\cal C}$~~ 
& ~($\Xi^0 K^-$)        & $-\frac{1}{\sqrt{2}}{\cal C}$~~ 
\\
& ~($n\pi^+$)           & $-\frac{1}{\sqrt{3}}{\cal C}$~~
& ~($\Xi^0 K^+$)        & $\frac{1}{\sqrt{6}}{\cal C}$~~
& ~($\Sigma^0 \overline{K}^0$) & $-\frac{~1}{\sqrt{12}}{\cal C}$~~
& ~($\Xi^- \overline{K}^0$) & $\frac{1}{\sqrt{2}}{\cal C}$~~
\\
& ~($\Sigma^+ K^0$)     & $-\frac{1}{\sqrt{6}}{\cal C}$~~
& ~($\Sigma^+ \pi^0$)   & $-\frac{~1}{\sqrt{12}}{\cal C}$~~ 
& ~($\Xi^{0}\pi^0$)     & $\frac{~1}{\sqrt{12}}{\cal C}$~~
&  & 
\\
& ~($\Sigma^0 K^+$)     & $\frac{1}{\sqrt{3}}{\cal C}$~~
& ~($\Sigma^0\pi^+$)    & $\frac{~1}{\sqrt{12}}{\cal C}$~~ 
& ~($\Xi^-\pi^+$)       & $-\frac{1}{\sqrt{6}}{\cal C}$~~
&  &  
\\
&  &  
& ~($\Sigma^+ \eta$)    & $\frac{1}{2}{\cal C}$~~
& ~($\Xi^0\eta$)        & $\frac{1}{2}{\cal C}$~~
&  &  
\\
&  &  
& ~($\Lambda \pi^+$)    & $-\frac{1}{2}{\cal C}$~~
& ~($\Lambda \overline{K}^0$) & $-\frac{1}{2}{\cal C}$~~
&  &  
\\ \hline
\multirow{6}{*}{($\bm{T'\phi}$)} 
& ~($\Delta^{++} \pi^-$)    & $\frac{1}{\sqrt{6}}{\cal H}$~~
& ~($\Delta^{++}K^-$)       & $\frac{1}{\sqrt{6}}{\cal H}$~~
& ~($\Xi^{*-}\pi^+$)        & $\frac{1}{\sqrt{18}}{\cal H}$~~
& ~($\Xi^{*-} \overline{K}^0$) & $\frac{1}{\sqrt{6}}{\cal H}$~~ 
\\
& ~($\Delta^{+}\pi^0$)      & $\frac{1}{6}{\cal H}$~~ 
& ~($\Delta^+\overline{K}^0$) & $\frac{1}{\sqrt{18}}{\cal H}$~~ 
& ~($\Xi^{*0}\pi^0$)        & $\frac{1}{6}{\cal H}$~~ 
& ~($\Xi^{*0} K^+$)         & $\frac{1}{\sqrt{6}}{\cal H}$~~ 
\\
& ~($\Delta^0\pi^+$)        & $\frac{\sqrt{2}}{3}{\cal H}$~~ 
& ~($\Xi^{*0} K^+$)         & $\frac{\sqrt{2}}{3}{\cal H}$~~ 
& ~($\Xi^{*0}\eta$)         & $-\frac{~1}{\sqrt{12}}{\cal H}$~~ 
& ~($\Omega^-\eta$)         & $-\frac{1}{\sqrt{3}}{\cal H}$~~ 
\\
& ~($\Delta^+ \eta$)        & $\frac{~1}{\sqrt{12}}{\cal H}$~~ 
& ~($\Sigma^{*+}\pi^0$)     & $\frac{1}{3}{\cal H}$~~ 
& ~($\Sigma^{*+}K^-$)       & $\frac{\sqrt{2}}{3}{\cal H}$~~ 
&  &  
\\
& ~($\Sigma^{*+} K^0$)      & $\frac{~1}{\sqrt{18}}{\cal H}$~~ 
& ~($\Sigma^{*0}\pi+$)      & $\frac{1}{3}{\cal H}$~~ 
& ~($\Sigma^{*0}\overline{K}^0$) & $\frac{1}{3}{\cal H}$~~ 
&  &  
\\ 
& ~($\Sigma^{*0} K^+$)      & $\frac{1}{3}{\cal H}$~~ 
& & 
& ~($\Omega^- K^+$)         & $\frac{1}{\sqrt{6}}{\cal H}$~~ 
&  &
\\ 
\end{tabular}
\label{tab:decuplet}
\end{table}

\section{Integral relations}
\label{app:integrations}

There are several useful integral relations involving propagators that occur in different expressions for self-energies that we summarize for convenience here.
The common integrals seen in the self-energy equations (\ref{eq:OctetOctetreuced}), (\ref{eq:octet-decuplet-reduced}), (\ref{eq:decuplet-octet-reduced}), and (\ref{eq:decuplet-decuplet-reduced}) generally contain some combination of the baryon propagator $D_{\cal B'}$, meson propagator $D_\phi$, and the $\Lambda$ propagator from the form factor $D_\Lambda$, where ${\cal B'}$ is understood to represent either an octet ($B'$) or decuplet ($T'$) baryon.
Terms containing both $D_{\cal B'}$ and $D_{\phi}$ or $D_\Lambda$ have poles in different half-planes, so that using Cauchy's integral formula one can choose a contour in either the upper or lower half-plane to perform the $k^-$ integration analytically.
Taking the pole in the baryon propagator, one can show that
\begin{equation}
\begin{aligned}
\label{eq:term1}
    &\int d^4k \frac{1}{D_\phi D_{\cal B'} D_\Lambda^4}
    = i\pi^2 \int dy\, dk_\perp^2  \frac{\bar{y}^4}{D_{{\cal B}\phi {\cal B'}}\, D_{{\cal B'}\Lambda {\cal B'}}^4}\, ,
\end{aligned} 
\end{equation}
and
\begin{equation}
\begin{aligned}
\label{eq:term2}
    &\int d^4k \frac{1}{D_{\cal B'} D_\Lambda^4} 
    = -i\pi^2 \int dy\, dk_\perp^2 \frac{\bar{y}^3}{D_{{\cal B}\Lambda {\cal B'}}^4}\, ,
\end{aligned} 
\end{equation}
where
\begin{equation}
\begin{aligned}
\label{eq:props at baryon pole}
    D_{{\cal B}\phi {\cal B'}}    &= k_\perp^2 - M_{\cal B}^2\, y\, \bar{y} + M_{\cal B'}^2\, y + m_\phi^2\, \bar{y}, \\
    D_{{\cal B}\Lambda {\cal B'}} &= k_\perp^2 - M_{\cal B}^2\, y\, \bar{y} + M_{\cal B'}^2\, y + \Lambda^2\, \bar{y}
\end{aligned}
\end{equation}
are the $\phi$ and $\Lambda$ propagators, respectively, taken at the $D_{\cal B'} = 0$ pole.

Terms with only $D_\phi$ and $D_\Lambda$ have poles on the same half-plane and the integral vanishes when $k^+ \neq 0$. When $k^+ = 0$, however, the integral is divergent and must be handled more carefully.
Using the Feynman parametrization, one can rewrite this term as
\begin{equation}
\begin{aligned}
\label{eq:DphiDLambdaSimp}
&\int\! d^4k\, \frac{1}{D_\phi D_\Lambda^4}
= \int\! d^4k\, \frac{1}{3!} \biggl(\frac{\partial}{\partial \Lambda^2}\biggr)^{\!3}
    \frac{1}{D_\phi D_\Lambda}
= \int\! d^4k\, \frac{1}{3!} \biggl(\frac{\partial}{\partial \Lambda^2}\biggr)^{\!3}
    \biggl(\frac{\partial}{\partial \Omega}\biggr)
    \int_0^1 \frac{dz}{(k^2- \Omega +i\epsilon)},
\end{aligned}
\end{equation}
where
\begin{equation}
\label{eq:Omega}
    \Omega = z\Lambda^2 + (1-z)m_\phi^2.
\end{equation}
Using the relation~\cite{Ji:2013}
\begin{equation}
    \int dk^- \frac{1}{k^2-\Omega+i\epsilon} 
    = 2\pi i \log{\biggl( \frac{k_\perp^2 + \Omega}{\mu^2} \biggr)}\, \delta(k^+),
\end{equation}
one can then reduce the integration in Eq.~(\ref{eq:DphiDLambdaSimp}) to yield
\begin{equation}
\begin{aligned}
\label{eq:DphiDL4}
\int d^4 k \frac{1}{D_\phi D_\Lambda^4}
& = -i\pi^2 \int dk^+dk_\perp^2 
    \int_0^1 dz \frac{z^3}{(k_\perp^2 + \Omega)^4}\delta(k^+).
\end{aligned}
\end{equation}
Since the techniques and steps are the same, one can also generalize these results for all powers of $D_\Lambda^n$, where $n > 1$. In this case by performing the $k^-$ integrations one obtains the relations
\begin{subequations}
\begin{eqnarray}
\int d^4k \frac{1}{D_\phi D_{\cal B'} D_\Lambda^n}
&=& i \pi^2 \int\!dy\, dk_\perp^2
    \frac{(-\bar{y})^n}{D_{{\cal B}\phi {\cal B'}} D_{{\cal B}\Lambda {\cal B'}}^n},   \\
\int d^4k \frac{1}{D_{\cal B} D_\Lambda^n}
&=& i \pi^2 \int\!dy\, dk_\perp^2 
    \frac{(-\bar{y})^{n-1}}{D_{{\cal B}\Lambda {\cal B'}}^n},            \\
\int d^4k \frac{1}{D_\phi D_\Lambda^n}
&=& i \pi^2 \int\!dk^+\, dk_\perp^2
    \int_0^1 dz \frac{(-z)^{n-1}}{(k_\perp^2 + \Omega)^n}\,
    \delta(k^+),                                      \\
\int d^4k \frac{1}{D_\Lambda^n}
&=& i \pi^2 \int\!dk^+\, dk_\perp^2
    \frac{(-1)^n}{(n-1)!}\frac{1}{(k_\perp^2 + \Lambda^2)^{n-1}}\,
    \delta(k^+).
\end{eqnarray}
\end{subequations}

\section{Example of decay rate derivation}
\label{decay-rate-derivation}

To demonstrate the relation between the decay rate and the imaginary part of the self-energy, we give here an explicit derivation for the decay rate of a spin-1/2 nucleon resonance to a nucleon and a pion, $N' \to N \pi$.
(A specific example of such a decay could be the Roper $N(1440)$ resonance.)
The invariant amplitude for this decay process is given by
\begin{equation}
    i{\cal M}  = \overline{u}(p_N, s_N) \frac{g_A}{2 f_\pi} \gamma^5 \slashed{p}_\pi \bm{\tau}\, u(p_{N'}, s_{N'}),
\end{equation}
where $u(p_{N'}, s_{N'})$ and $u(p_{N}, s_{N})$ are the spinors of the initial resonance and the nucleon, respectively, $p_\pi = p_{N'} - p_N$ is the pion momentum, and $\bm{\tau}$ represents the isospin matrices.
Computing the spin average of the amplitude squared $|{\cal M}|^2$, we get
\begin{equation}
\label{eq:amplitude2}
    \left< |{\cal M} |^2\right>  
    =  \frac{g_A^2}{4 f_\pi^2} \overline{M}_{NN'}^2 \big( \Delta_{N'N}^2-m_\pi^2 \big),
\end{equation}
where the mass difference between the resonance $N'$ and the nucleon $N$ is larger than the pion mass, $\Delta_{N'N} = M'-M > m_\pi$, which ensures that $\left< |{\cal M} |^2\right> > 0$. 
Since the decay rate is given by the phase space integration, 
\begin{equation}
\begin{aligned}
\label{eq: decay rate}
    \Gamma = &\frac{1}{2M'} \frac{1}{(2\pi)^2} 
    \int d^3 p_N \frac{1}{E_N} \int d^3 p_\pi 
    \frac{1}{2E_\pi}\, \delta^4(p_{N'} - p_{N} - p_\pi)
    \left< |{\cal M}|^2 \right>     \\
    = &\frac{1}{16 \pi M'^3}
    \sqrt{(M'^2+M^2-m_\pi^2)^2 - 4 M'^2 M^2}
    \left< |{\cal M}|^2 \right>,
\end{aligned}
\end{equation}
using Eq.~(\ref{eq:amplitude2}) we arrive at
\begin{equation}
\label{decayrate}
    \Gamma = 
    \frac{3 g_A^2}{64\pi f_\pi^2}
    \frac{\overline{M}_{\!NN'}^2}{M^3} \big( \Delta_{N'N}^2 - m_\pi^2 \big) \sqrt{(\overline{M}_{\!NN'}^2-m_\pi^2)(\Delta_{N'N}^2-m_\pi^2)}.
\end{equation}

To relate the decay rate with the imaginary part of the self-energy, we consider the self-energy of the octet $\to$ octet transition given by Eq.~(\ref{eq:OctetOctetreuced}), since the state $N'$ has spin 1/2.
Here we note that the imaginary contribution to the self-energy arises from the $1/(D_\phi D_{\cal B'} D_\Lambda^4)$ type term.
Specifically, the $k^-$ and $k_\perp$ integration of $1/(D_\phi D_{\cal B'} D_\Lambda^4)$ produces the logarithmic term,
\begin{equation}
\label{eq: negative Log}
    \int_0^1 dy \, \log(-\widetilde{D}),
\end{equation}
where
$\widetilde{D} = y\bar{y}\, M_{\cal B}^2 - y\, M_{\cal B'}^2 - \bar{y}\, m_\phi^2$.
It should be noted that terms in the self-energies proportional to just $1/(D_\phi D_\Lambda^4)$ or $1/(D_{\cal B'} D_\Lambda^4)$ cannot give imaginary contributions because the $k^-$ and $k_\perp$ integration produce no such negative logarithm term. 
This feature of each term with respect to the imaginary contribution is independent of the regularization method used for the loop calculation, and depends only on whether the mass difference $\Delta_{BB'}$ between the initial and intermediate baryons is larger or smaller than the mass of the intermediate meson $m_\phi$.
To see the characteristic of each term in Eq.~(\ref{eq:OctetOctetreuced}) more transparently, we take the point-like limit $\Lambda \to \infty$ of  
Eq.~(\ref{eq:OctetOctetreuced}) with 
$M_{\cal B}=M'$, $M_{\cal B'}=M$ and $m_\phi = m_\pi$, and obtain 
\begin{equation}
\label{RoperReduced}
\Sigma = 
    -i\frac{3 g_A^2}{8 f_\pi^2 M'}\int\! \frac{d^4p_\pi}{(2\pi)^4}
    \bigg[ 
        \frac{\overline{M}_{\!NN'}^2 \big( \Delta_{N'N}^2-m_\pi^2 \big)}{D_N D_\pi}
      + \frac{\overline{M}_{\!NN'}^2}{D_N}
      + \frac{\overline{M}_{\!NN'} \Delta_{N'N}}{D_\pi}
    \bigg],
\end{equation}
where the $p\cdot k/D_\pi$ term is odd under $k \leftrightarrow -k$ and vanishes after integration.
(Note also the sign flip 
    $-\Delta_{{\cal B'}{\cal B}} = \Delta_{{\cal B}{\cal B'}} = \Delta_{N'N}$
for the last term with respect to the corresponding last term in Eq.~(\ref{eq:OctetOctetreuced}) due to the notation correspondence.)
As the numerators in Eq.~(\ref{RoperReduced}) are now constants, for convenience we can use dimensional regularization to compute each of the terms individually.
It is easy to verify that neither the $1/D_N$ term nor the $1/D_\pi$ term can yield an imaginary part of $\Sigma$, while the $1/(D_N D_\pi)$ term provides the characteristic nonvanishing $\Im {\rm m} \Sigma$ when $\Delta_{N'N} > m_\pi$. 
In fact, from the dimensional regularization with $D=4-2\epsilon$ dimensions, we can write
\begin{eqnarray}
    \int \frac{d^4 p_\pi}{(2\pi)^4}\frac{1}{D_\pi}
    &=&\mu^{4-D}\int \frac{d^D p_\pi}{(2\pi)^D}\frac{1}{D_\pi} \nonumber \\
    &=&\frac{i m_\pi^2}{16 \pi^2}\left[\frac{1}{\epsilon}+1-\gamma-\log\frac{m_\pi^2}{\mu^2}+{\cal O}(\epsilon)\right],
\end{eqnarray}
where the factor $\mu^{4-D}$ is introduced to keep the dimension of the integral the same as in four dimensions.
Similarly, shifting the integration variable from $p_\pi$ to $p_N = p_{N'} - p_\pi$, we have
\begin{eqnarray}
    \int \frac{d^4 p_\pi}{(2\pi)^4}\frac{1}{D_N}
    &=&\mu^{4-D}\int \frac{d^D p_N}{(2\pi)^D}\frac{1}{D_N} \nonumber \\
    &=&\frac{i M^2}{16 \pi^2}\left[\frac{1}{\epsilon}+1-\gamma-\log\frac{M^2}{\mu^2}+{\cal O}(\epsilon)\right],
\end{eqnarray}
while the $1/(D_N D_\pi)$ term provides
\begin{eqnarray}
\label{DNDpiTerm}
    \int\frac{d^4p_\pi}{(2\pi)^4}\frac{1}{D_N D_\pi} 
    &=&\mu^{4-D}\int\frac{d^Dp_\pi}{(2\pi)^D}\frac{1}{D_N D_\pi} \nonumber \\
    &=&\frac{i}{16\pi^2}\bigg[\frac{1}{\epsilon}-\gamma-\log\pi-\int_0^1 dy\log\bigg(\frac{-D_{\text{cov}}}{\mu^2}\bigg)+{\cal O}(\epsilon)\bigg],
\end{eqnarray}
where $D_{\rm cov} = y\bar{y} M'^2 - y M^2 - \bar{y} m_\pi^2$ corresponds to $\widetilde{D}$ in Eq.~(\ref{eq: negative Log})
with the replacements
$M_{\cal B} \to M'$, $M_{\cal B'} \to M$ and $m_\phi \to m_\pi$.
This confirms that neither the $1/D_N$ term nor the $1/D_\pi$ term can yield an imaginary part of $\Sigma$, with the imaginary part coming from the $1/(D_N D_\pi)$ term for the region of $y$ integration in Eq.~(\ref{DNDpiTerm}) where $D_{\text{cov}}>0$. 
The condition for $D_{\text{cov}}>0$ is given by $y_{\rm min} < y < y_{\rm max}$, where 
\begin{equation}
y_{\rm min} = \frac{-\sqrt{(\overline{M}_{\!NN'}^2-m_\pi^2)
(\Delta_{N'N}^2-m_\pi^2)}  +M'^2-M^2 +m_\pi^2}{2M'^2}
\end{equation}
and
\begin{equation}
y_{\rm max} = \frac{\sqrt{(\overline{M}_{\!NN'}^2-m_\pi^2)
(\Delta_{N'N}^2-m_\pi^2)}  +M'^2-M^2 +m_\pi^2}{2M'^2}. 
\end{equation}
The imaginary part of $\Sigma$ therefore arises from the region 
of the $y$ integration for the $\log(-1) = i\pi$ term in Eq.~(\ref{DNDpiTerm}),
\begin{equation}
\label{eq: SE im cont Appended}
\int_{y_{\rm min}}^{y_{\rm max}} dy \, \log(-1) 
= i\pi \frac{\sqrt{(\overline{M}_{\!NN'}^2-m_\pi^2)
     (\Delta_{N'N}^2-m_\pi^2)}}{M'^2}.
\end{equation}
This contribution amounts to 
\begin{equation}
\begin{aligned}
\Im {\rm m}\, \Sigma 
= \frac{3g_A^2}{128 f_\phi^2} \frac{\overline{M}_{\!NN'}^2}{M'^3}
\big( m_\pi^2-\Delta_{N'N}^2 \big) \sqrt{(\overline{M}_{\!NN'}^2-m_\pi^2)(\Delta_{N'N}^2-m_\pi^2)}.
\end{aligned}
\end{equation}
Comparing this result with the decay rate given by Eq.~(\ref{decayrate}) confirms the relation
\begin{equation}
    \Gamma = -2\, \Im {\rm m}\, \Sigma.
\end{equation}
%


\end{document}